\documentclass[useAMS,usenatbib,aas_macros]{mn2e}
\usepackage{aas_macros}
\usepackage{amsmath}
\usepackage{graphicx}
\usepackage{color}

\newcommand{\equ}[1]{eq.~(\ref{eq:#1})}

\newcommand{\se}[1]{\S\ref{sec:#1}}
\newcommand{\fig}[1]{Fig.~\ref{fig:#1}}
\newcommand{\figs}[1]{Figs.~\ref{fig:#1}}
\newcommand{\Fig}[1]{Figure~\ref{fig:#1}}
\newcommand{\Figs}[1]{Figures~\ref{fig:#1}}
\newcommand{\tab}[1]{Table~\ref{tab:#1}}
\newcommand{\be}{\begin{equation}}
\newcommand{\ee}{\end{equation}}
\newcommand{\ba}{\begin{align}}
\newcommand{\ea}{\end{align}}
\newcommand{\bad}{\begin{equation} \begin{aligned}}
\newcommand{\ead}{\end{aligned} \end{equation}}
\newcommand{\bea}{\begin{eqnarray}}
\newcommand{\eea}{\end{eqnarray}}
\def\ra{\rangle}
\def\la{\langle}

\def\ssim{\!\sim\!}
\def\seq{\!=\!}
\def\ssimeq{\!\simeq\!}

\def\sgt{\!>\!}
\def\slt{\!<\!}

\def\sdash{\!-\!}

\newcommand{\msun}{M_\odot}
\newcommand{\Msun}{M_\odot}

\newcommand{\ifm}[1]{\relax\ifmmode#1\else$\mathsurround=0pt #1$\fi}
\newcommand{\kms}{\ifmmode\,{\rm km}\,{\rm s}^{-1}\else km$\,$s$^{-1}$\fi}

\newcommand{\Mpc}{\,{\rm Mpc}}
\newcommand{\kpc}{\,{\rm kpc}}
\newcommand{\pc}{\,{\rm pc}}

\newcommand{\Gyr}{\,{\rm Gyr}}

\newcommand{\Myr}{\,{\rm Myr}}

\newcommand{\cmc}{\,{\rm cm}^{-3}}
\newcommand{\cms}{\,{\rm cm}^{-2}}

\newcommand{\ltsima}{$\; \buildrel < \over \sim \;$}
\newcommand{\lsim}{\lower.5ex\hbox{\ltsima}}
\newcommand{\gtsima}{$\; \buildrel > \over \sim \;$}
\newcommand{\gsim}{\lower.5ex\hbox{\gtsima}}
\newcommand{\prop}{\propto}
\newcommand{\dd}{\rm d}

\newcommand{\rar}{\rightarrow}

\def\omm{\Omega_{\rm m}}
\def\oml{\Omega_{\Lambda}}
\def\omb{\Omega_{\rm b}}

\def\Rv{R_{\rm v}}
\def\Vv{V_{\rm v}}

\def\Mg{M_{\rm g}}
\def\Ms{M_{\rm s}}
\def\Re{R_{\rm e}}

\def\Sig1{\Sigma_1}

\def\Rd{R_{\rm d}}
\def\fg{f_{\rm g}}

\def\Mc{M_{\rm c}}
\def\Rc{R_{\rm c}}

\def\Vc{V_{\rm c}}
\def\Vrot{V_{\rm rot}}

\def\Ri{R_{\rm i}}

\def\Rf{R_{\rm f}}

\def\Ki{K_{\rm i}}
\def\Kf{K_{\rm f}}

\def\brho{\bar\rho}

\def\brhov{\bar\rho_{\rm v}}

\def\urho{\rho_{\rm u}}

\def\Rd{R_{\rm d}}

\def\fg{f_{\rm g}}

\def\rhog{\rho_{\rm g}}

\def\eps2{\epsilon_{-2}}

\def\tv{t_{\rm v}}
\def\Mv{M_{\rm v}}

\def\M11{M_{\rm v,11}}

\def\fb{f_{\rm b}}
\def\f16b{f_{\rm b,0.16}}
\def\sv25{\phi_{11,-2.5}}

\def\mv{m_{\rm v}}
\def\mc{m_{\rm c}}
\def\ms{m_{\rm s}}

\def\mt{m_{\rm t}}
\def\ft{f_{\rm t}}

\def\ellv{\ell_{\rm v}}
\def\bsig{\bar{\rho}_{\bf sat}}

\def\bsigc{\bar{\rho}_{\rm sat,c}}
\def\tvir{t_{\rm v}}
\def\tdyn{t_{\rm dyn}}
\def\Kc{K_{\rm c}}
\def\Wc{W_{\rm c}}
\def\Ws{W_{\rm c}}
\def\xc{x_{\rm c}}
\def\ch{c_{\rm 2}}
\def\Rh{R_{\rm h}}

\def\ch{c_{\rm h}}
\def\ah{\alpha_{\rm h}}
\def\c2h{c_{2{\rm h}}}
\def\s1h{s_{1{\rm h}}}

\def\Rm{R_{\rm max}}
\def\fdm{f_{\rm dm}}

\title[Cores by dynamical friction and AGN]
{Core Formation in High-z Massive Haloes: 
Heating by Post Compaction Satellites and Response to AGN Outflows}

\author[Dekel et al.]
{\parbox[t]{\textwidth}
{Avishai Dekel$^{1,2}$\thanks{E-mail: dekel@huji.ac.il},
Jonathan Freundlich$^{1,3}$,
Fangzhou Jiang$^{1,4,5}$,
Sharon Lapiner$^{1}$,
Andreas Burkert$^{6}$,
Daniel Ceverino$^{7}$,
Xiaolong Du$^{5}$,
Reinhard Genzel$^{8}$,
Joel Primack$^{9}$
}
\\ \\
$^1$Racah Institute of Physics, The Hebrew University, Jerusalem 91904 Israel\\
$^2$SCIPP, University of California, Santa Cruz, CA 95064, USA\\  
$^3$Universit\'e de Strasbourg, CNRS, Observatoire astronomique de Strasbourg, 
UMR 7550, F-67000 Strasbourg, France\\ 
$^4$TAPIR, California Institute of Technology, Pasadena, CA 91125, USA\\ 
$^5$Carnegie Observatories, 813 Santa Barbara Street, Pasadena, CA 91101, USA\\ 
$^6$Ludwig-Maximilians Universitat Munchen, Department fur Physik,
Scheinerstr. 1, D-81679 Munchen, Germany\\ 
$^7$Departamento de Fisica Teorica, Facultad de Ciencias, Universidad Astronomia
de Madrid, Cantoblanco, 28049 Madrid, Spain\\
$^8$Max Planck Institute for Extraterrestrial Physics,
Giessenbachstrasse 1, 85738 Garching, Germany\\
$^9$Physics Department, University of California, Santa Cruz, Santa Cruz, CA
95064, USA
}

\begin{document}

\large

\pagerange{\pageref{firstpage}--\pageref{lastpage}} \pubyear{2002}

\maketitle

\label{firstpage}

\begin{abstract}
Observed rotation curves in star-forming galaxies indicate a puzzling dearth of
dark matter in extended flat cores within haloes of mass $\geq\!10^{12}\msun$ 
at $z\ssim 2$. %$z\seq 0.65\sdash 2.5$. 
This is not reproduced by current cosmological simulations,
and supernova-driven outflows are not effective in such massive
haloes. % 41 
We address a hybrid scenario where post-compaction merging satellites heat 
up the dark-matter cusps by dynamical friction, allowing AGN-driven outflows 
to generate cores.  Using analytic and semi-analytic models (SatGen), we 
estimate the dynamical-friction heating as a function of satellite compactness 
for a cosmological sequence of mergers. % 48
Cosmological simulations (VELA) demonstrate that satellites of initial  
virial masses $>\! 10^{11.3}\msun$, that undergo wet compactions,
become sufficiently compact for significant heating.
Constituting a major fraction of the accretion onto haloes  
$\geq\!10^{12}\msun$, these satellites heat-up the cusps
in half a virial time at $z\ssim 2$. %43
Using a model for outflow-driven core formation (CuspCore),
we demonstrate that the heated dark-matter cusps develop extended cores in 
response to removal of half the gas mass, 
while the more compact stellar systems remain intact. 
The mergers keep the dark matter hot,
while the gas supply, fresh and recycled, is sufficient for the AGN outflows. 
AGN indeed become effective in haloes $\geq\!10^{12}\msun$,
where the black-hole growth is no longer suppressed by supernovae
and its compaction-driven rapid growth is maintained by a hot CGM. %83
For simulations to reproduce the dynamical-friction effects, they should 
resolve the compaction of the massive satellites and avoid 
artificial tidal disruption.  
AGN feedback could be boosted by clumpy black-hole accretion  
and clumpy response to AGN. % 35
\end{abstract}


%%%%%%%%%%%%%%%%%%%%%%%%%%% 1
\section{Introduction}
\label{sec:intro}

%Cusps in sims.
As seen in cosmological gravitating N-body simulations with no baryons,
dark-matter (DM) haloes robustly produce cuspy 
inner density radial profiles with a central
negative log slope $\alpha\! \sim\! 1$ \citep[][NFW]{nfw97}. 
%Observed cores in low mass haloes.  SN feedback
In contrast, dwarf galaxies are observed kinematically to have inner flat
DM cores,
$\alpha\! \sim\! 0$ 
\citep[e.g.][]{flores94,burkert95,deblok01,deblok08, oh11, oh11_sim, oh15}.
This is commonly understood both analytically and in simulations in terms of 
bursty
supernova (SN) feedback 
\citep[e.g.][]{pontzen12,dutton16b,freundlich20_cuspcore},
which is capable of effectively ejecting the gas from the central regions
in haloes below a critical halo virial velocity of $\Vv\!\sim\!100\kms$, 
where the supernovae energy deposited in the ISM is comparable to the binding 
energy of the gas in the central halo potential well \citep{ds86}. 
The inner DM halo responds to the shallowing of the gravitational potential 
well due to the central mass loss by expanding and producing a flat core. 
%\adr{elaborate on models?}

\smallskip
%Observed extended cores in massive galaxies at high z above SN scale.
Very surprisingly, pioneering kinematic {\it observations} 
by \citet{genzel21},
% wyts16, genzel17 xx
of 41 massive star-forming disc galaxies at $z\seq 0.65 \sdash 2.5$,  
indicate the occurance of low central DM fractions ($\fdm$)
and DM cores that extend beyond the stellar effective radii 
to $\sim\! 10\kpc$. 
The data include near-infrared observations with SINFONI and KMOS at the
ESO-VLT, as well as LBT-LUCI, and sub-millimeter observations from IRAM-NOEMA.
The kinematics is determined from ionized gas traced by H$\alpha$ and molecular
gas traced by CO.
The baryons in the selected galaxies are rotationally supported 
($\Vrot/\sigma\sgt 2.3$),
lying in the Main Sequence of star-forming galaxies (SFGs), with stellar masses 
$\Ms\seq 10^{9.8-11.4}\msun$. 
There is a correlation between low $\fdm$ and a high bulge-to-disc ratio.
By properly inverting the relations based on abundance matching of observed
galaxies and simulated $\Lambda$CDM DM haloes 
\citep{behroozi13,moster18}, % xx 
the corresponding halo virial mass range is $\Mv\seq 10^{11.6-12.9}\msun$. 
% From Fig. 7
In the high part of the redshift range, $z\!\sim\!1.2\sdash 2.5$, 
more than two thirds of the galaxies show low DM fractions within the effective
radius ($\fdm \slt 0.3$), of which about one half are indicated to
have extended cores. 
The cores tend to appear in this sample for $\Mv\!\geq\!10^{12}\msun$, 
with the median at $\Mv\seq 10^{12.4}\msun$, while less massive galaxies
tend to have higher DM fractions.
At the lower redshift band, $z\!\sim\!0.65\sdash 1.2$, 
the portion of low DM fraction galaxies is reduced to about a third, 
but their vast majority are with extended cores.
At these redshifts, the DM fractions are typically not as low as at higher 
redshifts, and the mass dependence of this phenomenon is clearer, with
a few galaxies of high DM fractions below $10^{12}\msun$.
This adds to earlier indications for low central DM fractions in massive
star-forming disks \citep[e.g.][]{wuyts16}.
The analysis is being extended to 100 galaxies 
(Price et al. 2021, in preparation), strengthening the appearance of
DM-deficient extended cores especially in massive SFGs at $z\ssim 2$.
The appearance of low DM fractions in extended cores is very surprising 
because in such massive 
haloes, and at such redshifts, the gravitational potential is likely
to be too deep for SN feedback to effectively remove central gas for
gravitationally pushing the DM out and generating cores \citep{ds86}.

\smallskip
%\adr{elaborate on Genzel's analysis: strengths and caveats.}
Beyond the unavoidable observational uncertainties,
the non-trivial dynamical analysis of the observed rotation curves, 
which reveals the apparent low DM fractions and extended cores, 
is based on various assumptions and is thus associated with further
uncertainties.
These include, for example, the assumption of an exponential disk and an
uncertain correction term to the centripetal force by a pressure gradient 
\citep[``asymmetric drift"][]{burkert10}. Furthermore,
a cuspy NFW profile is assumed in the fit of a DM halo to the
data, and when the corresponding central DM fraction conflicts with the 
observations, the conclusion is that the assumed NFW cusp should be replaced 
by a core. These uncertainties are expected to be reduced in a 
subsequent analysis, e.g., by using a more flexible model profile that allows 
a core.
However, the uncertainties that have been carefully considered so far 
do not seem to systematically
weaken the general need for a dearth of central DM and extended cores. 
Thus, despite the uncertainties, the intriguing implications of the pioneering 
results call for a special theoretical effort to explore potential processes 
that may push the DM out and generate cores in massive galaxies, while the
stellar systems remain intact.

\smallskip % sharma21
One should mention that
\citet{sharma21} report from the KROSS survey somewhat higher $fdm$ values, 
but in fact their results are largely consistent with \citet{genzel21}.
They obtain higher DM fractions because because of multiple reasons:
(a) they refer to larger radii of $(1.9 \sdash 3)\Re$ as opposed to $\Re$,
(b) their KROSS sample is at $z \ssim 1$, while the low $\fdm$ values 
are observed by \citet{genzel21} mostly at $z \ssim 2$, with a typical
reduction of $0.2\sdash 0.3$ between these redshifts, and
(c) their galaxies are typically of lower masses than the galaxies 
of \citet{genzel21}, while it is shown in Price et al. (2021, in preparation) 
that $\fdm$ is strongly decreasing with mass.   

\smallskip % failure in simulations
Several different mechanisms could in principle work toward generating 
cores in massive galaxies.
These include, for example,
mergers of SN-driven cored building blocks \citep{dekel03},
dynamical friction (DF) by the DM on merging satellites \citep{elzant01}
% tonini06, romano-diaz08, goerdt10 xx
and AGN-driven outflows \citep[e.g.][]{martizzi12,peirani17}, 
which may work in massive galaxies in analogy to supernova-driven outflows 
that generate cores in low mass galaxies.
The puzzling issue is that none of the currently available cosmological 
simulations seem to give rise to such low central DM fractions and 
extended cores in massive haloes, despite the fact that all the above mentioned
physical elements, including mergers, dynamical friction, supernovae and AGNs, 
are supposed to be incorporated in the simulations.
Examples showing cusps in massive galaxies are in the TNG simulation 
\citep{wang20} and in the FIRE-2 simulations \citep{lazar20}.

\smallskip % AGN feedback does not work  
The effect of {\it AGN feedback} on the DM profile in the central regions of 
clusters of galaxies has been studied using cosmological simulations
\citep[e.g.][]{martizzi12} 
and isolated-halo simulations \citep{martizzi13}.
It has been found that the repeating episodes of gas ejection and recycling 
can indeed generate a core of radius $\sim\!10\kpc$,
but with the simulated halo masses of $10^{14}\msun$ and 
$1.4\times10^{13}\msun$ in these studies, the virial radii are 
$\sim\!1.2\Mpc$ and $\sim\!0.6\Mpc$ respectively, 
so the core radii extend only to $1\sdash 2\%$ of the virial radii
as opposed to the desired $\sim\!10\%$.
No cores, or cores of limited extent, have also been found in massive galaxies 
in the Horizon-AGN simulation \citep{peirani17,peirani19} and in the NIHAO-AGN 
simulations \citep{maccio20}, both including AGN feedback. 
A preliminary inspection of the haloes of massive galaxies in the
TNG50 cosmological simulation, 
where a relatively strong AGN feedback is implemented \citep{weinberger18}, 
indicates cores that extend to only  
$1\sdash 2\%$ of the virial radius
(Sandro Tacchella, private communication).
% xx peirani17 H-AGN  DM flattened at z=3-1.6
%
It seems that AGN feedback in cosmological simulations, as implemented so far,
fails to reproduce cores of $\sim\!0.1\Rv$ extent as deduced from observations
by \citet{genzel21}.
Potential ways to remedy this could be to have the DM halo pre-heated
and thus be closer to escape before the mass ejection,
and/or to boost the strength of the AGN feedback itself and tighten its 
coupling with the gas in the inner halo.  
%\adr{Try analytic estimate}.

\smallskip % DF
%DF alone does not work (sims. El Zant).
The effect of {\it dynamical friction} on the cusps of massive haloes has been
discussed by \citet{elzant01}. They used semi-analytic simulations
with hundreds of satellites that sum up to $10\%$ of the host mass of
$10^{12}\msun$ during $2\sdash 3\Gyr$, which is roughly a halo virial time at
$z\seq 0$. 
This led to almost-flat cores that extend to $(4\sdash 6)\kpc$, 
which at $z\seq0$, with $\Rv\ssim 300\kpc$, corresponds to only 
$(0.01\sdash 0.02)\Rv$. 
In \citet{elzant04}, they used N-body simulations with multiple satellites that
amount to $3\sdash 20\%$ of the host-halo mass. Semi-cores of log slope $-0.35$
that extend out to $\Rc \ssim 0.06\Rv$ were generated in $\sim\!6\Gyr$.
However, in both cases, 
as well as in \citet{elzant08}, where a universal profile is maintained under
satellite sinkage,
the models assumed point-mass satellites and ignored 
their tidal stripping, thus overestimating the effect of DF on the host cusp.
While this has been a stimulating proof of concept, 
the situation studied is not a representative cosmological population of
satellites that could reproduce the realistic effect of DF on the host cusps.
The impression is that DF by itself is not likely to generate flat cores as 
extended as observed.

\smallskip % DF
%\adr{heating vs profile change: Burkert vs El Zant}
Clearly, efficient satellite penetration that may result in DM heating requires
that the incoming {\it satellites} 
would be massive, compact, and on relatively radial orbits,
such that tidal stripping would be minimal prior to reaching the host central 
regions where the DF is to increase the energy and angular momentum of the 
DM cusp.
The questions to be verified are whether the realistic satellites that merge 
with the host haloes of the relevant masses at the relevant
redshifts are sufficiently massive, compact and on proper orbits for effective 
DF heating or flattening of the host DM cusps.
It is also interesting to find out how the deposited energy is divided
between heating the cusp and partially flattening the profile into a core.

\smallskip
%A combination of DF and outflows 
We study here a promising {\it two-staged scenario}
where DF heating or flattening and AGN feedback are both boosted, 
and they are put together in a hybrid 
scenario where the DF that acts on compact merging satellites heats 
up or flattens the inner DM halo, making it more susceptible to expansion 
under clumpy AGN-driven outflows.

\smallskip % cold cusp
We note that standard NFW cusps are kinematically cold, as Jeans equilibrium
implies that the (isotropic) velocity dispersion profile is decreasing 
toward the center like $\sigma^2 \!\propto\! r$, while the gravitational
potential gets deeper. This means that the dark matter is
more tightly bound in the inner cusp, which implies that it would be hard 
to unbind it by central mass ejection. This is why kinematically 
heating the DM particles by DF prior to the mass ejection, 
bringing them closer to the escape velocity from the cusp, 
could be a key to DM expansion and core formation.

\smallskip  % BN threshold mass
We demonstrate below that
in such a scenario the observed preferred host-halo masses and 
preferred redshifts for DM cores are natural outcomes.
% mass
We note that for host haloes of $\Mv\! \geq\! 10^{12}\msun$ 
at $z\!\sim\!1 \sdash 3$, the likely massive merging satellites are with  
haloes of $\mv\! \geq\! 10^{11.3}\msun$, 
which are above the mass threshold for drastic compaction events into
blue nuggets
\citep{zolotov15,tomassetti16,tacchella16_prof,tacchella16_ms}.
This {\it golden mass} \citep{dekel19_gold} is determined by the combined 
effect of supernova feedback at 
lower masses \citep{ds86} and virial shock heating of the circum-galactic
medium (CGM) aided by AGN feedback at higher masses \citep{bd03,db06}.
The satellites above the golden mass are therefore likely to be more 
{\it compact}
than lower-mass satellites that typically feed lower-mass hosts. 
These compact satellites could be optimal for deep penetration into the host 
cusp with less tidal stripping, allowing effective DF heating or flattening
in the host cusp.
The highly compact baryonic cusp of the satellite should allow the buildup of
a central stellar system that would remain intact under the AGN-driven 
gas outflows that push the less-concentrated DM outward.
%AGN mass
AGN feedback is indeed expected to be most effective above the golden mass
of $\Mv\ssim 10^{12}\msun$, 
where the black-hole (BH) growth is no longer suppressed by supernova feedback
\citep{dubois15},
the BH growth is made possible by the hot-CGM
\citep{bower17,angles17},
and the rapid BH growth is triggered by a wet-compaction event
\citep{dekel19_gold,lapiner21}.
This is confirmed by observations of luminous AGN fractions above the golden
mass \citep{forster19}
% bulge
The preferred appearance of massive bulges in low $\fdm$ galaxies  
is indeed consistent with the association of cores with the bulge-forming 
compaction events that tend to occur above the golden mass.

\smallskip % z dependence
The preferred high redshift for DM cores 
could naturally emerge from the higher merger rate,
higher gas fraction, and therefore more effective compaction events at higher 
redshifts. Furthermore, at higher redshifts, the dark-matter cusps are 
maintained "hot" under the dynamical friction by a cosmological sequence of 
satellites, and are thus more susceptible to core formation by AGN outflows, 
while at lower redshifts the cusps cool and expand and thus make the AGN 
outflows less effective.

\smallskip % Burkert+
A companion paper (Burkert et al. 2021) addresses the same
problem of DM core formation emphasizing somewhat different angles and a 
similar hybrid scenario of DF heating followed by response to outflows.
This version of the scenario appeals instead to DF heating by 
inward migrating clumps in violently unstable discs, and it bases the 
feasibility tests of the two parts of the scenario on simple  
N-body simulations tailored for the purpose.  

\smallskip % outline
The paper is organized as follows.
In \se{vela} we derive from the VELA simulations the fiducial mass profiles 
of pre-compaction and post-compaction satellites below and above the golden 
mass.
In \se{toy} we use a toy model to make analytic estimates of the DF heating 
as a function of satellite compactness.
In \se{satgen} we utilize SatGen semi-analytic simulations
to explore the DF heating as a function of satellite compactness and orbit,
as well as for a cosmological sequence of satellites.
In \se{relax} we address the post-heating relaxation of the cusp, 
cooling and expansion, and use simplified N-body simulations to study 
the interplay between DF heating and density flattening,
and evaluate the cooling timescale.
In \se{agn} we use the CuspCore analytic model to study the response of the
pre-heated haloes to gas ejection by AGN feedback.
In \se{disc} we discuss our results and refer to the prospects of reproducing
them in cosmological simulations.
In \se{conc}, we summarize our conclusions.
In appendices A \sdash J (available as supplementary material online),
we bring supportive descriptions of the analytic models and simulations used,
and add figures relevant to cases that are complementary to our fiducial 
cases.

%%%%%%%%%%%%%%%%%%%%%%%%%%%%%%%%%%%%% 2
\section{Compact Satellites above a Threshold Mass in Cosmological Simulations}
\label{sec:vela}

\begin{figure*} % 1
\centering
\includegraphics[width=1.00\textwidth]
%{figs/MOMv_vs_rORv_profiles_1x4panels_v9.pdf}
{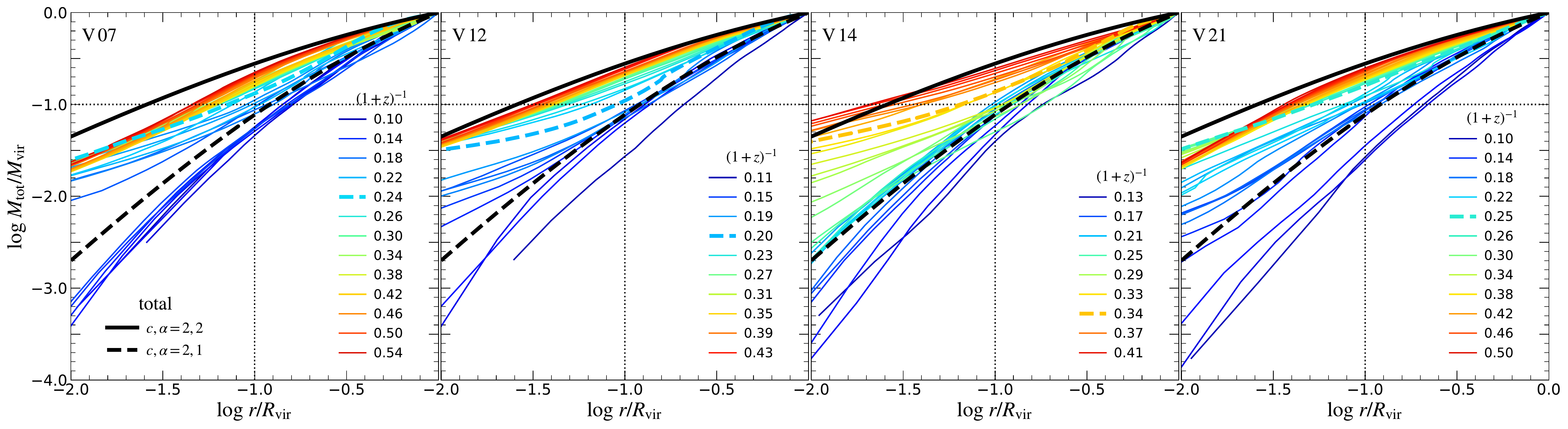}
\includegraphics[width=1.00\textwidth]
%{figs/MdmOMv_vs_rORv_profiles_1x4panels_v9.pdf}
{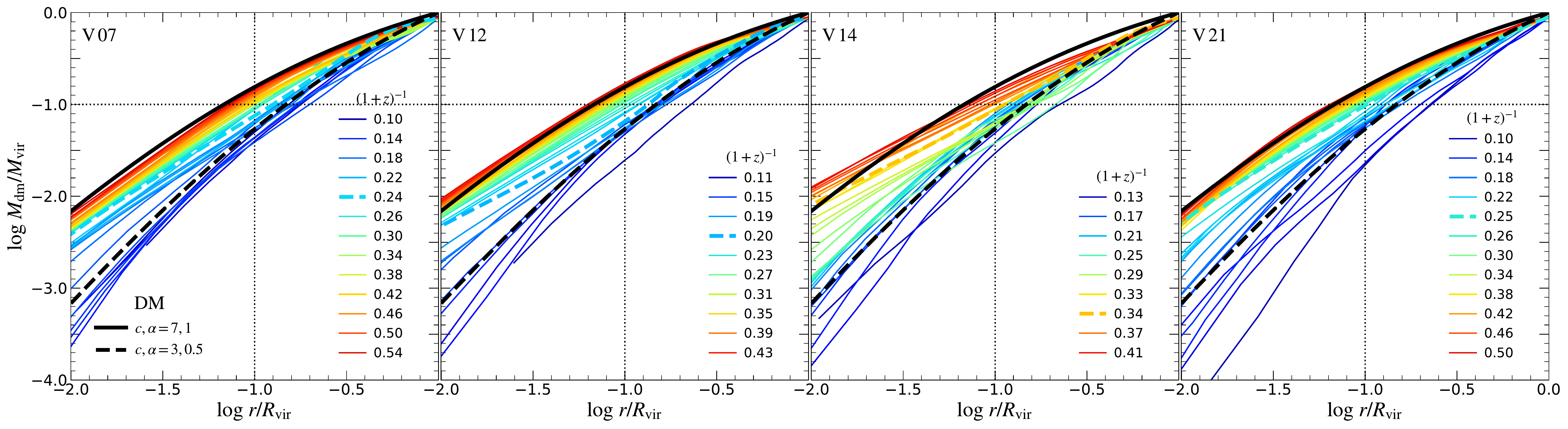}
\caption{
Total mass profiles for representative VELA galaxies at a sequence of times,
before and after the major compaction events.
{\bf Top:} for the total mass of dark matter, stars and gas.
{\bf Bottom:} for the dark matter alone.
Marked are the corresponding expansion factors $a \seq (1+z)^{-1}$,
separated by $\Delta a = 0.03$ (which roughly corresponds to
$\Delta t \ssim 400 \Myr$ at $a \seq 0.25$), with the blue-nugget
highlighted by a thick dashed curve.
We see universal shapes of the profiles pre-compaction and
post-compaction, with a significant increase in global compactness during the
compaction process, both for the total mass and for the DM alone.
The black curves mark our fiducial DZ fits for the pre-compaction (dashed)
and post-compaction (solid) profiles, as deduced from \fig{vela_fits}.
We conservatively adopt the DM profiles for our merging satellites.
}
\label{fig:vela_prof}
\end{figure*}

\begin{figure*} % 2
\centering
\hskip -0.1cm
\includegraphics[width=0.245\textwidth,trim={0.6cm 0.4cm 0.6cm 0.6cm},clip]
%{figs/vela_aBN_cparam_all_dm.pdf}
{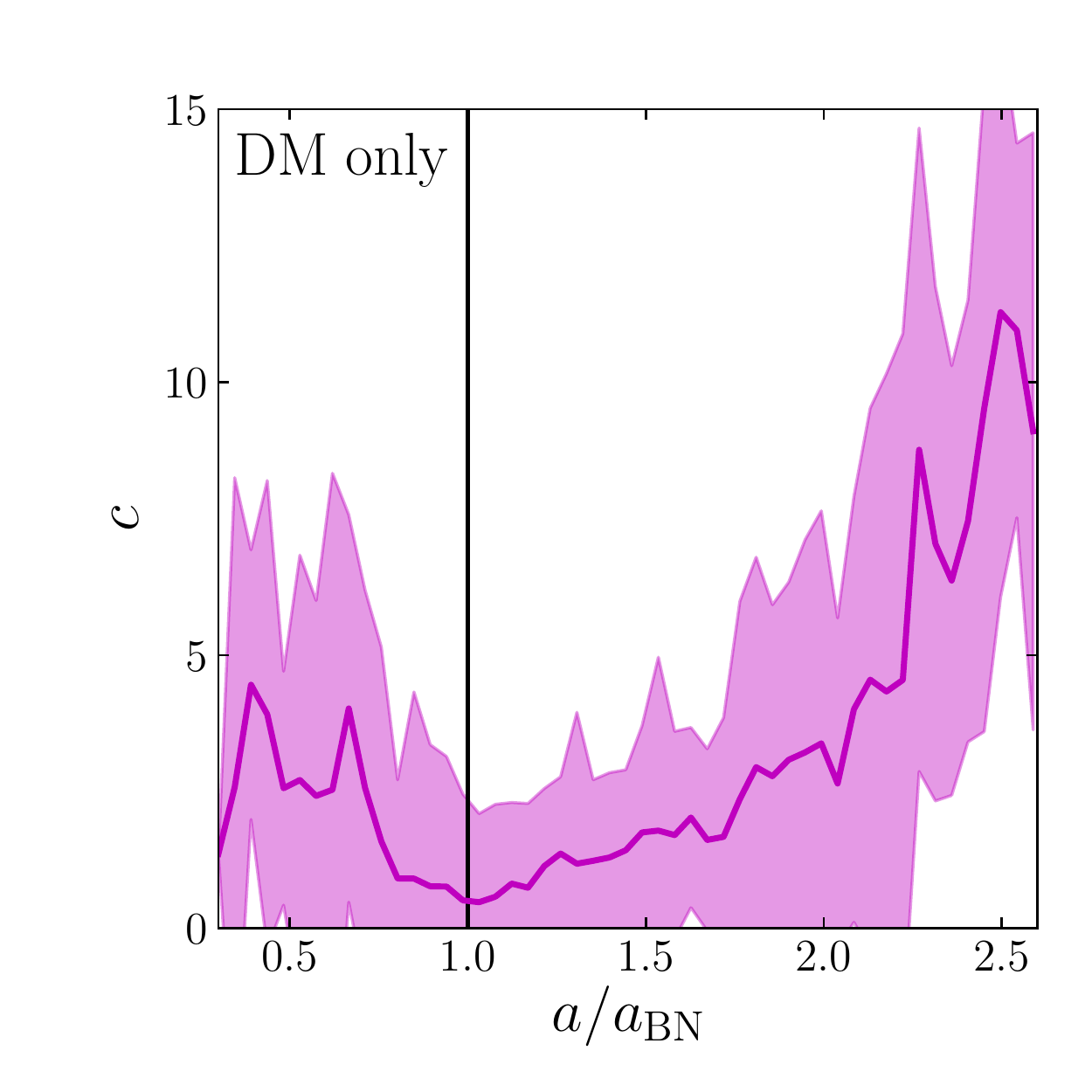}
\includegraphics[width=0.245\textwidth,trim={0.6cm 0.4cm 0.6cm 0.6cm},clip]
%{figs/vela_aBN_aparam_all_dm.pdf}
{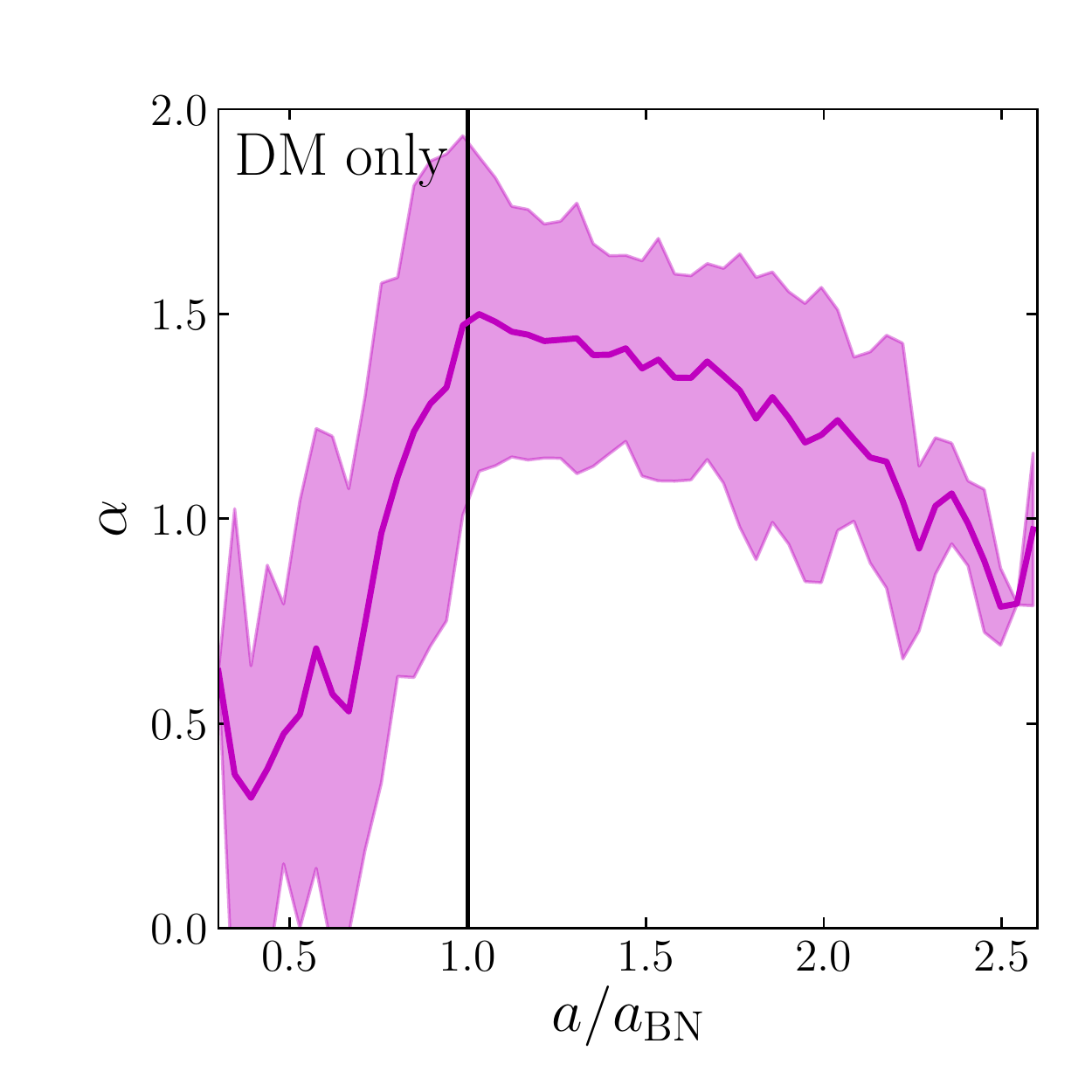}
\hskip 0.1cm
\includegraphics[width=0.245\textwidth,trim={0.6cm 0.4cm 0.6cm 0.6cm},clip]
%{figs/vela_aBN_cparam_all_tot.pdf}
{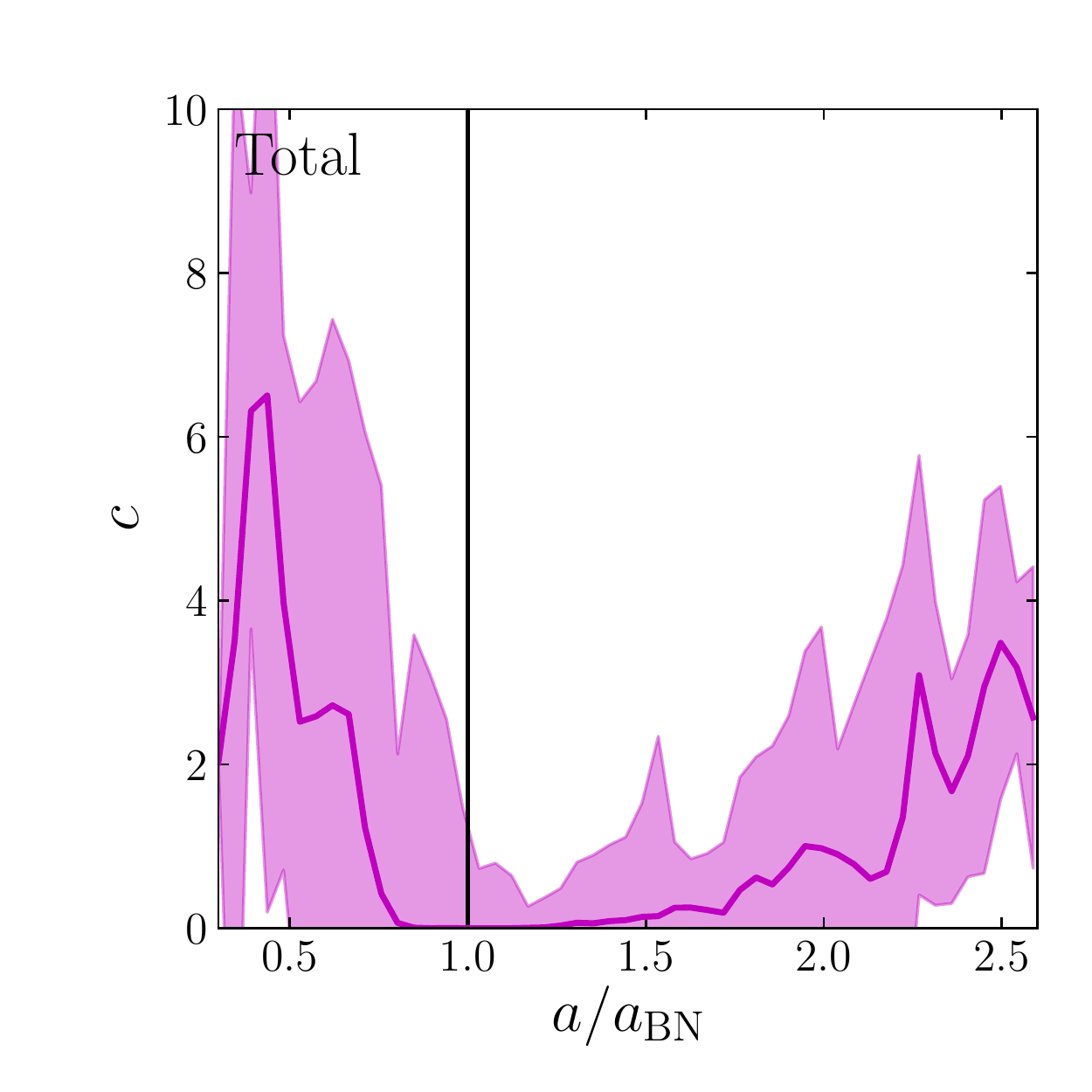}
\includegraphics[width=0.245\textwidth,trim={0.6cm 0.4cm 0.6cm 0.6cm},clip]
%{figs/vela_aBN_aparam_all_tot.pdf}
{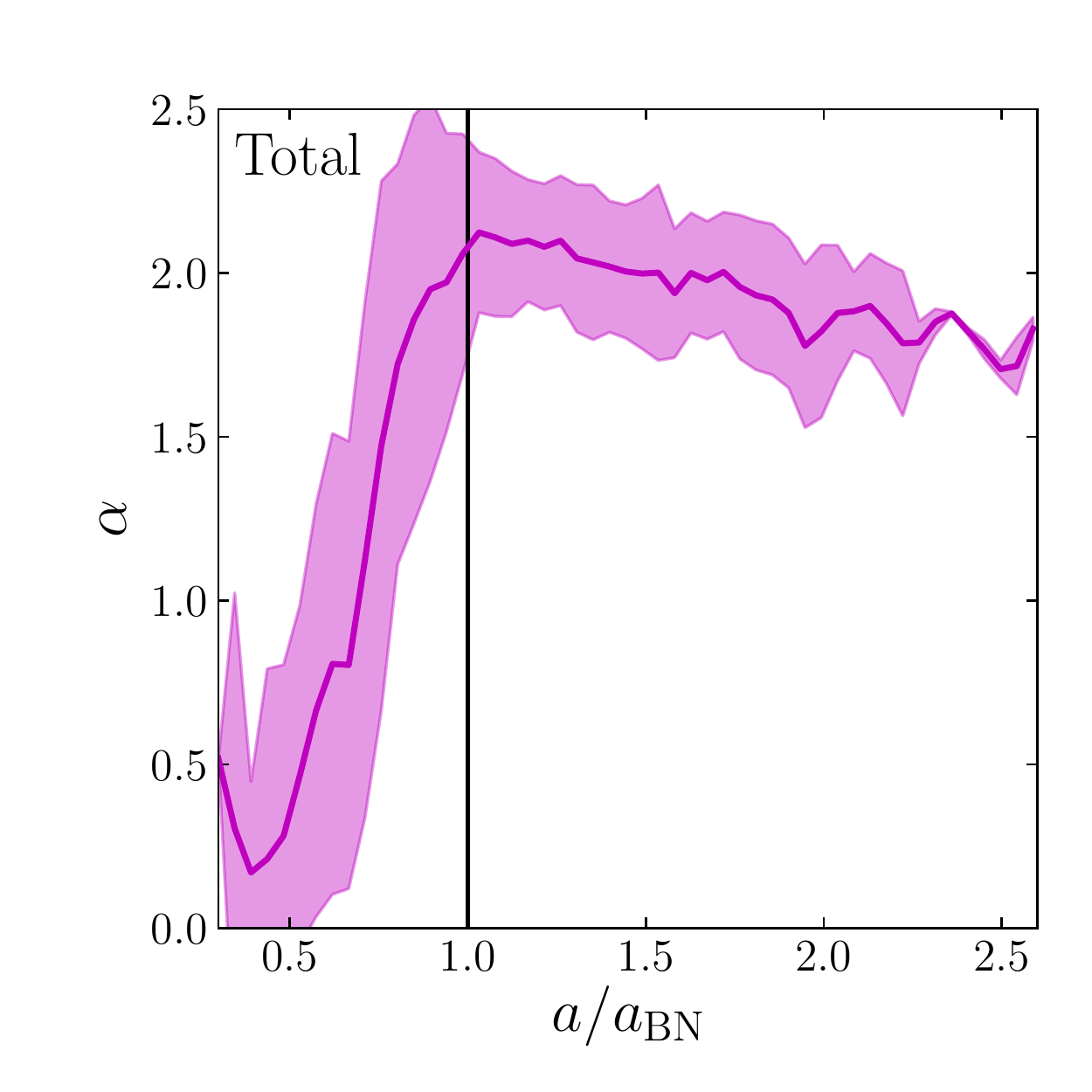}
\caption{
Dekel-Zhao profile fits to the DM mass profiles
in all 34 VELA simulated galaxies.
Shown are the median and standard deviation
%$(16\%,84\%)$ percentiles 
of the parameters $c$ and $\alpha$ as a function of time
(expansion factor $a\seq (1+z)^{-1}$)
with respect to the time of the blue nugget at the end of the major compaction.
This quantity is a proxy for mass, below and above a halo mass of
$\mv \seq 10^{11.3}\msun$.
We see different typical profiles well before and after the compaction event,
namely below and above the critical mass.
{\rm Left:} for the DM mass profiles, where
we crudely adopt $(c,\alpha) \seq (3,0.5)$ and $(7,1)$ as our fiducial
profiles pre-compaction and post-compaction, respectively. 
{\rm Right:} for the total mass profiles, 
where we crudely adopt $(c,\alpha) \seq (5,0.5)$ and $(3,1.8)$, respectively.
The corresponding profiles are shown in \fig{vela_prof}.
%\adr{Consider versus mass, and only good BNs}
}
\label{fig:vela_fits}
\end{figure*}

A key for the strength of the cusp heating by dynamical friction
is the initial compactness of the merging satellite, which determines the mass 
with which it penetrates into the host-halo cusp.
A mass threshold for compact satellites would be translated to
a host-halo mass threshold for the formation of extended cores in the hosts. 
Indeed,
cosmological simulations reveal that most galaxies undergo a major event
of wet compaction to a ``blue nugget" when they are near a golden halo mass of
$\mv\!\simeq\!10^{11.3}\msun$ \citep{zolotov15}.
Following \citet{tacchella16_prof},
we use galaxies from the VELA zoom-in cosmological simulations
(briefly described in Appendix \se{app_vela}, 
available as supplementary material online),
in order to derive the mass profiles of haloes pre compaction versus 
post compaction, namely below and above the golden mass, respectively
\citep{zolotov15,tomassetti16,huertas18}.
These will serve us in evaluating the compactness of the merging satellites
as a function of their mass.

\smallskip  % universal profiles
\Fig{vela_prof} shows the evolution of mass profiles 
for four representative simulated VELA galaxies, at a sequence
of output times before and after the major compaction events,
which occur in different galaxies at different redshifts but
typically when the halo mass is near $\mv\ssim 10^{11.3}\msun$.
The figure shows the profiles of the total mass, consisting of dark matter, 
stars and gas,
in comparison with the mass profiles of the DM alone.
In both cases, we see rather universal profiles both before and after the
compaction events, showing that the overall compactness is 
growing significantly during the wet-compaction process, both for the DM and
for the total mass.

%==============================
%\subsection{Two-parameter Dekel+ profile}
%--------------------
%\subsubsection{The Dekel-Zhao density profile}

\smallskip
In order to quantify the universal mass profiles, relevant here in particular
to the merging satellites, we use the Dekel-Zhao profile  
\citep[][DZ]{dekel17,freundlich20_prof}\footnote{Available~for~implementation
~in\\ https://github.com/JonathanFreundlich/Dekel\_profile .}
(summarized in Appendix \se{app_Dekel-Zhao}, 
available as supplementary material online).
This is a two-parameter functional form, with a flexible inner slope,
and with analytic expressions for the profiles of density, mass and velocity 
as well as potential and kinetic energies (and lensing properties).
It has been shown to fit DM haloes in the NIHAO cosmological 
simulations with baryons better than the other commonly used two-parameter
profiles (such as the generalized-NFW and the Einasto profiles). 
This is being confirmed in the Auriga, Apostle and EAGLE simulations 
(Marius Cautin, private communication).
In the DZ profile,
the mean density within a sphere of radius $r$, $\brho(r)$ 
(denoted later ${\brho}_{\rm sat}(\ell)$ for the satellites), 
with a virial mass $\Mv$ ($\mv$) inside a virial radius $\Rv$
($\ellv$), is characterized by two shape parameters, an inner slope 
$\alpha$ and a concentration $c$,
\be
\brho(r) = \frac{{\brho}_{\rm c}}{x^{\alpha}\,(1+x^{1/2})^{2(3-\alpha)}}\, ,
\quad x= \frac{r}{\Rv}\, c \, ,
\label{eq:barrho}
\ee
in which the constant is related to the parameters $(c,\alpha)$ by
\be
{\brho}_{\rm c} = c^3 \mu(c,\alpha)\, \brhov , \quad
\brhov = \frac{\Mv}{(4 \pi/3) \Rv^3} \, ,
\label{eq:rhoc}
\ee
\be
\mu(c,\alpha) = c^{\alpha-3}\, (1+c^{1/2})^{2(3-\alpha)} \, .
\ee
The mean density contrast of the haloes with respect to the cosmological
background in the EdS cosmological regime (approximately valid at $z \!>\! 1$)
is $\brhov/\urho \ssim 200$,
and by definition $\brhov$ is the same for the host and for the
satellites when they were still isolated.
To be used below, 
the associated mass encompassed within a sphere of radius $r$ is
\be
\frac{M(r)}{\Mv} = \frac{1}{c^3 \brhov} x^3 \brho(x)
=\frac{\mu}{{\brho}_{\rm c}}\, x^3\, \brho(x)\, ,
\label{eq:M}
\ee
and the log slope of the mass profile is
\be
\nu(r) = \frac{d\log M}{d\log r} = \frac{3-\alpha}{1+x^{1/2}} \, .
\label{eq:nu}
\ee

\smallskip % c and a
We fit a Dekel-Zhao profile with the free parameters $(c,\alpha)$ to each of
the VELA simulated mass profiles, separately for the DM and for the total mass. 
The best-fit is obtained by minimizing
residuals in equally spaced log bins in the range $(0.01\sdash 1)\Rv$
and $(0.03\sdash 1)\Rv$ for the DM and total mass respectively, assuming that
the most relevant mass profile is near $(0.03\sdash 0.1)\Rv$.
\Fig{vela_fits} shows the median and standard deviation
%$(16\%,84\%)$ percentiles 
for the two profile parameters over all the 34 simulated galaxies as a 
function of time, via the expansion factor $a\seq(1+z)^{-1}$,
with respect to the time of the blue-nugget peak
(BN) at the end of the wet compaction process.
%\adr{Consider mass instead?}.
Based on this figure,
we adopt for the fiducial pre-compaction DM profile the DZ parameters
$(c,\alpha)\seq(3,0.5)$,
and for the fiducial post-compaction profile $(c,\alpha)\seq(7,1)$.
Other choices, relevant at different times before or after the compaction,
make only little differences to the effective diffuseness and compactness of
the mass distribution in these two distinct phases of evolution.
For the total mass we adopt 
$(c,\alpha)\seq(5,0.5)$ and $(3,1.8)$ for the diffuse and compact satellites, 
respectively.

\smallskip  % DM vs total satellites
While the DZ best-fit parameters are quite different for the DM and for the
total mass profiles, the actual mass profiles turn out not to be that different.
The diffuse haloes roughly match an NFW profile with $c_{\rm NFW}\ssimeq 5$,
with the total mass about 50\% higher than the DM mass interior to $0.067\Rv$.
The compact haloes, interior to the same radius, have more mass  
than the diffuse haloes by a factor of a few,
with the total mass twice as large as the DM mass.
We find below that the effect of using the DM mass in the merging satellites
is not very different from the effect of using the total mass.
In order for our analysis to be on the conservative side in terms of the DF
heating, we adopt as our fiducial satellite profile the typical profile of
the dark matter alone.

\smallskip % c2 and s1
As described in \equ{c2} to \equ{alpha} of Appendix \se{app_Dekel-Zhao}
(available as supplementary material online),
based on equations 11-14 of \citet{freundlich20_prof},
an alternative, sometimes more accessible pair of shape parameters is
the concentration $c_2$, referring to the radius where the log local density
slope is $-2$ (as in the NFW profile), and the negative log local density slope
$s_1$ at some small radius $r_1$ (e.g. at 1\% of the virial radius).
While there is a valid DZ profile for any values
of $c$ ($>\! 0$) and $\alpha$ ($<\!3$), a valid profile is not guaranteed for
all arbitrary values of $c_2$ and $s_1$. For example, $c_2$ is not defined for
an isothermal sphere or a steeper density profile.

\smallskip % c2 and s1
The alternative DZ parameters corresponding to the universal profiles are
$(c_2,s_1) = (3, 0.94)$ and $(15.6, 1.52)$ for the fiducial diffuse and compact
satellite haloes respectively, using the DM mass alone.
The inner slopes $s_1$ are respectively somewhat flatter and steeper than an 
NFW cusp 
(which has $s_1\seq 1.06$ and $1.27$ for $c_{\rm NFW}\seq 3$ and $15.6$ 
respectively), with a smaller and larger concentration compared to the 
$c_{\rm NFW}\ssimeq 5$ that fits typical simulated haloes in the relevant 
mass range at $z\ssim 2$.
For the total mass, the alternative DZ parameters are
$(c_2,s_1) = (4.5, 1.31)$ and $(>\!100, 2.2)$, the latter
indicating that the slope is steeper than $2$ in the whole relevant radius
range.  

\smallskip  % mass dep
%\adr{check redundancy}
The above implies that deeper satellite penetration, that are associated with
more significant host-cusp heating by dynamical friction,
are expected once the host halo is 
more massive than $\Mv\sim\!10^{12}\msun$, such that a significant fraction of 
the accreting mass is in post-compaction nuggets above 
$\mv \sim\!10^{11.3}\msun$.
On the other hand, much less penetration and heating is expected for hosts of
$\Mv \!<\! 10^{11}\msun$, where most of the merging satellites tend to be 
pre-compaction.
This will be studied next.

%%%%%%%%%%%%%%%%%%%%%%%%%%%%%%%% 3
\section{Dynamical-Friction Heating by Satellites: A Toy Model}
\label{sec:toy}

We start with a simple analytic toy model in order to obtain a crude estimate 
for the
energy deposited in the host-halo cusp as a result of the dynamical friction 
exerted on penetrating satellites as a function of their compactness.
These estimates will be improved in \se{satgen} using semi-analytic 
simulations.

%===================
\subsection{Satellite penetration - tidal stripping}

In order to evaluate the energy deposited in the host cusp by dynamical
friction, we should first estimate the satellite mass as it penetrates
the cusp after it had suffered tidal stripping along its orbit.  
We first evaluate the mass within the satellite stripping radius when 
the satellite is at the cusp half-mass radius, crudely assuming an onion-shell 
outside-in stripping outside the stripping radius. 
When doing so, we neglect changes in the density and kinematics
inside the tidal radius (crudely referred to as ``tidal heating") 
that may affect the stripping. 
These are incorporated in the treatment by SatGen in \se{satgen}.
We then apply a simple correction to the mass within the tidal-radius, 
to be calibrated by
simulations, in order to take into account effects such as the deviations of 
the tidal force from spherical symmetry about the satellite and for the fact 
that the stripping is not instantaneous but rather takes a dynamical time.

%-----------
\subsubsection{The tidal radius}

The tidal radius \citep{king62}, where self-gravity balances the tidal and
centrifugal forces along the line connecting the centers of host and satellite, obeys the equation 
\be
%\boxed{
\bsig(\ell_{\rm t}) = \brho (r)\, [2 - \nu(r) +\epsilon^2(r)]\, ,
%}
\label{eq:tidal}
\ee
where $\bsig(\ell)$ and $\brho(r)$ are the mean density profiles
of the satellite (before it entered the host virial radius $\Rv$)
and the host, respectively.
Here $\nu(r)$ is the local log slope of the host mass profile $M(r)$,
namely $\nu(r) \seq d\ln M/d\ln r$.
It is $\nu(r)\seq 3\!-\!\alpha(r)$, where $\alpha(r)$ is minus 
the local log slope of $\bar\rho(r)$.
For example, $\nu \seq 3,2$ and $1$ for a flat core,
an NFW cusp and an isothermal-sphere,
for which $\alpha\seq 0,1$ and $2$,
respectively.

\smallskip
The quantity $\epsilon(r)$, representing the centrifugal force that helps
the stripping, is the local circularity of the satellite orbit at $r$,
namely the ratio of tangential to circular velocity
\be
\epsilon(r) = \frac{V_{\rm tan}(r)}{V_{\rm circ}(r)} \, ,
\ee
with $\epsilon \seq 0$ and $1$ for local radial and circular orbits, 
respectively.
This local circularity is typically not the same as its value at $\Rv$, 
upon entry to the halo,
which commonly serves as one of the two parameters characterizing the orbit.
The value of the local $\epsilon$ may vary as the satellite is orbiting within 
the host halo, and we do not know a priori its effective range of values.
We therefore use $\epsilon\seq 0.5$ as a fiducial value,
recalling that it could range from zero to above unity.

%--------------------
\subsubsection{The penetrating satellite mass}

We use for the satellites the Dekel-Zhao profile, \equ{barrho}.
As shown in \equ{app_frac} of Appendix \se{app_Dekel-Zhao}
(available as supplementary material online), 
the fraction $f\seq m/\mv$ of satellite mass 
within a sphere of radius $\ell$ about the satellite center
with respect to the initial total satellite mass, 
obeys  
\be
(f/\mu)\, [ (f/\mu)^{-[2(3-\alpha)]^{-1}} -1 ]^6
 = \bsig(\ell)/\bsigc \, .
\label{eq:fraca}
\ee
Using the tidal condition, \equ{tidal}, we obtain an equation for the mass
fraction $f_{\rm t}(r)$ within the tidal radius,
\be
%\boxed{
\left( \frac{f_{\rm t}}{\mu} \right)
\left[ \left( \frac{f_{\rm t}}{\mu} \right)^{\!\! -[2(3-\alpha)]^{-1}} 
\!\!\!\! -1 \right]^6
\!\!\! = \frac{[2\!-\!\nu(r)\!+\!\epsilon^2(r)]\,
\brho(r)}{\mu\,c^3\,\brhov} \, .
%}
\label{eq:frac}
\ee
Recall that $\mu$ is a function of the satellite parameters $c$ and $\alpha$.
For a given host mass profile $M(r)$, and its log slope
$\nu(r)$, and for a given satellite orbit circularity $\epsilon(r)$, 
this equation can be solved numerically for $f_{\rm t}(c,\alpha)$.
We note that the solution depends only on the profiles, and it does not depend
explicitly on the total masses or the ratio $\mv/\Mv$.
 
\smallskip % analytic solutions
One can see that for a given $\alpha$ ($<\!3$), 
in the limit $c \rar \infty$, where $\mu \rar 1$, 
the equation yields $f_{\rm t} \rar 1$.  
Similarly, for a given $c$, in the limit $\alpha \rar 3$ (from
below), again $\mu \rar 1$ and $f_{\rm t} \rar 1$.
Thus, the mass fraction within the tidal radius is increasing with either 
$c$ or $\alpha$,
approaching no stripping as $c \rar \infty$ or $\alpha \rar 3$.
Analytic solutions of \equ{frac} can be obtained for certain values of 
$\alpha$, e.g., there is a simple explicit solution for $\alpha=0$, 
and solutions for $\alpha\seq 1$ and $\alpha \seq 2$ via solutions of cubic 
polynomial equations.

\smallskip % correction to bound mass
The actual bound satellite mass at $r$, for the purpose of dynamical friction,
may deviate from the mass within the tidal radius. Among other
effects, this is because the tidal stripping is not spheri-symmetric, 
because it is not instantaneous, occurring over a dynamical timescale at $r$,
and because of the %tidal-heating 
effects on the inner satellite.
For example, \citet{green21} assumed that the actual mass stripping rate is 
crudely modeled as $\dot{m} = A m(>\ell_{\rm t})/t_{\rm dyn}$, 
where $t_{\rm dyn}$ is approximated by the dynamical time of the halo at $r$, 
and found a best fit to N-body simulations
with $A\!\simeq\! 0.55$.\footnote{In these
simulations, the satellite bound mass is defined using an iterative un-binding 
algorithm \citep{bosch18}, which considers the energy in the 
satellite frame, with the gravitational potential energy involving 
satellite particles only.}
We can therefore consider the mass within the tidal radius to be
an underestimate of the actual satellite mass, 
possibly by a factor of $\sim\!2$ or more at the relevant radii within the 
cusp if the modeling above is valid.  
We crudely model this correction in the cusp as
\be
f = B\, f_{\rm t} \, . 
\label{eq:f}
\ee
The value of $B$ can be calibrated by matching the results of the
toy model to the SatGen simulations described in \se{satgen}, where
any evolution of the inner satellite is included.
We adopt $B\seq 2$ as our fiducial value, but this choice has no qualitative
effect on the results.
%\adr{For NFW compact match need $B=1.6$. For compact steep 2.6. 
%For NFW diffuse 3.9. For steep diffuse 15.}
% B=2.5 to match steep compact.  Get much lower NFW diffuse.
% B=1.6 to match NFW compact.  Get much lower NFW diffuse.

%Motivated by this, we may model the integrated correction over the orbit, 
%from tidal mass to bound mass at $r$, by
%\be
%(1-f) = B (1-f_{\rm t})
%f = f_{\rm t} + (1-B)\, (1-f_{\rm t}) \, ,
%f = B f_{\rm t} + (1-B)
%\label{eq:f}
%\ee 
%where $B \!\lsim\! 1$ represents the fraction of the mass outside the tidal
%radius that is instantaneously unbound for the purpose of dynamical
%friction, to be calibrated by simulations.
%Based on a match with a fiducial run of SatGen (\se{satgen}), 
%with a steep-cusp halo, and with a compact satellite of $\mv/\Mv\seq 0.1$ and 
%$\epsilon\seq 0.4$ at $\Rv$, 
%we adopt here $B \seq 0.9$.
% B=0.77 to match steep compact. But get much too high steep diffuse.
% B=0.85 to match NFW compact. But need B=0.932 to match NFW diffuse

%-----------------------
\subsubsection{The satellite mass in a given host cusp}

%\adr{Appendix NFW, same as Dekel-Zhao}

The host halo is also described by a 
Dekel-Zhao profile, \equ{barrho} and \equ{app_barrho},
with parameters $(\ch,\ah)$.
The log slope of the mass profile, $\nu(r)$, to be used in \equ{frac}, 
is given for the DZ profile by \equ{nu}.
%\be
%\nu(r) = \frac{d\log M}{d\log r} = \frac{3-\ah}{1+x^{1/2}} \, .
%\label{eq:nu}
%\ee
We evaluate $\bar{\rho}(r)$ and $\nu(r)$ at the 
half-mass radius of the cusp of radius $\Rc\seq10\kpc$, which is
roughly $\Rh\!=\!7\kpc$, and insert them in \equ{frac} in order to solve
for the satellite mass fraction there.

\smallskip % NFW
Our fiducial case is an initial host halo of $\Mv \!=\! 10^{12.5}\msun$ 
at $z\!=\!2$ (with virial radius and velocity
$\Rv\!=\! 150\kpc$ and $\Vv\seq 300\kms$).
Such a halo, if unperturbed by baryons, is well fit by an NFW profile,
\equ{app_nfw} \citep{nfw97}, with a moderately steep cusp, 
and an NFW concentration $c_{\rm NFW}\seq 5$ \citep{bullock01_c}. 
This NFW profile is matched best across $r\seq (0.01 \sdash 1)\Rv$
by a DZ profile of 
$(\c2h,\s1h)\seq(5.04,0.908)$, or $(\ch,\ah)\seq(7.13,0.216)$.\footnote{The
%$(\c2h,\s1h)\seq(5.036,0.9076)$, or $(\ch,\ah)\seq(7.126,0.2156)$.\footnote{The
very inner slope at $0.01\Rv$, $\s1h$, is apparently slightly smaller than 
unity, but the fit is overall good across the cusp and in the most relevant
region of $r\!\lsim\!0.1\Rv$. 
Different DZ fits, that enforce a perfect fit at $0.01\Rv$ and have
an almost similar overall quality, are possible, but they do not 
significantly affect the results.}
In this case we neglect earlier baryonic effects on the DM halo,
either steepening by adiabatic contraction or partial flattening by outflows. 

\smallskip % cusp radius
We note that the assumed ``cusp radius" of $\Rc\seq10\kpc$ is rather crude, 
motivated by the desired ``core radius" based on the observational
estimates.
Just for reference, in the fiducial NFW host, the local negative log slope
of the density profile at this radius is $\simeq\!1.56$.

\smallskip % steep cusp
As an alternative case with an initially steeper cusp we use
%$(\c2h,\s1h)\seq(5.036,0.9076)$, or $(\ch,\ah)\seq(1.13,1.29)$.
$(\ch,\ah)\seq(1.13,1.29)$ or $(\c2h,\s1h)\seq(5.04,0.91)$.
This steep-cusp profile is typical to haloes of $\Mv \!\gsim\! 10^{12}\msun$ 
from the NIHAO cosmological simulations with baryons, 
as analyzed in \citet{freundlich20_prof}. 
We will mention the results for the steep-cusp halo below, but 
the figures referring to this case are presented in Appendix \se{app_steep}
(available as supplementary material online).

\smallskip % fig 3
The left panel of \fig{frac} shows the bound mass fraction $f$ as a       
function of the DZ parameters $(c,\alpha)$ for the satellite profile, 
as obtained by solving \equ{frac} (with the correction of \equ{f} assuming
$B\seq2$).
This is shown here for the fiducial NFW host and in \fig{app_frac_steep}
(available as supplementary material online)
for the steep-cusp host.

\smallskip % vela fiducial
As described in \se{vela}, we learn from the VELA simulations that the 
DM profiles of typical diffuse and compact haloes, below and above 
$\mv\ssim 10^{11.3}\msun$, are well fit by a DZ profile with
$(c,\alpha)\seq(3,0.5)$ and $(c,\alpha)\seq(7,1)$, respectively.
These parameters for the satellites are marked in the figure.
Also marked are the parameters for the total mass profiles,
$(c,\alpha)\seq(5,0.5)$ and $(3,1.8)$,
for the diffuse and compact satellites
respectively.

\smallskip % f NFW xxxx B=2
For the NFW host halo, with the fiducial DM satellite profiles, we obtain 
$f\seq m/\mv \!\simeq\! 0.38$ and $0.05$ for the compact and diffuse 
satellites, respectively.
Thus, the fiducial massive, post-compaction satellite penetrates into the cusp
with a significant fraction of its original mass, which promises significant 
cusp heating.
On the other hand, the fiducial low-mass, pre-compaction satellite is more
significantly stripped and it penetrates to the halo cusp with a small fraction 
of its mass, such that it is not expected to provide much cusp heating. 

\smallskip % total satellite
For the NFW halo but with the somewhat more compact total satellite profiles, 
we obtain $f\seq m/\mv \!\simeq\! 0.74$ and $0.12$.
This is a deeper penetration, as expected from the larger satellite
compactness due to the baryons, but the qualitative difference between the
compact and diffuse satellites remains the same.

\smallskip % f steep xxxx B=2
For the steep-cusp halo, we obtain        
$f\!\simeq\! 0.23$ and $0.01$ for the fiducial compact and diffuse 
satellites, respectively.
Here, as expected from the steeper cusp,
the satellites penetrate to the cusp with a somewhat smaller fraction of their
mass. 
Nevertheless, the penetration of the 
compact satellite to the steep cusp is still with a noticeable fraction of its
original mass, so it still has the potential to provide significant heating, 
to be estimated next.

%\smallskip
%\adr{Derive the distribution of local circularity from the distribution of
%circularity at $\Rv$.}

%===========================
\subsection{Energy deposited in the host cusp}

Given the penetrating mass fraction $f$,
we wish to estimate the energy deposited in the host cusp
by dynamical friction, with respect to the kinetic energy within this cusp. 
This will tell whether dynamical friction is capable of significantly heating 
the cusp, and possibly partly flatten the density profile, in preparation for 
core formation by AGN feedback.

%---------
\subsubsection{A single satellite}

%Work by Dynamical Friction in the Cusp}
We crudely approximate the work done by dynamical friction 
on a satellite within the cusp by
\be
\Ws= s\,\Rc\, F\, ,
\ee
where $F$ is the typical DF force exerted by the dark matter
on the satellite in the cusp, $\Rc$ is the characteristic cusp radius, 
encompassing mass $\Mc$,
and $s \!\gsim 1$ is a factor that characterizes the effective length of the 
path of the satellite inside the cusp.
The DF force is approximated by the Chandrasekhar formula
\citep{chandrasekhar43} (as long as
the cusp has not flattened to a core),
\be
F = 4\pi \ln\Lambda\, G^2\, \frac{\rho\, m^2}{V^2}\, H(u) \, ,
\label{eq:DF}
\ee
where $\rho$ is the typical DM density within the cusp, $m$ is the bound
satellite mass when in the cusp, $V$ is the satellite orbital velocity,
and $\ln\Lambda$ is the Coulomb logarithm.
The latter is commonly assumed to be approximately $\ln (\Mv/m) \ssim 3$. 
The correction factor $H(u)$ that limits the effect to particles that move 
slower than the satellite, assuming a Maxwellian distribution of velocities, is
\be
H(u) = {\rm erf}(u) - \frac{2u}{\sqrt{\pi}}\, e^{-u^2}\, , \quad
u=\frac{V}{\sqrt{2}\sigma}\, ,
\label{eq:H}
\ee
where $\sigma$ is the velocity dispersion in the cusp.
For example, $H(u\!=\!0.55)\!=\!0.1$ and $H(u\!=\!0.7)\!=\!0.2$.

\smallskip % caveats (to discussion?)
We note that 
the use of the Chandrasekhar formula involves certain inaccuracies, beyond the
uncertainties in the values adopted for the parameters, which could 
either strengthen or weaken the force. One is an uncertainly in the satellite
mass that is relevant for the purpose. For example, when the mass is measured
in a given way, \citet{green21} indicate that a multiplicative 
factor $\beta\ssim 0.75$ in \equ{DF} yields best fit to results deduced from 
the Bolshoi 
cosmological N-body simulation. This should be better calibrated using
carefully designed simulations where the dynamical friction is studied in more
detail.
On the other hand, self-friction within the satellite is ignored, and so is 
the binding effect of the baryons within the satellite. 
Dynamical friction is known to be significantly reduced when the
satellite is orbiting in a very flat host core, but here we limit ourselves to
affecting a steep cusp in its development toward a somewhat flatter cusp.
In summary, our toy-model treatment of dynamical friction should be considered
as a crude approximation only, good for an order-of-magnitude estimate. 

\smallskip
Adopting the mean density within $\Rc$  and the circular velocity at $\Rc$,
$\rho \!=\! (4\pi/3)^{-1} \Mc/\Rc^3$ and
$V\!=\!\Vc\!=\!(G \Mc/\Rc)^{1/2}$, we get
\be
\Ws = 1.8\, s_2\, (\ln\Lambda)_3 \, H(u)_{0.1}\, \frac{G m^2}{\Rc} \, ,
\label{eq:w0}
\ee
where the subscripts denote the fiducial values assumed for the
different quantities in this expression ($s_2\seq s/2$ etc.).

%-------------
%\subsubsection{Work compared to the cusp energy}
\smallskip
The DF work estimated in \equ{w0}
is to be compared to the kinetic energy in the cusp,
$\Kc\!\simeq\! (1/2)\,\Mc\,\sigma^2$.
We assume for the host $\Mv\!=\!10^{12.5}\msun$, whose
virial radius is $\Rv\! =\! 150\kpc\,M_{12.5}^{1/3}\,(1+z)_3^{-1}$,
in which $M_{12.5} \seq \Mv/10^{12.5}\msun$ and $(1+z)_3\seq(1+z)/3$.
For an NFW profile with $c_{\rm NFW}\!=\!5$ at $z\!=\!2$,
a cusp of $\Rc\!=\!10\kpc$ corresponds to $r\!=\!0.067\Rv$ and $\xc\!=\!0.33$,
so from \equ{app_nfw} the cusp mass is $\Mc\!=\!0.04\Mv$.
This implies that $\Vc \seq 0.77\Vv$.
We read from Figure 1 of \citet{freundlich20_prof} that at $r\!=\!0.067\Rv$
we have $\sigma\!\simeq \!\Vv$ (and that indeed $\Vc\!\simeq\!0.78\Vv$).
These imply for the NFW host that\footnote{Note that $1 \kpc\Gyr^{-1}
\!\simeq\! 0.955 \kms$.}
\be
\Kc \!\simeq\! 0.5\,\Mc \Vv^2 \!\simeq\! 0.83\, \Mc \Vc^2 
\!\simeq\! 6\!\times\!10^{15} \Msun\kpc^2\Gyr^{-1} \, .
\label{eq:kc}
\ee
We also learn that $u\!\simeq \!0.54$, such that $H(u)\!\simeq \!0.1$ is a
reasonable estimate.
For the steep-core host, we find that $\Kc$ is $1.25$ times larger.

\smallskip
Using \equ{w0} and \equ{kc}, we obtain
\be
\frac{\Ws}{\Kc} = 2.2\, s_2\, (\ln\Lambda)_3 \, H(u)_{0.1}\, 
\frac{m^2}{\Mc^2} \, .
\label{eq:w}
\ee
Thus, for a single satellite of initial mass $\mv$ in the NFW host of mass 
$\Mv$, we obtain 
\be
\frac{\Ws}{\Kc} = 0.55\, s_2\, f_{0.2}^2\, 
\left( \frac{\mv}{\Mv} \right)_{0.1}^2 \, .
%}
\label{eq:w_single}
\ee
The value of $s$ here absorbs the uncertainties in the three parameters
$s$, $\ln\Lambda$ and $H(u)$.
Here, $f$ ($\equiv\!0.2 f_{0.2}$) is to be obtained, e.g.,
by solving \equ{frac} given the profile shapes of the halo and the satellite,
with the possible correction of \equ{f}.
Note that $f$ in this model is independent of $\mv/Mv$.
We learn that a single satellite of $\mv \!\geq\! 0.13 \Mv$
may be capable of heating the cusp, with $\Ws \! \geq \! \Kc$, once
more than $0.2$ of its mass survives tidal stripping before entering the cusp.
For the steep-core host, the numerical factor in \equ{w_single}
is $1.25$ times lower because of the higher $\Kc$.

\begin{figure*} % 3
\centering
\includegraphics[width=1.01\textwidth,trim={0.5cm 0.5cm 0.0cm 0.0cm},clip]
%{figs/a_vs_c_2Dhist_colored_f_ws_wc_NFW_host_v11.pdf}
{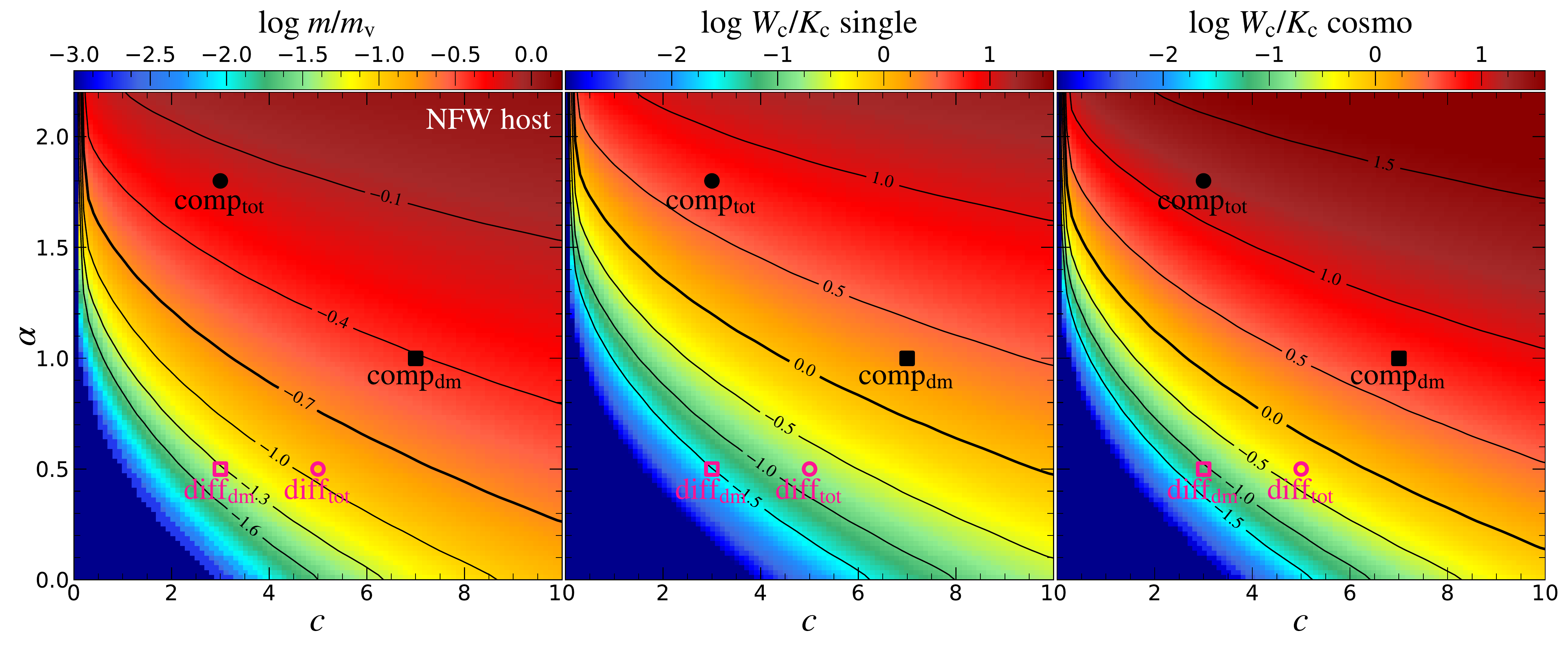}
\caption{
Toy-model estimates for satellite penetration and energy deposited in the 
host cusp by dynamical friction, as a function of the satellite compactness
via the Dekel-Zhao profile parameters of concentration and inner slope
$(c,\alpha)$. 
The host-halo profile is NFW with a moderate
inner slope $s_1\!\simeq\!1$ and $c_{\rm NFW}\seq 5$.
\Fig{app_frac_steep} 
(available as supplementary material online)
shows the same for
or a steeper cusp with DZ parameters $s_1\seq1.5$ and $c_2\seq 5$.
We consider the cusp radius to be $\Rc\seq 10\kpc \ssimeq 0.067\Rv$.
The effective circularity of the satellite orbit in the cusp is assumed to be
$\epsilon\!=\!0.5$.
{\bf Left:} 
The fraction of bound satellite mass $f\seq m/\mv$ when at the half-mass
radius of the host cusp, $r\seq 0.7\Rc\seq 7\kpc$,
as obtained by solving \equ{frac}, corrected by \equ{f} with $B\seq2$.
{\bf Middle:}
The energy deposited in the host cusp of $\Rc\seq 10\kpc$ by dynamical 
friction acting on a single satellite with $\mv/\Mv\seq 0.1$, assuming 
$s\seq 2$ in \equ{w_single}. 
The energy is normalized by the cusp initial kinetic energy, $\Wc/\Kc$.
{\bf Right:}
The energy deposited in the cusp by a cosmological sequence 
of satellites during one virial crossing time, as obtained from \equ{wvir} 
with $\tau\seq 1$.
The fiducial DZ profiles of pre-compaction and post-compaction 
satellites, as determined from the VELA simulations in \se{vela}, below and
above the golden mass of $\mv\!\sim\!10^{11.3}\msun$, are marked
by open-magenta and filled-black symbols respectively.
The squares refer to the fiducial profiles of DM only, with
$(c,\alpha) \seq (3,0.5)$ and $(7,1)$ respectively.
For comparison, the circles refer to the slightly more compact
total mass profiles, with 
$(c,\alpha)\seq (5,0.5)$ and $(3,1.8)$ respectively.
For an NFW host we read for the fiducial diffuse and compact satellites 
respectively
$m/\mv \ssim 0.05, 0.39$, $\Wc/\Kc \ssim 0.03, 2.05$ for the single satellite,
and $\Wc/\Kc \ssim 0.08, 5.59$ for the sequence of satellites. 
This implies that typical low-mass, pre-compaction satellites are not expected 
to heat up the host cusps haloes 
while the massive, post-compaction satellites are expected to significantly
heat up the host cusp over a non-negligible fraction of a virial crossing time.
Using the satellite total-mass profiles, the three quantities are slightly
larger, but the qualitative result remains the same. 
For the steep-cusp host, \fig{app_frac_steep}
(available as supplementary material online), 
the satellite stripping is stronger due to the steeper cusp, but the heating 
by compact satellites is still significant during half a virial time.
%\adr{Sharon: change symbol locations for tot to
%$(c,\alpha) \seq (5,0.5)$ and $(3,1.8)$ for diffuse and compact.
%This is $(c_2,s_1) \seq (5,1.05)$ and $(170,2.05)$}
}
\label{fig:frac}
\end{figure*}

\smallskip % fig 3 single 2 2 xxxx 
The middle column of \fig{frac} shows the toy-model estimates
of $\Wc/\Kc$, based on \equ{w_single}, for a single satellite
of $\mv\seq 0.1\Mv$ as a function of the satellite's DZ profile parameters
$(c,\alpha)$, and for the NFW host, assuming $s\seq 2$
(and $\epsilon\seq0.5$ and $B\seq2$ as before).
The same for the steep-cusp host is shown in \fig{app_frac_steep}
(available as supplementary material online). 
The contours $\log \Wc/\Kc \seq 1$ mark the boundaries for significant cusp
heating.
The fiducial diffuse and compact satellites, as evaluated from the
DM in the VELA simulations, \se{vela}, are marked by open and filled circles
at $(c,a)\seq (3,0.5)$ and $(7,1)$, respectively.
We read for the NFW host 
$\Wc/\Kc \ssimeq 2.05$ and $0.03$ for the fiducial compact and diffuse
satellite, respectively.
For the fits to the more compact total satellite mass, we obtain
$\Wc/\Kc \ssimeq 7.5$ and $0.2$, respectively. 
This is more heating, but with a similar difference between the compact and
diffuse satellites.
For the steep-cusp host, and the DM satellite profiles,
the respective results are $\Wc/\Kc \ssimeq 0.56$ and $0.001$, namely less
heating, as expected.
Overall, we learn quite robustly
that a single compact satellite of $\mv\ssim 0.1\Mv$ is clearly
capable of significantly heating up the cusp in an NFW host, 
and is expected to do so also in a steep-cusp host.
In contrast, a fiducial diffuse satellite is expected to have only 
a negligible effect on the host cusp in the two types of hosts, even if the
satellite compactness is evaluated via its total mass profile. 
According to \equ{w_single},
the energy deposited by a more massive satellite, $\mv = 0.1\,\mu\,\Mv$, 
is expected to provide more heating by a factor $\mu^2$.

\smallskip % AM
A note of caution is that the deposited kinetic energy may not all be in the 
form of ``heat", e.g., it could partly be in the form of bulk motion and in 
particular rotational energy.
In this case the analysis should involve angular momentum, both in the
estimation of the effect of DF on the DM cusp and in the evaluation of the
DM response to outflows (\se{agn}).
For satellite entry to the cusp on a rather tangential orbit, 
it can be shown that the angular-momentum deposit by DF with respect to the
maximum angular momentum of the cusp, had it been a disk with circular orbits,
is comparable to the relative energy deposit of \equ{w_single}.
This implies a likely significant effect of the compact satellite
on the cusp angular momentum.
We defer an analysis involving angular momentum to future work.

%-----------
\subsubsection{A cosmological sequence of satellites}

For a cosmological sequence of satellites, we assume that all satellites of
mass $\mv\!>\!10^{11.3}\msun$ are post-compaction 
\citep{zolotov15,tomassetti16},
with a high concentration and a cusp (\se{vela}), 
which make then penetrate to the host cusp with a non-negligible mass fraction
$f$.
For a host halo of $\Mv \seq 10^{12.3}\msun$, 
these are mergers of mass ratio $\mv/\Mv\!\geq\! 1\!:\!10$. 
We read from Figure 7 of \citet{neistein08_m}, based on EPS theory, 
that $\sim\!60\%$ of the total accretion rate is in such satellites (at all
redshifts).
For the average total accretion rate we adopt the analytic estimate of
\citet{dekel13} in the EdS cosmological regime 
(a good approximation at $z \sgt 1$),
\be
\dot{M} = 0.47\,\Gyr^{-1}\,\Mv\,(1+z)_3^{5/2} \, .
\label{eq:Mdot}
\ee
%\adr{This is for 0.03. Can use 0.04, namely 0.62, for top $1/3$.}
The rate of total energy deposited in the cusp by dynamical friction
involving all these satellites can be computed by integration over their EPS
conditional satellite mass function normalized to a total that equals
$0.6\dot{M}$, convolved with the energy per satellite as given by 
\equ{w_single}. 
The mass function of first-order subhaloes at infall, could be approximated 
by $dn/dm \!\prop\! m^{-1.5}$ 
at $m \ssimeq 0.02 \sdash 0.2 \Mv$,
with a truncation at $m_{\rm max} \!\lsim\! 0.5 \Mv$
\citep{jiang14}.
For a crude conservative estimate, we assume here that each of the relevant 
satellites above $0.1\Mv$ is of a given mass $\mv$ somewhat above $0.1 \Mv$,
and write the accreted mass rate in these satellites
as $\dot{N} \mv = 0.6 \dot{M}$, where $\dot{N}$ is the number of relevant
satellites that merge during one Gigayear. This gives
\be
\dot{N} = 1.4\Gyr^{-1}\, (\mv/\Mv)_{0.2}^{-1}\, (1+z)_3^{5/2} \, .
\ee
%\adr{Use 0.15? or 0.1?}
Multiplying the result for a single satellite from \equ{w_single} by $\dot N$,
we obtain an approximation for the rate of relative work done on the host 
cusp by all the compact satellites
\be
\frac{\dot{W}}{\Kc} 
= 3\,\Gyr^{-1}\, f_{0.2}^2\, s_2\, (\mv/\Mv)_{0.2}\, (1+z)_3^{5/2} \, .
\label{eq:wdot}
\ee
In $\tau$ halo virial crossing times, each given by
$\tvir\!\simeq\!0.5\Gyr\,(1+z)_3^{-3/2}$ \citep{dekel13},
the relative DF work becomes
\be
%\boxed{
\frac{\Wc(\Delta t\seq \tau\tvir)}{\Kc} 
= 1.5\, \tau\, f_{0.2}^2\, s_2\,(\mv/\Mv)_{0.2}\,  (1+z)_3 \, .
%}
\label{eq:wvir}
\ee
\smallskip % mv
Based on Figure 7 of \citet{neistein08_m}, we adopt $\mv/\Mv=0.2$ as our
fiducial value.

\smallskip % tau
An upper limit for the total energy deposited could be estimated by 
integrating over a whole Hubble time,
$t\!\simeq\! 6.6\tvir \!\simeq\! 3.3\Gyr(1+z)_3^{-3/2}$, 
namely about a factor of $6.6$ times the energy in \equ{wvir}.
During a time window of $0.5\tvir$, the timescale for the ``heat" deposited by
dynamical friction to ``dissipate" as estimated below in \se{N-body} 
from an N-body simulation, we estimate $W/\Kc \ssim 0.75\, (1+z)_3$.

\smallskip % fig 3  2 2 1 xxxx
The right column of \fig{frac} shows the toy-model estimates for $\Wc/\Kc$ 
for a cosmological sequence of satellites 
as a function of the satellite's profile parameters
$(c,\alpha)$, for the NFW host, based on \equ{wvir}.
This is assuming $\mv\seq0.2\Mv$, at $z\seq 2$, and a duration of one virial
time $\sim\!0.5\Gyr$, namely $\tau\seq1$ 
(with $s\seq 2$, $B\seq 2$, and $\epsilon\seq0.5$ when solving for $\ft$ in
\equ{frac}, as before).
The same for the steep-cusp host is shown in \fig{app_frac_steep}
(available as supplementary material online).
We read for the NFW host, using the satellite DM mass profiles,
%$\Wc/\Kc \ssimeq 5.59$ and $0.079$ 
$\Wc/\Kc \ssimeq 5.6\,\tau$ and $0.08\tau$ for the fiducial compact and diffuse
satellite, respectively.
When using the satellite total mass, the results are
$\Wc/\Kc \ssimeq 20\,\tau$ and $0.5,\tau$.
For the steep-cusp host, with the satellite DM profiles,
the respective results are
%$\Wc/\Kc \ssimeq 1.53$ and $0.003$.
$\Wc/\Kc \ssimeq 1.5\,\tau$ and $0.003\,\tau$.
We learn that the median energy deposited by a sequence of satellites 
in one virial time is crudely estimated to be comparable to and slightly 
larger than the energy deposited by a single satellite of $\mv \ssim 0.1\Mv$.
During a period of $\sim 0.5\tvir$, the satellites are expected to 
significantly heat up the cusp in an NFW host, and they are estimated to do
so also in a steep-cusp host. 
In contrast, a sequence of diffuse satellites is expected to have only
a negligible effect on the host cusp for the two types of hosts during a period
of order a virial time.

\smallskip % z-dep
Based on the proportionality to $(1+z)$ in \equ{wvir},
a somewhat higher satellite compactness is needed for
heating the cusp to the same level over a similar period at lower redshifts, 
while a somewhat lower compactness would be sufficient at higher redshifts.

\smallskip % fig c2 s1
\Fig{app_frac} 
(available as supplementary material online)
complements \fig{frac} by showing the same
toy-model estimates for satellite penetration and the energy
deposited in the host cusp by dynamical friction,
but in the plane of the alternative DZ parameters
$(c_2,s_1)$ instead of the natural DZ parameters $(c,\alpha)$.

\smallskip % xxx dominance of massive satellites
We note that,
with a satellite initial mass function that declines flatter than
$\mv^{-2}$ up to a drop at $m_{\rm max}$, the
dependence on $\mv^2$ of the energy deposited in the cusp, \equ{w_single},
indicates that the total heating of the cusp is dominated by the more 
massive satellites.
%\adr{check with SatGen}
%
One could comment that 
this dominance of massive satellites may be strengthened because,
beyond the crude tidal-stripping approximation of \equ{tidal},
the actual satellite mass loss also depends on the time the satellite 
spends in the
strong-tide regime before entering the cusp. In particular,
satellites that make it to the cusp in the first or second pericenter of their
orbit are expected to bring in more mass than satellites that hang around
and are subject to stripping for a longer time (see \fig{satgen_single} below).
One can estimate that, for a given satellite density profile,
the rate of decay of the satellite orbit is increasing with $\mv/\Mv$.
This can be obtained by relating the decay rate to either the
ratio of AM loss by the DF torque to the orbital AM of the satellite,
or the ratio of work done on the satellite to its energy.
These ratios are both proportional to $m/M$,
because the DF acceleration is proportional to $m$, while the
specific angular momentum or energy are proportional to $M$, and the tidal
stripping is not strongly dependent on either.

%%%%%%%%%%%%%%%%%%%%%%%%%%%% 4
\section{Dynamical-Friction Heating: Semi-Analytic SatGen Simulations}
\label{sec:satgen}

% satgen
In order to explore more accurately the satellite properties required for 
efficient penetration
and strong DM heating by dynamical friction, we utilize the
dedicated Semi-Analytic Satellite Generator, 
SatGen \citep{jiang21_satgen},\footnote{Available~for~
implementation~in\\ https://github.com/shergreen/SatGen .}
(described in Appendix \se{app_SatGen},
available as supplementary material online).
It simulates tidal 
% and ram-pressure 
stripping while the satellite of an assumed initial density profile
orbits in the potential well of the host halo and the central galaxy
and penetrates to the inner halo subject to dynamical friction by the host dark
matter.
The code utilizes EPS merger trees and physical prescriptions for the
relevant mechanisms.
% FJ:
The tidal mass loss and the structural evolution of the satellites depend on
the initial structure of the satellites, following \citet{errani18},
such that a dense, cuspy satellite at infall is more resistant to tidal
disruption than a cored satellite.
In the version used here, we model the satellite with non-dissipative
particles, without explicitly treating the gas component. 

\smallskip % energy deposit
The energy deposited in the radius shell $(r\pm dr)$ of the host
is computed either by the work done by dynamical friction on the satellite
along its orbit,
or by the orbital energy loss of the satellite, with similar results.
Dynamical friction is modeled using \equ{DF} with $\ln\Lambda \seq \ln(\Mv/m)$.
In both cases, rather crudely, the energy is assumed to be deposited locally.

%========================
\subsection{A single satellite}
\label{sec:satgen_single}

\begin{figure*} % 4 
\centering
\includegraphics[width=0.80\textwidth]
%{figs/test_evolve_CompactVersusDiffuse_NFWhost.pdf}
{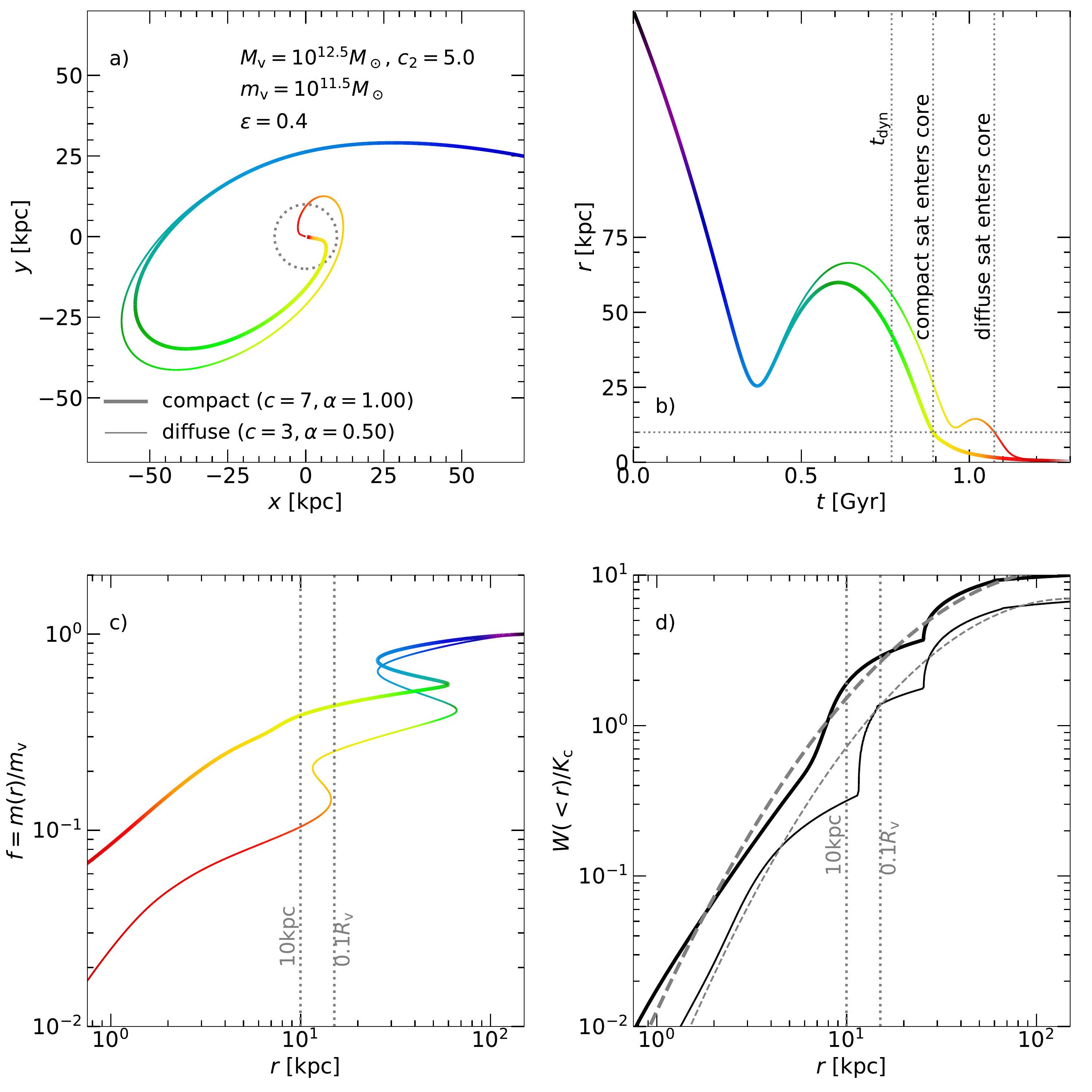}
\caption{
Semi-analytic SatGen simulations of single satellites, diffuse and compact,
marked by thin lines and thick lines respectively.
The Dekel-Zhao mass profile parameters of the satellites
are $(c,\alpha)=(3,0.5)$ and $(7,1)$, respectively,
consistent with the VELA profile pre and post compaction in \fig{vela_prof}
and \fig{vela_fits}.
The initial satellite mass at $\Rv$ is $\mv\seq 10^{11.5}\msun$ in a host of
$\Mv\seq 10^{12.5}\msun$, with an NFW profile of $c_{\rm NFW}\seq 5$.
The same for a steep-cusp halo is shown in \fig{app_satgen_steep_single}
(available as supplementary material online).
The orbit has a typical circularity at $\Rv$ of $\epsilon\seq 0.4$.
{\bf Top:}
The satellite orbits, a face on projection (left) and
radius within the host halo as a function of time, with the
times of entry into the $10\kpc$ cusp marked.
{\bf Bottom left:}
The satellite bound mass fraction $f\seq \mt/\mv$ as a function of radius.
{\bf Bottom right:}
The energy deposited by dynamical friction in the host, interior to radius $r$,
with respect to the kinetic energy within the cusp
($K_{\rm c}\seq 0.5 M(r\!<\!10\kpc) \Vv^2$).
The dashed curves refer to the fits used in the CuspCore analysis of \se{agn},
listed in \tab{fits}.
We read $\mc/\mv \ssimeq 0.1,0.4$ and $W_{\rm c}/K_{\rm c}\ssimeq 0.3,1.8$
for the diffuse and compact satellites respectively.
We learn that the bound gas fraction of the compact satellite is $\sim\!10$
times larger than for the diffuse satellite.
As a result the DF on the compact satellite generates a significant energy
change in the host cusp, while the effect of the diffuse satellite is
negligible, consistent with the analytic estimate of \se{toy}.
%\adr{Fangzhou: Please generate separate files for top and bottom panels.}
}
\label{fig:satgen_single}
\end{figure*}

\begin{figure*} % 5
\centering
\includegraphics[width=0.85\textwidth,trim={0.0cm 0.0cm 0.0cm 0.0cm},clip]
%{figs/compare_DFheating_CompactionVersusControl_WrKc.pdf}
{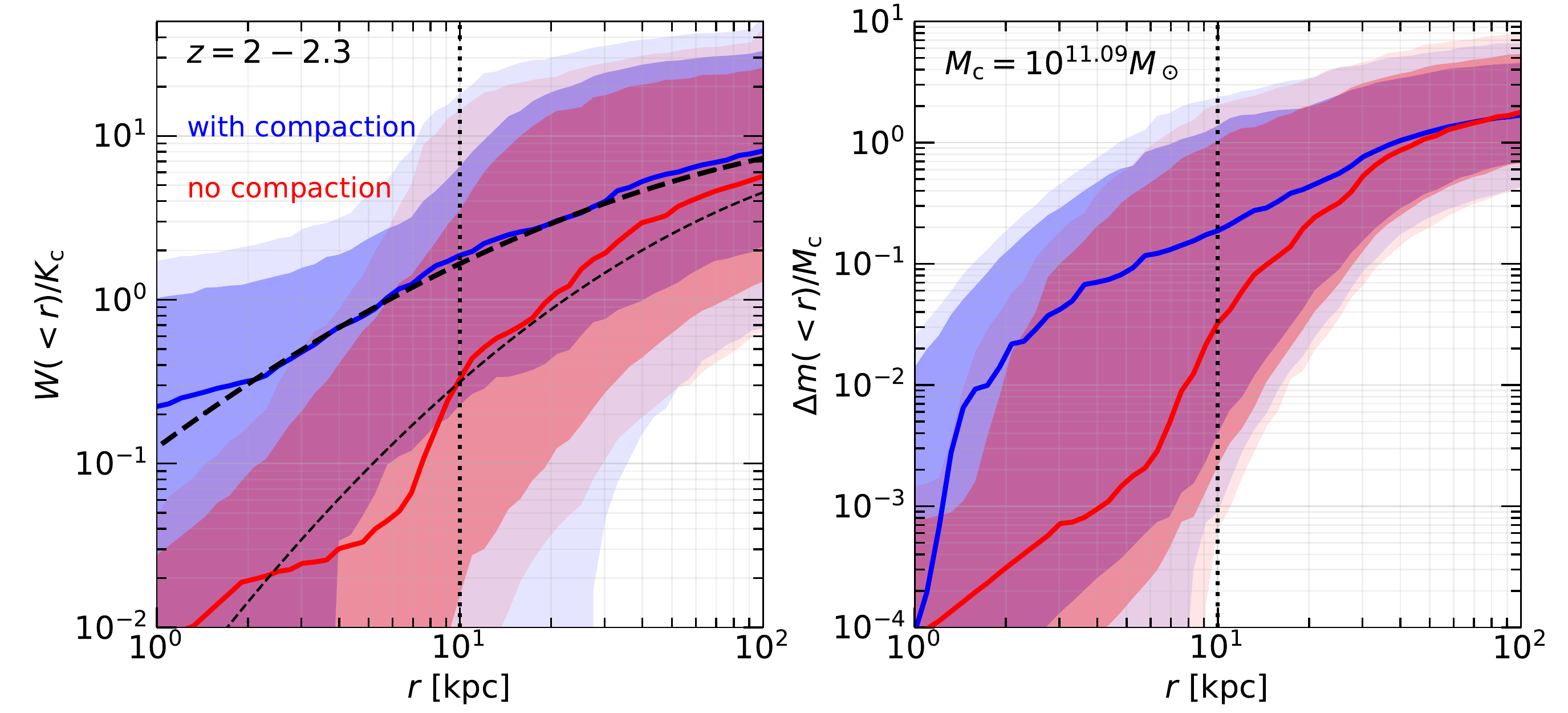}
\caption{SatGen simulations of a cosmological sequence of satellites during 
one halo virial time at $z\ssim 2$, $\tvir \!\simeq\! 0.5\Gyr$, 
corresponding to a redshift interval $\Delta z \! \simeq \! 0.3$.
{\bf Left:} the energy deposited within radius $r$ with respect to the kinetic
energy in the $10\kpc$ cusp, $W(<r)/\Kc$ (left), with the dashed lines
representing the fit used in \se{agn} and listed in \tab{fits}.
{\bf Right:} the satellite mass deposited inside $r$ with respect to the 
host cusp mass within $10\kpc$, $\Delta m(<\!r)/\Mc$.
The host halo mass is $\Mv \seq 10^{12.5}\msun$, with an NFW profile of
$c_{\rm NFW} \seq 5$. 
The same for the steep-cusp profile %of $(c_2,s_1)\seq(5,1.5)$
is shown in \fig{app_satgen_steep_cosmo}
(available as supplementary material online).
The fiducial run (blue) assumes compact satellites of 
$m_{\rm v} \!>\! 10^{11}\msun$, while the reference run (red) assumes that all
satellites are diffuse.
Shown are the medians and the 68\% and 95\% percentiles over 600 random merger 
trees.
The DZ profiles for the compact and diffuse satellites have the
fiducial parameters $(c,\alpha) \seq (7,1)$ and $(3,0.5)$, respectively.
We see significant mass and energy deposit in the cusp for the compact 
satellites (valid for hosts of $\Mv \!>\! 10^{12}\msun$ at $z\!>\!1$), 
and low mass and energy deposit for the case of diffuse satellites 
(approximating the situation in host haloes of 
$\Mv \!<\! 10^{11.5}\msun$ or at low redshifts).
}
\label{fig:satgen_cosmo}
\end{figure*}

\Fig{satgen_single} summarizes the results of running SatGen with a single
satellite, our fiducial compact or diffuse, penetrating into an NFW host of
$c_{\rm NFW}\seq 5$.
The DZ mass profile parameters of the initial satellites at the host virial
radius are the fiducial $(c,\alpha)=(7,1)$ and $(3,0.5)$, respectively,
consistent with the DM VELA profiles after and before the compaction 
in \fig{vela_prof} and \fig{vela_fits}.
The initial satellite mass at $\Rv$ is $\mv\seq 10^{11.5}\msun$ in a host of
$\Mv\seq 10^{12.5}\msun$. 
The orbit has a typical circularity at $\Rv$ of $\epsilon\seq 0.4$.

\smallskip
The orbit (top) shows that
the compact satellite enters the $10\kpc$ cusp near its second pericenter,
while the diffuse satellite enters only near its third pericenter. 
The satellite bound mass fraction at the half-mass radius of the core 
($7 \kpc$) is $m/\mv \!\simeq\! 0.32$ and $0.09$ respectively.
These are to be compared to the toy model predictions of 
$m/\mv \!\sim\! 0.39$ and $0.046$.
The energy deposited in the cusp is $\Wc/\Kc \!\simeq\! 2.0$ and $0.32$ for the
compact and diffuse satellite respectively.
In comparison,
substituting the mass fractions from SatGen in \equ{w_single} of the toy
model yields $\Wc/\Kc \!\simeq\! 2.05$ and $0.029$, 
emphasizing the qualitative difference between the effects of the
compact and diffuse satellites.
The semi-analytic simulation thus confirms the qualitative estimates
from the toy model that the fiducial compact satellite is expected to
significantly heat up the host cusp, while the fiducial
diffuse satellite is not expected to generate major heating.

\smallskip % total satellite
In order to evaluate the effect of including the central baryons in the
satellites, we re-ran SatGen with our fiducial NFW host halo but with the 
satellites following the more compact DZ-profile fits to the VELA simulated 
galaxies using the {\it total} mass rather than the dark matter alone.
The DZ profiles are here 
$(c,\alpha)\seq(5,0.5)$ and $(3,1.8)$
for the diffuse and compact satellites respectively, 
following the VELA pre and post compaction 
galaxies from \fig{vela_prof}.
We learn from \fig{app_satgen_single_tot}
(available as supplementary material online), which summarizes this SatGen run
compared to the fiducial case of \fig{satgen_single}, that the more compact 
satellites indeed penetrate to the host cusp with a higher mass and deposit 
more energy there accordingly, as expected.
However, the difference is rather small,
with $m/\mv \ssimeq 0.5$ compared to $0.4$,
and $\Wc/\Kc \ssimeq 2.5$ compared to $2$, for the compact satellites.
This is a smaller difference than expected based on the toy model of \se{toy}.
These results indicate that our main analysis using the fiducial satellite 
profiles as 
derived from the DM mass in the VELA simulated galaxies should provide good,
conservative estimates for the cusp heating by dynamical friction.

\smallskip % steep cusp
In order to evaluate the effect of a {\it steep-cusp} host halo, 
we re-ran SatGen with the fiducial DM satellite profiles but with the 
steep-cusp host instead of the fiducial NFW. 
We read from \fig{app_satgen_steep_single} 
(available as supplementary material online) 
that for the steep-cusp host  
the same satellites penetrate with $m/\mv \!\simeq\! 0.30$ and $0.075$ for
the compact and diffuse satellite, respectively.
These values are only slightly smaller than for the NFW host, probably 
because the two host haloes are not very different outside the cusp.
The satellite mass becomes significantly smaller only in the inner cusp, inside
$\sim\! 5\kpc$, where the tidal stripping is more efficient due to the higher
density of the steep-cusp halo.
The corresponding toy-model estimates are $m/\mv \!\simeq\! 0.23$ and $0.01$.
The energy deposited in the steep cusp according to SatGen is 
$\Wc/\Kc \ssimeq 2.0$ and $0.32$.
These values are rather similar to those in the NFW host.
Apparently, the lower satellite mass that has penetrated into the cusp is 
balanced by the higher density in the cusp, making the work by dynamical 
friction almost the same in the two cases.
The toy-model estimates are $\Wc/\Kc \ssimeq 0.56$ and $0.001$, indicating a
stronger dependence on the host-cusp properties than is actually produced by 
SatGen, but emphasizing again that the
effects of the compact and diffuse satellites are very different, and that the
compact satellite is expected to provide non-negligible heating also in the 
steep-cusp halo.

%========================
\subsection{A cosmological sequence of merging satellites}
\label{sec:satgen_cosmo}
 
We next run SatGen with a cosmological distribution of satellites over time
$\tau \tvir$, where $\tau \sim 1$, from $z\seq 2.3$ to $2$.
We actually generate random merger trees for a target halo of of a given mass
at $z\seq 2$ starting at $z\seq 20$, evolve these cosmological sequences of
satellites in time, and keep track of the DF heating induced by them
over the last virial time at each given halo-centric radius.
The satellite masses are drawn from the EPS conditional mass function
\citep[e.g.,][]{lacey93,neistein08_m,parkinson08,benson17}.
We assume that the orbital energy at infall through $\Rv$ is the same as that 
of a circular orbit of radius $\Rv(t)$, and draw the circularity $\epsilon$ at
random from the distribution $\dd P/\dd \epsilon = \pi\,\sin(\pi\epsilon)/2$, 
which approximates the distribution measured in cosmological simulations 
\citep[e.g.,][]{wetzel11,bosch17}, where 
$\la\epsilon\ra \!\simeq\! 0.45$ at $z\seq 2$.
We consider the infall locations to be isotropically distributed on the virial 
sphere. 
In the fiducial run, the host halo is again of mass $\Mv \seq 10^{12.5}\msun$
with an NFW profile and $c_{\rm NFW}\seq 5$. Every satellite with 
$\mv \!>\!  10^{11}\msun$ is assumed to be post-compaction, with the fiducial
compact DZ 
profile of $(c,a) \seq (7,1)$, while all the less massive satellites are
assumed to be diffuse, pre-compaction, with $(c,a) \seq (3,0.5)$. 
In a reference run, with the same distribution of masses and initial
orbits, all the satellites are assumed to be diffuse.
This reference case is supposed to approximate the real situation for less 
massive hosts, say $\Mv \!<\! 10^{11.5}\msun$, where no compact satellites (of
$\mv\!>\!10^{11}\msun$) are expected. 

\smallskip
\Fig{satgen_cosmo} shows the energy deposited within the sphere of radius $r$
about the halo center, during a period of one virial time at $z\seq 2$,
$\tvir \!\simeq\! 0.5\Gyr$, corresponding to a redshift interval 
$\Delta z \!\simeq\! 0.3$.\footnote{In the EdS regime, approximately valid at
$z\!>\! 1$, the relation is 
$\Delta t \!\simeq\! -1.68 \Gyr\,1+z)_3^{-5/2} \Delta z$.}
We read for the NFW host halo
$\Wc/\Kc \!\simeq\! 1.8$ and $0.35$ for the run with and without 
post-compaction satellites, respectively. 
This is compared to the toy model predictions of
%$\Wc/\Kc \ssimeq 5.59$ and $0.079$.
$\Wc/\Kc \ssimeq 5.6$ and $0.08$.
They both point to a similar qualitative difference
where the post-compaction satellites generate
significant cusp heating while the diffuse satellites provide only minor
heating during a time interval comparable to the virial time at $z\ssim 2$.

\smallskip
For the steep-cusp halo run of SatGen with a sequence of satellites, 
shown in \fig{app_satgen_steep_cosmo}
available as supplementary material online),
we obtain $\Wc/\Kc \ssimeq 1.8$ and $0.35$, respectively.
This is very similar to the SatGen results for the NFW host, 
consistent with the similarity of the results obtained for a single satellite.
The corresponding toy-model predictions are
$\Wc/\Kc \ssimeq 1.5$ and $0.003$, consistent with the SatGen result for the
compact satellite, though estimating less energy deposit in the
steep-cusp halo compared to the NFW halo.
We learn that the cosmological sequence of compact satellites during half a 
virial time is expected to provide significant cusp heating in the
steep-cusp host halo as well.

%%%%%%%%%%%%%%%%%%%%%%%%%%%%%%%%%%%%%%%%%%%% 5
\section{Post-heating Relaxation}
\label{sec:relax}

\begin{figure*} % 6
\centering
\includegraphics[width=0.33\textwidth,trim={0.8cm 0.9cm 0.4cm 0.8cm},clip]
%{figs/DF_0404NFW_rhof.pdf}
{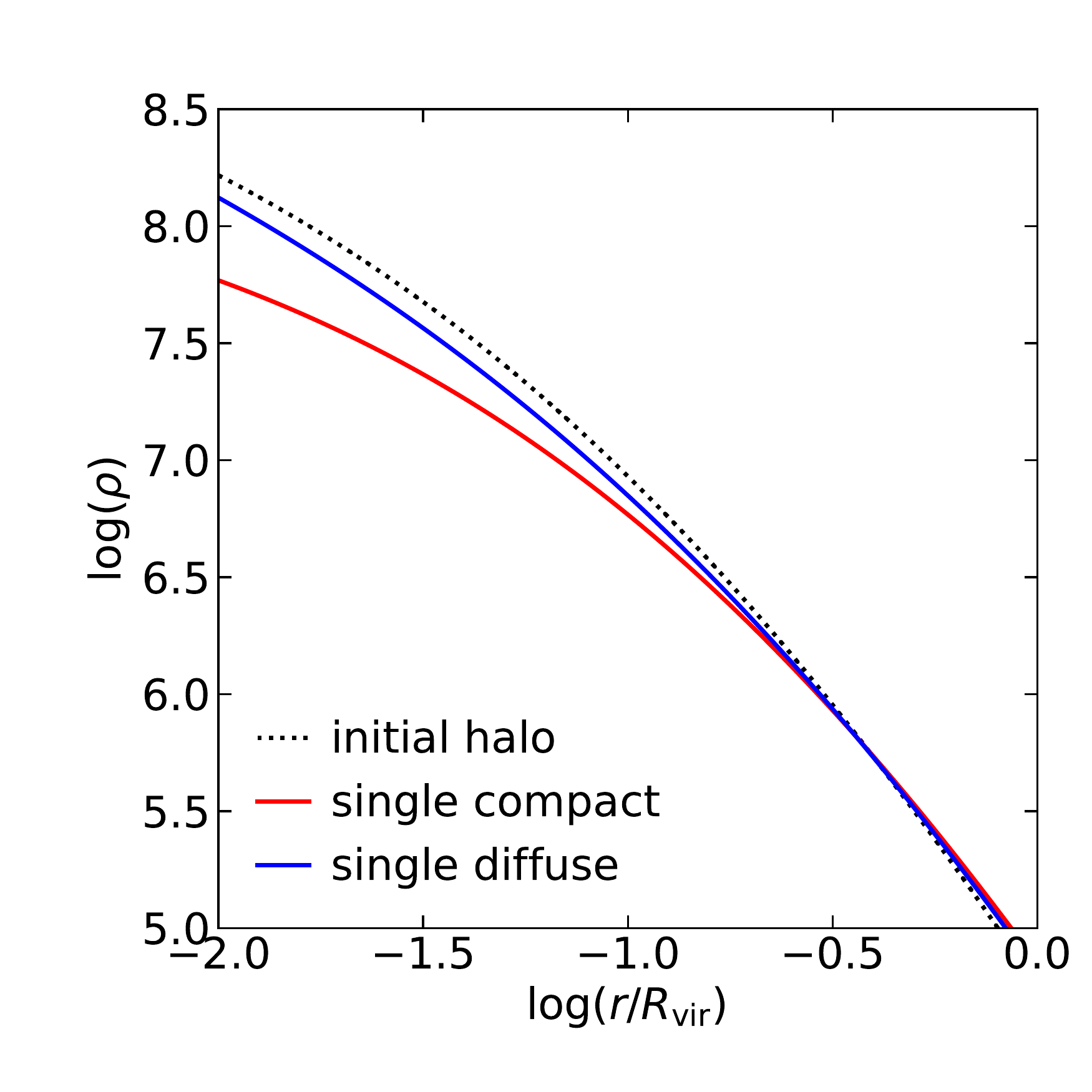}
\includegraphics[width=0.33\textwidth,trim={0.8cm 0.9cm 0.4cm 0.8cm},clip]
%{figs/DF_0404NFW_Lagrangian_dErf_W_compact.pdf}
{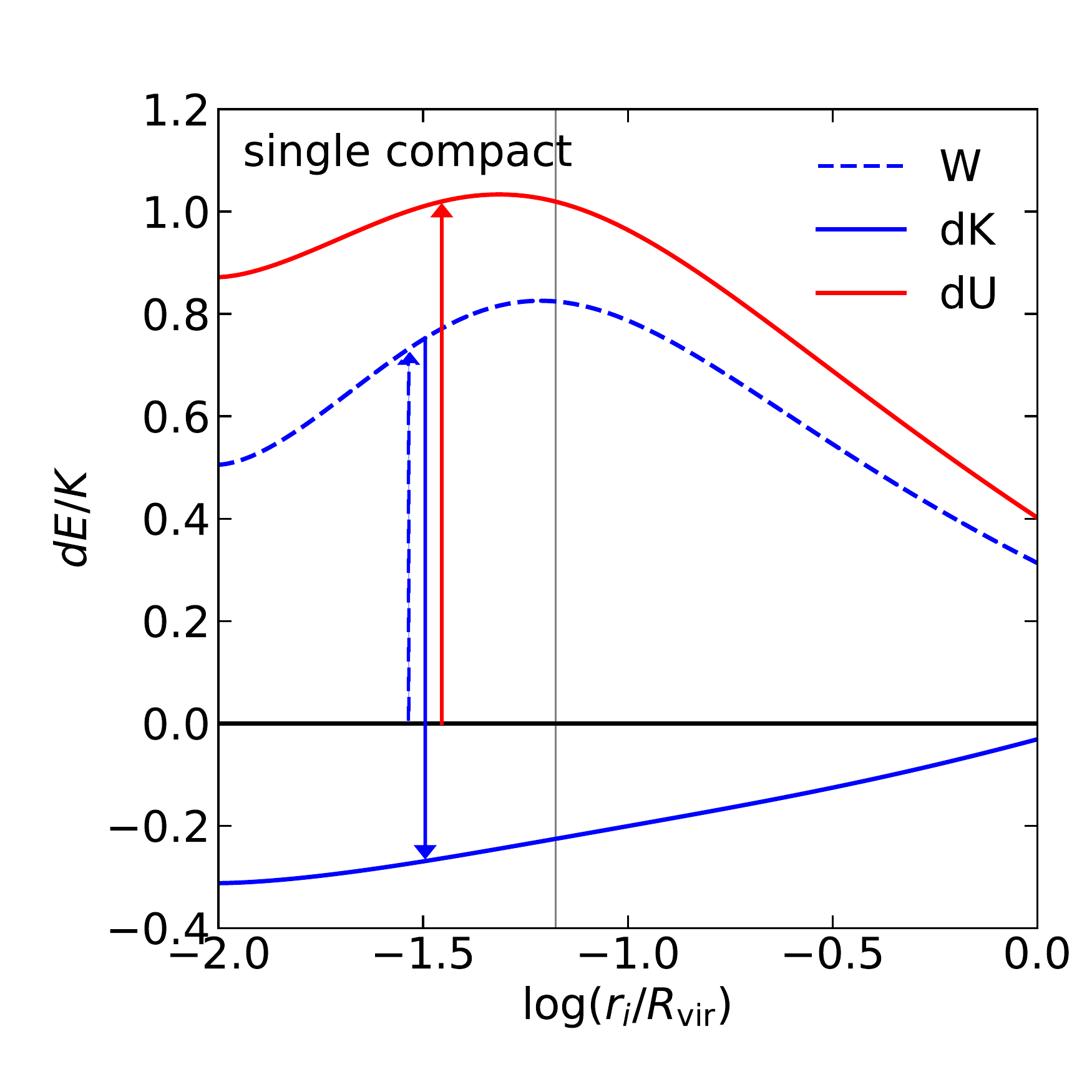}
\includegraphics[width=0.33\textwidth,trim={0.8cm 0.9cm 0.4cm 0.8cm},clip]
%{figs/DF_0404NFW_Lagrangian_dErf_W_diffuse.pdf}
{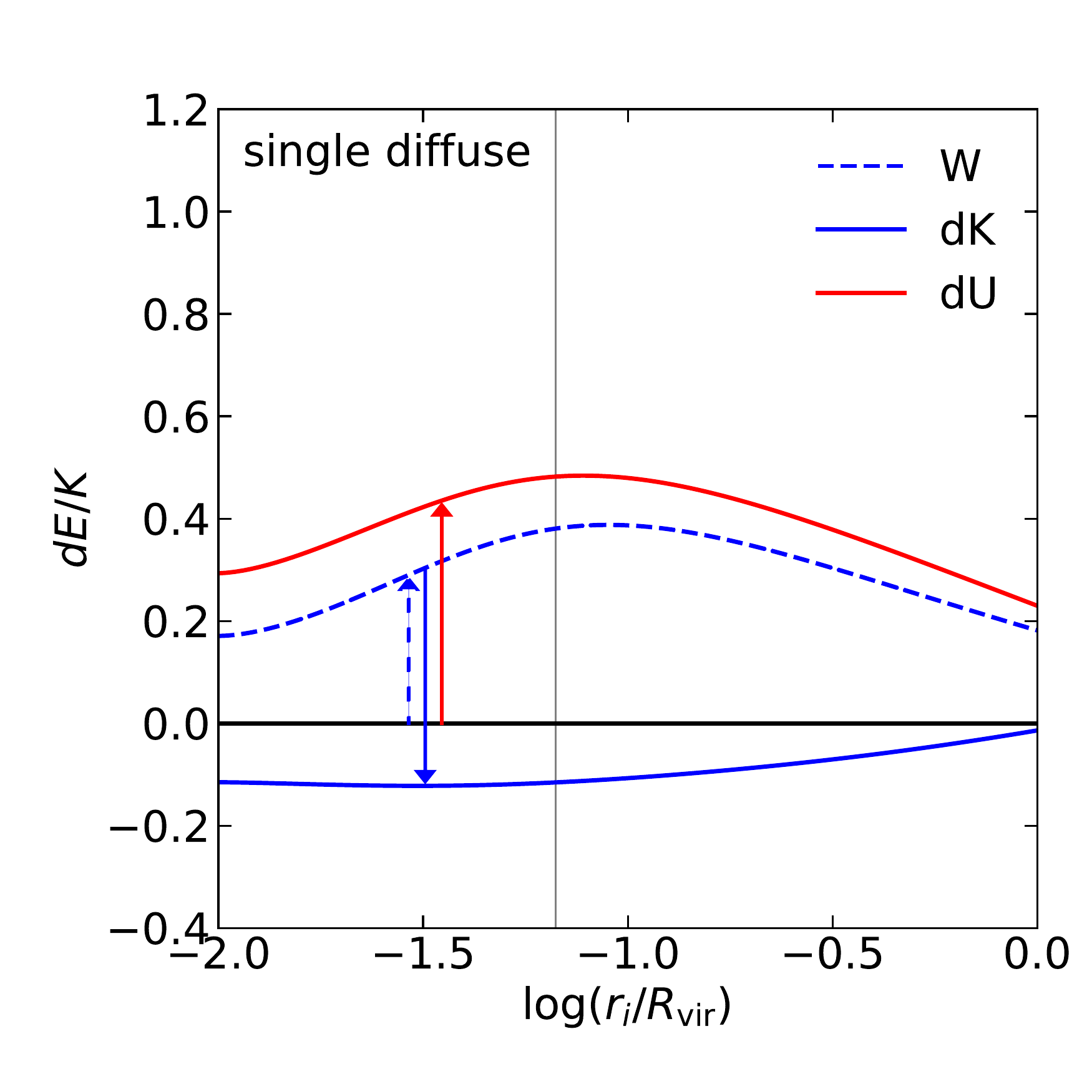}
\caption{
The relaxation of a DM halo
from an initial to a final configuration in Jeans equilibrium,
based on the CuspCore model,
after a deposit of kinetic energy that mimics dynamical friction by our
fiducial compact and a diffuse satellites.
The initial halo is NFW with a central point mass of $8\times 10^{10}\msun$
resenting baryons.
The energy deposited as a function of radius in the host is adopted from the
SatGen simulations, the bottom-left panel of \fig{satgen_single}.
Shown are the density profile (left)
and the changes in the kinetic and potential energies interior to radius $r$
($dK$ and $dU$, blue and red, respectively),
normalized by the kinetic energy $K(<r)$.
We see that after the initial heating of $K$ by the work of dynamical friction
$W$ (dashed), the changes are the same in amplitude,
$\Delta K \!\simeq\! -\Delta U$, as the cusp cools and flattens.
The compact satellite causes a much stronger effect.
%
%\adr{The normalization here at $\Rc$ is $\Kc=1.7\times 10^{16}$
%compared to $6\times 10^{15}$ in SatGen and toy
%(which should be $8\times 10^{15}$).}
}
\label{fig:DF-CuspCore}
\end{figure*}

Before we proceed to the response of the heated DM to AGN
outflows in \se{agn}, we address the behavior of the cusp after the DF heating.
In the spirit of our simple toy modeling,
we first assume that the work done by dynamical friction on the satellite
is instantaneously deposited in the dark matter as kinetic energy
(``heating"),
in terms of increased velocity dispersion, 
while the potential energy and the associated density
profile remain unchanged at that instant.
In the case of a single satellite, after it has disrupted or settled in the
center, we expect the host halo to relax over a certain time window $\tau\tvir$
into a new Jeans equilibrium. In this process the dark matter ``cools" while it
expands, namely the potential energy grows (becoming less negative) 
at the expense of the kinetic energy, and the cuspy density profile partly
flattens.
We first address this process in \se{toy_relax} via simplified toy models,
and then in \se{N-body} using an N-body simulation.

%================
\subsection{Toy models}
\label{sec:toy_relax}

For a first crude qualitative impression of this process,
we consider the cusp to be an isolated shell of mass $M$
in virial equilibrium, where the kinetic and potential energies are related via
$K\seq -U/2\seq GM^2/(2R)$,
both at the initial radius $R_{\rm i}$ and at the final radius $R_{\rm f}$,
after a deposit of energy $W$ and a relaxation process.
Conservation of energy implies
\be
-\frac{G M^2}{2 \Ri} +W = -\frac{G M^2}{2 \Rf} \, ,
\ee
so
\be
\frac{\Kf}{\Ki} = \frac{\Ri}{\Rf} = 1 -\frac{W}{\Ki} \, .
\label{eq:KfKi}
\ee 
The system expands, while it cools, from a peak kinetic energy of
$K_{\rm peak} \seq \Ki + W$, immediately after the action of dynamical
friction, to below the initial kinetic energy, $\Kf \seq \Ki - W$.
In the relaxation process, the kinetic and potential energy changes
are $\Delta K \seq -\Delta U \seq -2W$  
%\smallskip
Thus, after an energy deposit due to DF by a single satellite, 
the system may be found in one of two configurations. During the time window
for expansion and cooling, $\tau \tvir$ (to be estimated below), 
the system is ``hotter" then before with the cuspy profile practically 
unchanged.  After this period, the system has a flatter density profile 
but it is ``cooler" than in its initial state.

\begin{figure*} % 7
\centering
\includegraphics[width=0.95\textwidth]
%{figs/NB_Kexcess.pdf}
{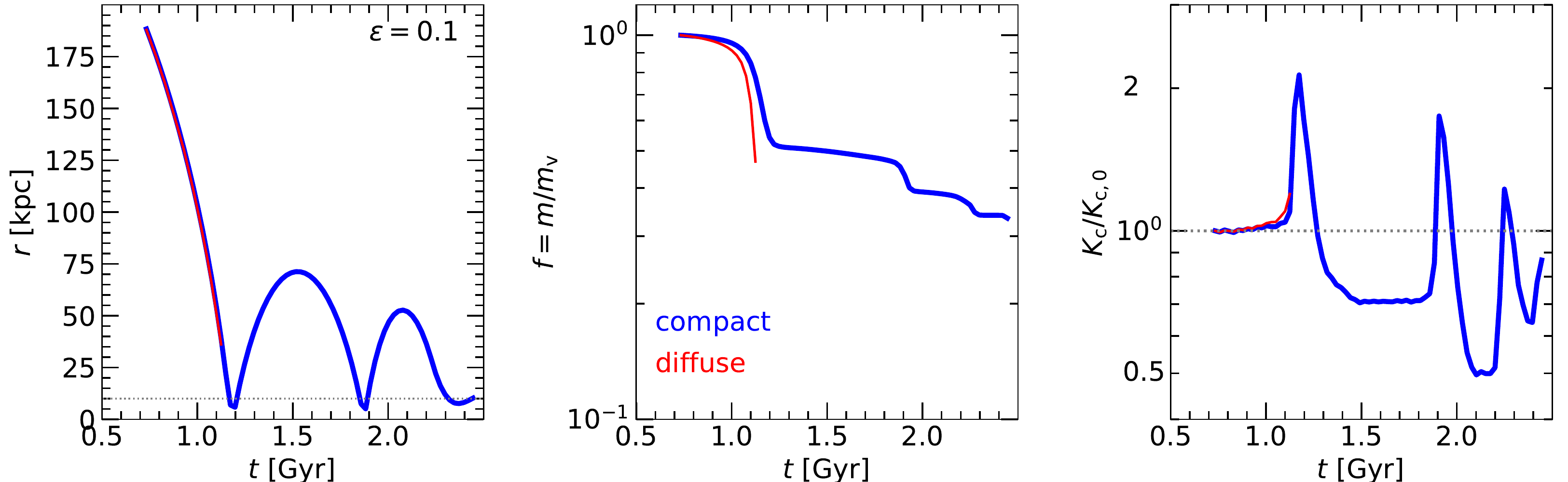}
\caption{
Time evolution in an N-body simulation of a compact satellite (blue)
and a diffuse satellite (red) of $\mv\seq 10^{11.5}\msun$ on a rather radial
orbit in a host halo of $\Mv\seq 10^{12.5}\msun$ at $z\seq 2$.
{\bf Left:} the satellite distance from the host centre.
{\bf Middle:} the bound mass of the satellite.
{\bf Right:} the kinetic energy within the host cusp with respect to its 
initial value.
We read that the heated episodes near the pericenters last a total of 
$\sim\! 0.25\Gyr$, which is $\sim\! 0.5\tvir$.
}
\label{fig:NB_K(t)}
\end{figure*}

\begin{figure*} % 8
\centering
\includegraphics[width=0.80\textwidth]
%{figs/HostProfileSequence.pdf}
{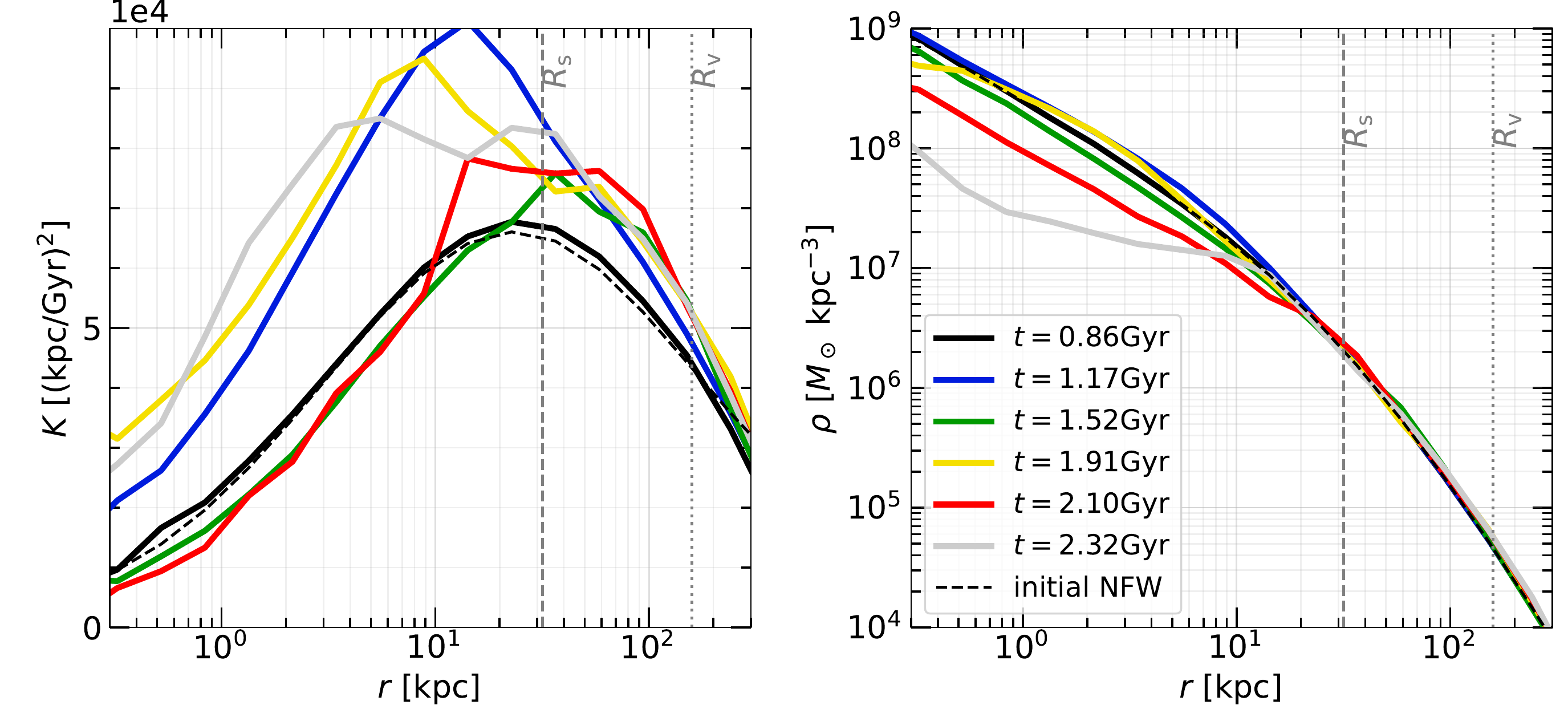}
\caption{
Evolution of the profiles of kinetic energy (left) and density (right) 
in an N-body simulation of a compact satellite in a massive halo, the same as
\fig{NB_K(t)}. The initial NFW profile is shown (dashed).
The times shown are:
(a) $t\seq 0.86\Gyr$ (black), virial crossing, cold, cuspy.
(b) $t\seq 1.17\Gyr$ (blue), first pericenter, hot, cuspy.
(c) $t\seq 1.52\Gyr$ (green), first apocenter, cold, cuspy.
(d) $t\seq 1.91\Gyr$ (yellow), second pericenter, hot, cuspy.
(e) $t\seq 2.10\Gyr$ (red), second apocenter, slightly hot, slightly flattened.
(f) $t\seq 2.32\Gyr$ (grey), third pericenter, hot, more flattened.
}
\label{fig:NB_sequence}
\end{figure*}

%\smallskip
%\adr{do a toy model for a full cusp?}

\smallskip %DF-CuspCore 
For a more quantitative estimate, we assume for the host halo
a Dekel-Zhao density profile both at the initial and at the final
configurations, both in Jeans equilibrium, as described 
in section 3.3 of \citet{freundlich20_cuspcore}.
The transition between the two states is assumed to start with an instantaneous 
addition of kinetic energy to the dark matter at every radius according to an 
assumed input profile $W(\slt r)$, mimicking the work by dynamical friction, 
while the mass distribution and therefore the potential energy are fixed.
This is followed by a relaxation to a new equilibrium while shells enclosing 
given DM masses are assumed to conserve energy.
This is analogous to the CuspCore model used in \se{agn} and described in 
Appendix \se{app_CuspCore} 
(available as supplementary material online)
for the response to outflows, except that the first 
instantaneous change is in the kinetic energy rather than in the potential 
energy. 
We assume here an initial host halo of $\Mv \seq 10^{12.5}\msun$
at $z\seq 2$, with an NFW profile of $c_{\rm NFW}\seq5$,
and adopt the corresponding input energy-deposit profile from the results 
of the SatGen simulations of a single satellite, the medians in the 
bottom-left panel of \fig{satgen_single}. 
This is for the fiducial compact and diffuse satellites, with 
$\mv/\Mv \seq 0.1$ and $\epsilon \seq 0.4$.
 
\smallskip
\Fig{DF-CuspCore} shows the resultant changes in the density profile
and in the kinetic and potential energies, $dK$ and $dU$,
interior to $r$, normalized by $K(<r)$, with respect to the initial 
configuration. The initial heating of $K$ through the work $W$ by
dynamical friction (dashed blue line) shows for the compact satellite 
maximum relative heating near the cusp radius of $10\kpc \ssimeq 0.067\Rv$. 
The subsequent relaxation to a new Jeans equilibrium (solid lines) 
is associated with cooling and expansion such that $\Delta K \ssim -\Delta U$, 
in qualitative agreement with the result from the crude shell toy model.
%The fact that $W/K$ is larger at smaller radii
%makes the expansion flatten the cusp toward a core. 
The effect of a compact satellite is naturally
much stronger than the effect of a diffuse satellite.

\smallskip % tau from N-body, and being maintained hot 
The N-body simulation described in \se{N-body}, of the effect of 
a single compact satellite on the cusp, confirms the expected two-stage
behavior of first heating the cusp without affecting the density profile,
and then relaxing to a cooler and extended configuration with a flatter
density profile.
We learn from this simulation that the time window for the ``hot" phase is
$t_{\rm hot} \ssim 0.5\tvir$. 
Substituting $\tau \seq 0.5$ in \equ{wvir}, we expect a total heating of
$\Wc/\Kc \!\simeq\! 0.75\, (1+z)_3$ during this hot phase.
This being of order unity at $z \ssim 2$ indicates that, at these high halo
masses and redshifts, the cusp can be maintained in the ``hot" phase, and thus
be more susceptible to core formation by AGN outflows (\se{agn}).
We will study in \se{agn} the DM response to AGN-driven outflows when
starting from either the ``hot" or the ``cold" configuration.

%=====================================
\subsection{Heating and flattening in N-body simulations}
\label{sec:N-body}

We realized in \se{toy_relax} that a halo cusp, after being heated by a
single satellite, would relax into a new Jeans equilibrium
in an energy-conserving process that involves expansion and cooling, 
yielding a cold cusp with a somewhat flatter density profile.
Here, we use an N-body simulation in order to verify these heating and cooling 
processes. 
We wish in particular to estimate the time it takes for the heated cusp to cool
down. This will be used as the effective duration for evaluating the heating 
effect of a cosmological sequence of satellites, expressed as $\tau$ in 
\equ{wvir} of the toy model and as the proper duration of the SatGen 
simulation in \se{satgen_cosmo}.
According to these estimates, if the cooling process takes a significant
fraction of a virial time or more, namely $\tau$ is not much smaller than
unity, then we can expect the cusp to be maintained hot by the sequence 
of mergers with compact satellites. If, on the other hand, $\tau \!\ll\! 1$,
the cusp will drop to a cool phase between the individual merger episodes.
The different responses of a hot cusp and a cooled-flattened cusp to
AGN-driven outflows will be discussed in \se{agn}.

\smallskip
The N-body simulation uses the public {\small GADGET}-2 code
\citep{springel05_gadget2},
with a particle mass of $m_{\mathrm{p}} \seq 3\!\times\! 10^5\msun$ 
and a softening length of 
$r_{\textrm{ref}}=0.0003\,(N_v/10^6)^{-1/3}\ellv$,
where $N_{\rm v} \seq \mv/m_{\mathrm{p}}$ is the number of particles
within $\ellv$ \citep[following][]{vankampen00,ogiya19},
namely $r_{\rm ref} \seq 20\pc$. 
We consider a satellite of $\mv\seq 10^{11.5}\msun$ orbiting a
host of $\Mv \seq 10^{12.5}\msun$ near $z \seq 2$ along a rather eccentric 
orbit of circularity $\epsilon \seq 0.1$ at the halo virial radius and a
velocity virial velocity at that radius.
We compare the cases of a compact and a diffuse satellite, where the compact 
satellite is initialized with an NFW profile of
$c_{\textrm{NFW}}=50$, and the diffuse satellite is initialized with a Burkert 
profile \citep{burkert95} of a scale radius $r_s \seq 12\kpc$, 
the same as the inner scale radius of an NFW profile with 
$c_{\textrm{NFW}}=6$.
These profiles are not the same as the fiducial profiles used 
in the rest of the paper, but they are rather close, and serve us for our
qualitative purpose here.
More details on this N-body simulation are provided in Appendix \se{app_N-body}
(available as supplementary material online).

\smallskip
\Fig{NB_K(t)} shows the time evolution of the relative excess of 
kinetic energy of the dark matter within the cusp with respect to its
initial kinetic energy.
It also shows the distance of the satellite from the host center, indicating
the successive pericenters and apocenters with the orbit decay into the cusp,
and the mass of the satellite at that time.
We see hot-phase periods near the first three orbit pericenters,  
relaxing to cold-phase periods in between as well as after the coalescence.
The overall time in the hot phase is $t_{\rm hot} \ssim 0.5\tvir$.

\smallskip
\Fig{NB_sequence} 
shows the density profile and kinetic-energy profile of
the host halo at different times through the merger process compared to
the initial cold NFW profile. 
In the kinetic energy,
we see hot phases in the cusp during periods about the first, second and third
pericenters, cooling down in between. The maximum heating is near and interior
to the cups radius of $\sim\!10\kpc$.
The density profiles show
that the initial profile hardly changes until after the first
pericenter. Very minor flattening starts appearing after the first pericenter,
and a more noticeable but still partial flattening develops after the second 
pericenter, becoming more significant after the third pericenter. 
The latter may be subject to uncertainties due to difficulties in identifying a
center for the halo at these stages of satellite coalescence with the host halo
center.

\smallskip
Substituting $\tau \seq 0.5$ in the toy-model estimate \equ{wvir}, 
we expect a total heating of
$\Wc/\Kc \!\simeq\! 0.75 (1+z)_3$ during this hot phase.
This being of order unity at $z \ssim 2$ indicates that, at these high halo
masses and redshifts, the cusp can be maintained for most of the time
in the ``hot" phase, and thus be more susceptible to core formation by AGN 
outflows, to be demonstrated in \se{agn}.

%%%%%%%%%%%%%%%%%%%%%%%%%%%%%%%%%%%%%%%%%%%%%%%% 6
\section{DM Response to AGN Outflow}
\label{sec:agn}

%\adr{Jonathan: please read carefully and improve this section}

%\adr{Either elaborate in Appendix CuspCore, or have here a subsection 
%on CuspCore.}

%\adr{May 3 fits:\\ 
%NFW single diffuse 16.61 0.71 compact 16.81 0.62 \\ 
%NFW cosmo diffuse 16.57 0.77 compact 16.74 0.41 \\
%Steep single diffuse 16.81 0.58 compact 16.57 0.65\\ 
%Steep cosmo diffuse 16.60 0.75 compact 16.78 0.40
%}

%===========================
\subsection{Modeling the response to outflows}

\smallskip % cuspcore and DZ
In order to model the dark-matter response in the inner halo
to an episode of AGN-driven gas outflow, we use
the {CuspCore} analytic model presented in \citet{freundlich20_cuspcore},
that has been originally implemented for supernova-driven outflows
in low-mass galaxies.
The model, as summarized in Appendix \se{app_CuspCore}\footnote{The simplest 
version of CuspCore is available online for application in 
https://github.com/Jonathanfreundlich/CuspCore.}
(available as supplementary material online),
consists of a two-stage response, starting from
an instantaneous change of potential due to an assumed central mass removal 
while the velocities are frozen,
and following by an energy-conserving relaxation to a new Jeans equilibrium.
The model assumes a spherical halo, isotropic velocities, Jeans equilibrium 
in the initial and the final stages, and energy conservation for shells 
enclosing a given DM mass.
This model proved successful when compared to cosmological simulations (NIHAO)
except in situations of major mergers.
It uses the DZ density profile with flexible concentration and inner
slope, utilizing its analytic expressions for the potential and kinetic 
energies, as described in Appendix \se{app_Dekel-Zhao} and 
\citet{freundlich20_prof}. 

\smallskip % initial halo
We assume here as before 
a halo of mass $\Mv\seq 10^{12.5}\msun$ with a given initial
density profile. We add a central point mass representing baryons
of $\Mg\seq\Ms\seq 4\times 10^{10}\msun$, following the average 
stellar-to-halo mass ratio for such a halo mass at $z\seq2$ \citep{behroozi19} 
and assuming a gas fraction of 0.5 for star-forming galaxies at that
redshift \citep{tacconi18}. The corresponding 
dark-matter velocity-dispersion profile is determined by requiring Jeans 
equilibrium \citep{freundlich20_prof}.
 
\smallskip % eta
The outflowing mass is assumed to be a fraction $\eta$ of the gas mass, ranging
from zero to unity, namely a fraction $0.5\eta$ of the central baryonic mass.
We estimate in \se{gas} that the typical gas mass in the galaxy of
%a galaxy within a halo of $\Mv \ssim 10^{12.5}\msun$ at $z \ssim 2$, 
$M_{\rm gas} \ssim 4\times 10^{10}\msun$
is comparable to the typical mass of the fresh cosmologically accreted gas 
during $\sim\! 0.5\tvir \sim 0.25\Gyr$. 
Taking into account accreting recycled gas that is three times more massive
than the accreting fresh gas,
the gas available for ejection by AGN feedback during that period is 
$\sim\! 1.2\times 10^{11}\msun$.
Recall that we learned from the N-body simulation of \se{N-body} that
$\sim\!0.5\tvir$ is the ``cooling" time for the cusp heated 
by dynamical friction, and we will show below that
a pre-heated cusp is most responsive to outflows,
so this is the time interval within which AGN feedback is most effective
in producing a core. 
We model here only a single episode of $\eta\seq 0.5$ or $\eta\seq 1$, 
and consider it to provide a lower limit for the effect of core formation.

\begin{figure*} % 9
\centering
\includegraphics[width=0.33\textwidth,trim={1.2cm 0.9cm 0.7cm 0.9cm},clip]
%{figs/DF_0525NFW_single_heat_vs_flat_rho_compact_1_1W.pdf}
{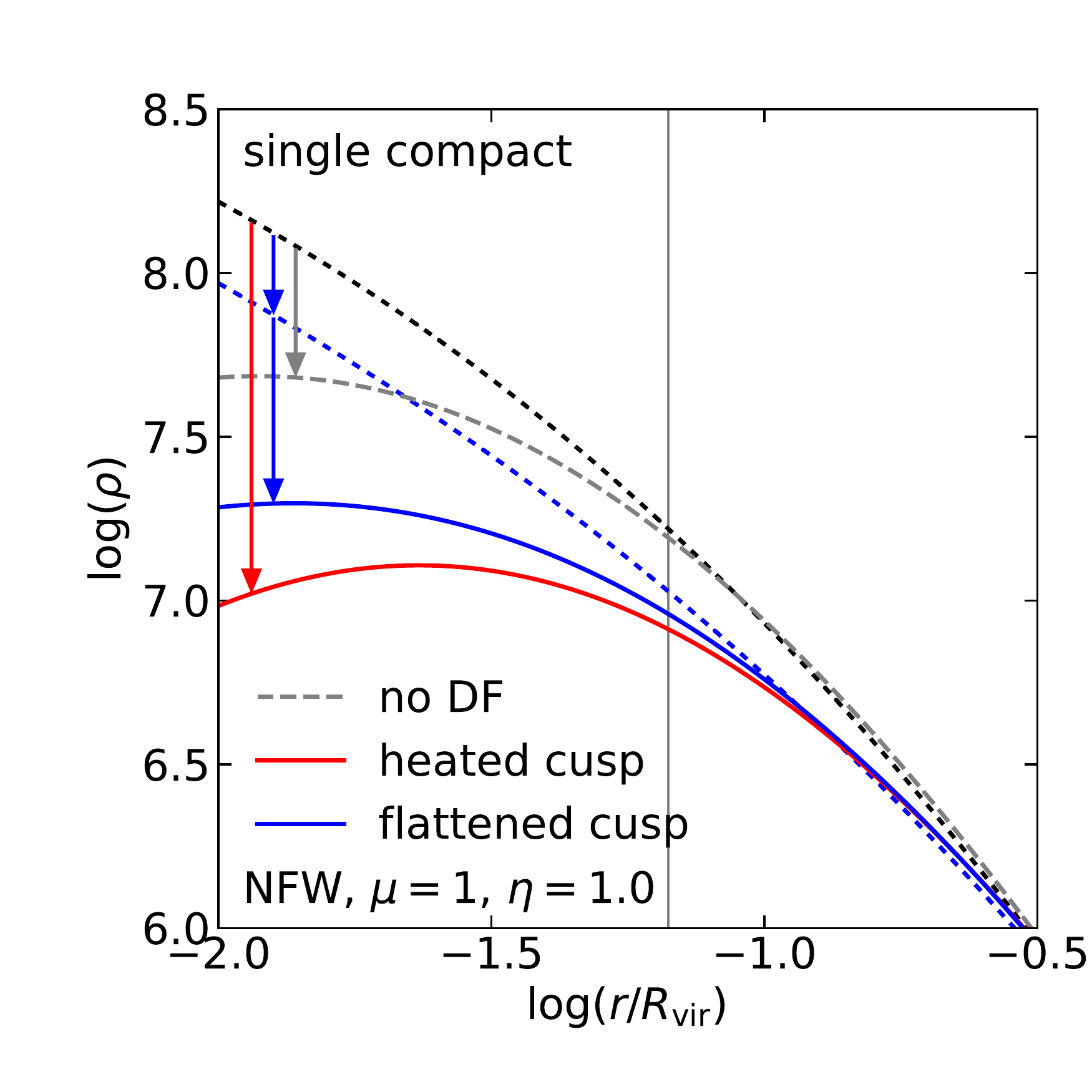}
\includegraphics[width=0.33\textwidth,trim={1.2cm 0.9cm 0.7cm 0.9cm},clip]
%{figs/DF_0525NFW_single_heat_vs_flat_rho_compact_05_2W.pdf}
{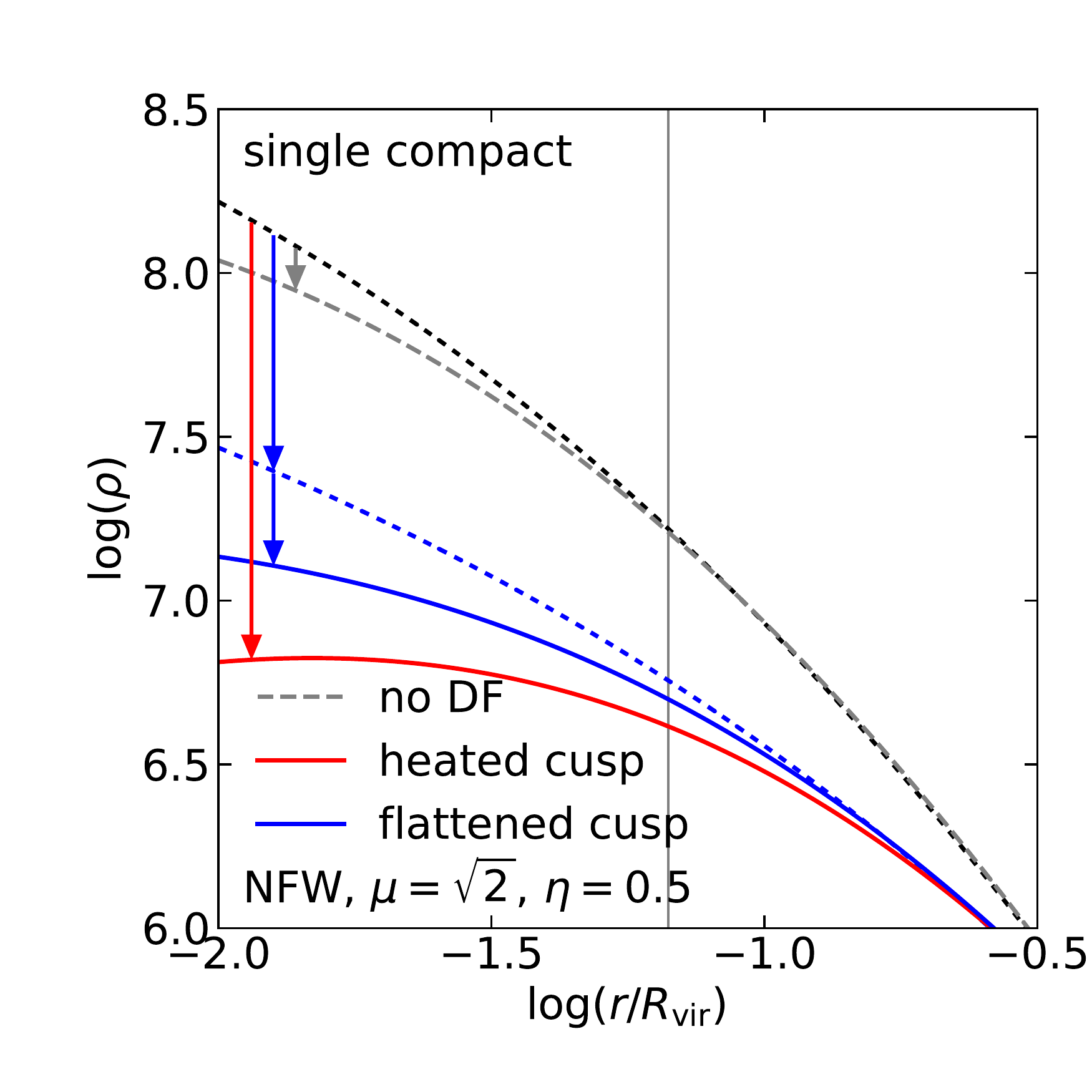}
\includegraphics[width=0.33\textwidth,trim={1.2cm 0.9cm 0.7cm 0.9cm},clip]
%{figs/DF_0525NFW_single_heat_vs_flat_rho_diffuse_05_2W.pdf}
{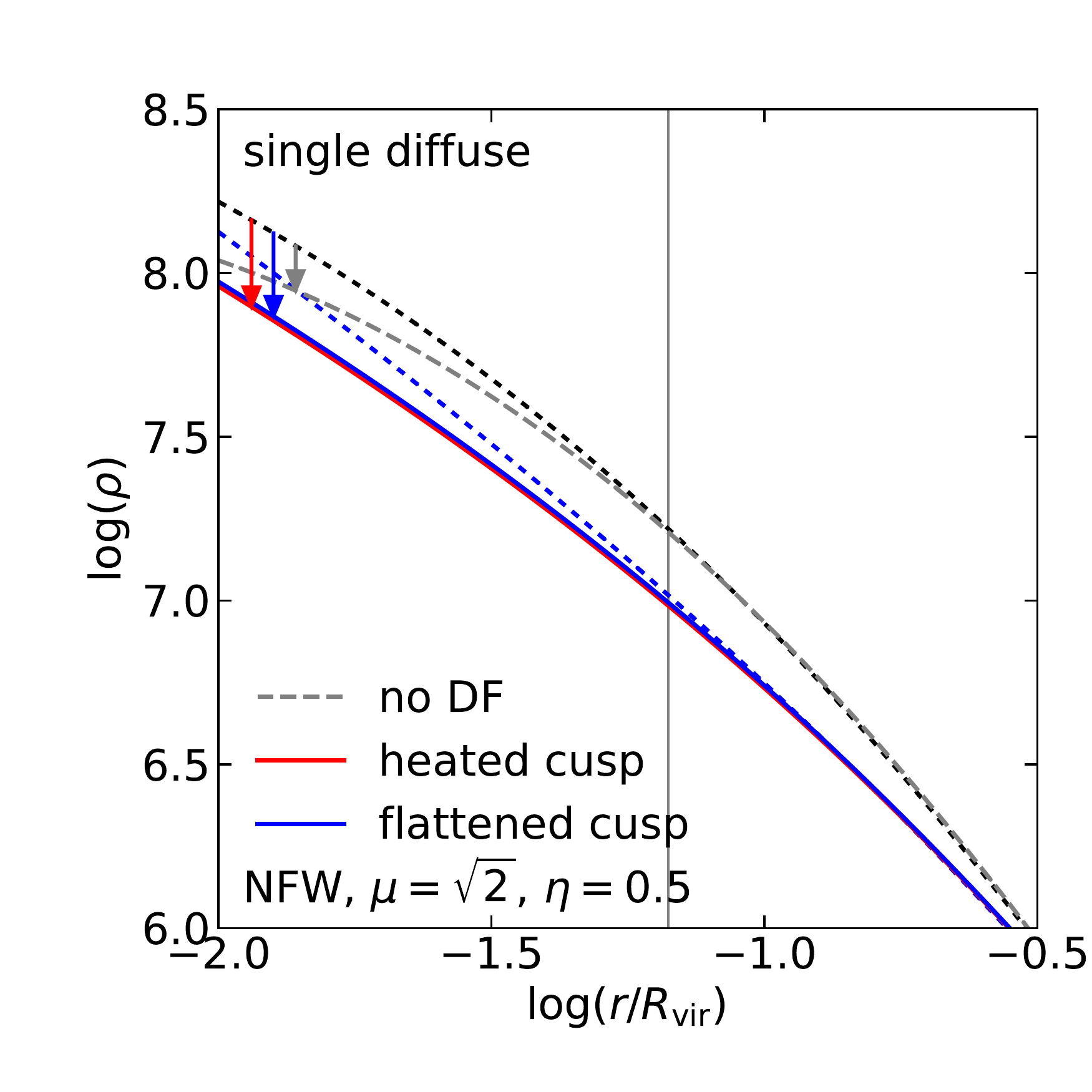}
\caption{
Dark-matter density-profile response to an outflow according to the CuspCore 
model.
The initial halo is NFW with $\Mv\seq10^{12.5}\msun$ and $c_{\rm NFW}\seq 5$
with $\Rv \seq 150\kpc$ at $z\seq 2$.
The DF heating is by a single fiducial satellite of initial mass $\mv$,
compact or diffuse, based on the energy-deposit 
profile as revealed by SatGen, using the fit shown in the bottom-left panel 
of \fig{satgen_single} and \tab{fits}.
The instantaneous outflow involves a fraction $\eta$ of the central gas
mass of $10^{10.6}\msun$.
{\bf Left:} a compact satellite $\mv\seq 0.1\Mv$ and $\eta\seq 1$.
{\bf Middle:} a compact satellite and $\mv\seq 0.1\sqrt{2}\Mv$ 
(obtained by doubling the energy deposited by DF according to 
\fig{satgen_single}) and $\eta=0.5$.
{\bf Right:} the same but for a diffuse satellite.
Three cases are shown in each panel as follows.
{\bf Grey}: the dotted line is the pre-DF initial cuspy DM profile, 
cold NFW in Jeans equilibrium.
It is assumed to be the pre-outflow profile,
with the dashed grey line referring to the corresponding post-outflow profile.  
{\bf Red:} the pre-outflow density profile is the same grey dotted 
line, but in this case the halo has been heated by dynamical friction, 
and the solid red line is the corresponding post-outflow profile, showing a
flat core extending to the core radius of $\sim\!10\kpc$ (vertical line).
{\bf Blue:} the dotted line is the somewhat flatter density profile late 
after the heating by dynamical friction, after relaxation involving 
expansion and cooling. With this cold pre-outflow density profile,
the solid blue line refers to the post-outflow profile, showing a core 
but not as flat as in the pre-heated case. 
Results for a steep-cusp halo are shown in \fig{app_outflow_steep_single}
(available as supplementary material online).
We learn that the diffuse satellite causes a much weaker effect than the
compact satellite, as predicted by the toy model.
We also learn that a heated cusp is more responsive to an outflow than a 
flattened, cold cusp, though both can end up as extended cores under 
pre-heating by the compact satellite.
%\adr{Consider adding $\mu=1$, $\eta=0.5$ that is not enough for a core}
}
\label{fig:outflow_single}
\end{figure*}

%=========================
\subsection{A single satellite}

\begin{table}
\centering
\begin{tabular}{@{}cccccc}
\multicolumn{6}{c}{{\bf Fitting function for the energy deposited by DF}} \\
\hline
Host & Satellite & Compactness & A & B & $\Rm$ \\
% & $10^{12}\msun$ & $10^{10}\msun$ & $10^{10}\msun$ & $\msun/\yr$ & kpc & kpc &
%kpc & km/s & km/s & & &\\

\hline
\hline
NFW & single & compact & 16.81 & 0.62  & $\Rv$ \\
NFW & single & diffuse & 16.61 & 0.71  & $\Rv$ \\
NFW & cosmo  & compact & 16.80 & 0.24  & $692$ \\
NFW & cosmo  & diffuse & 16.70 & 0.45  & $610$ \\
\hline
steep & single & compact & 16.81 & 0.58  & $\Rv$ \\
steep & single & diffuse & 16.57 & 0.65  & $\Rv$ \\
steep & cosmo  & compact & 16.70 & 0.70  & $\Rv$ \\
steep & cosmo  & diffuse & 16.50 & 1.00  & $\Rv$ \\
\hline
\end{tabular}%\newline
\caption{
We use as input to the outflow model CuspCore the 
energy deposited in the host cusp by the dynamical friction exerted on
satellites as obtained from the corresponding SatGen simulations.
We actually use a functional fit (shown in the SatGen figures) of the form
$\log_{10} W(\slt r) \seq A-B\,[\log_{10}(r/\Rm)]^2$,
with the parameters listed in this table, 
where $A$ and $B$ are log energy in units of $\msun\kpc^2\Gyr^{-2}$,
and $\Rm$ is in kpc, with $\Rv\seq 150\kpc$.
}
\label{tab:fits}
\end{table}

%\smallskip % DF halo, sats, W, outflow
\Fig{outflow_single} shows the change in the DM density profile
due to an instantaneous outflow episode based on the CuspCore model. 
The initial DM halo is our standard, with $\Mv \seq 10^{12.5}\msun$  
and an NFW profile of $c_{\rm NFW}\seq 5$ at $z\seq 2$.
%\footnote{It is actually a Dekel-Zhao
%profile with $(c,\alpha)\seq(7.126,0.2156)$ or $(c_2,s_1)\seq(5.035,0.9076)$ 
%which is a best fit to NFW using uniformly spaced log radii in the range 
%$\log(r/\Rv) = (-2,0)$.}
The initial kinetic energy obeys Jeans equilibrium 
\citep[following][]{freundlich20_cuspcore},
with an additional presence of a $10^{10.9}\msun$ central point mass 
representing a baryonic component, half gas and half stars.
The pre-outflow configuration is the result of DF heating by a
single satellite of mass $\mv$, %\seq 10^{11.5}\msun$, 
either the compact or the diffuse fiducial cases deduced
from the cosmological simulations described in \se{vela}. 
The energy deposited by dynamical friction is read from
the SatGen simulation for a satellite of $\mv\seq 0.1\Mv$,
as shown in \fig{satgen_single}. We actually use 
the functional fit that is shown in the figure and presented in \tab{fits}.
%$\log_{10} W(\slt r) \seq A-B\,[\log_{10}(r/\Rm)]^2$.
%For the NFW host and single satellite the parameters are
%$(A,B,\Rm) \seq (16.81,0.62,\Rv)$ and $(16.61,0.71,\Rv)$  % May 3 by Jonathan
%for the compact and diffuse satellites respectively, where the energy is  
%in $\msun\kpc^2\Gyr^{-2}$.}
%logE_cosmo_compact=16.51-0.51*(logr-log10(Rvir))^2
%logE_cosmo_diffuse=16.65-1.1*(logr-log10(Rvir))^2
%logE_single_compact=16.81-0.58*(logr-log10(Rvir))^2
%logE_single_diffuse=16.57-0.65*(logr-log10(Rvir))^2
%\adr{$1 \kpc \Gyr^{-1} = 0.955 \kms$}
%
A satellite mass of $0.1 \mu \mv$ is mimicked by multiplying the deposited
energy from \fig{satgen_single} by $\mu^2$, following \equ{w_single}.
The mass removed by the outflow is assumed to be a fraction $\eta$
of the $10^{10.6}\msun$ gas (a fraction $0.5\eta$ of the baryonic mass), 
and the halo expands differentially while conserving energy into its final
Jeans equilibrium according to the CuspCore model.
The two cases shown, both ending up with an extended core,
are for $\mu\seq 1$ with $\eta \seq 1$
and for $\mu \seq \sqrt{2}$ with $\eta \seq 0.5$.

\begin{figure*} % 10
\centering
\includegraphics[width=0.33\textwidth,trim={1.2cm 0.9cm 0.7cm 0.9cm},clip]
%{figs/DF_0525NFW_cosmo_heat_vs_flat_rho_compact_05_02W.pdf}
{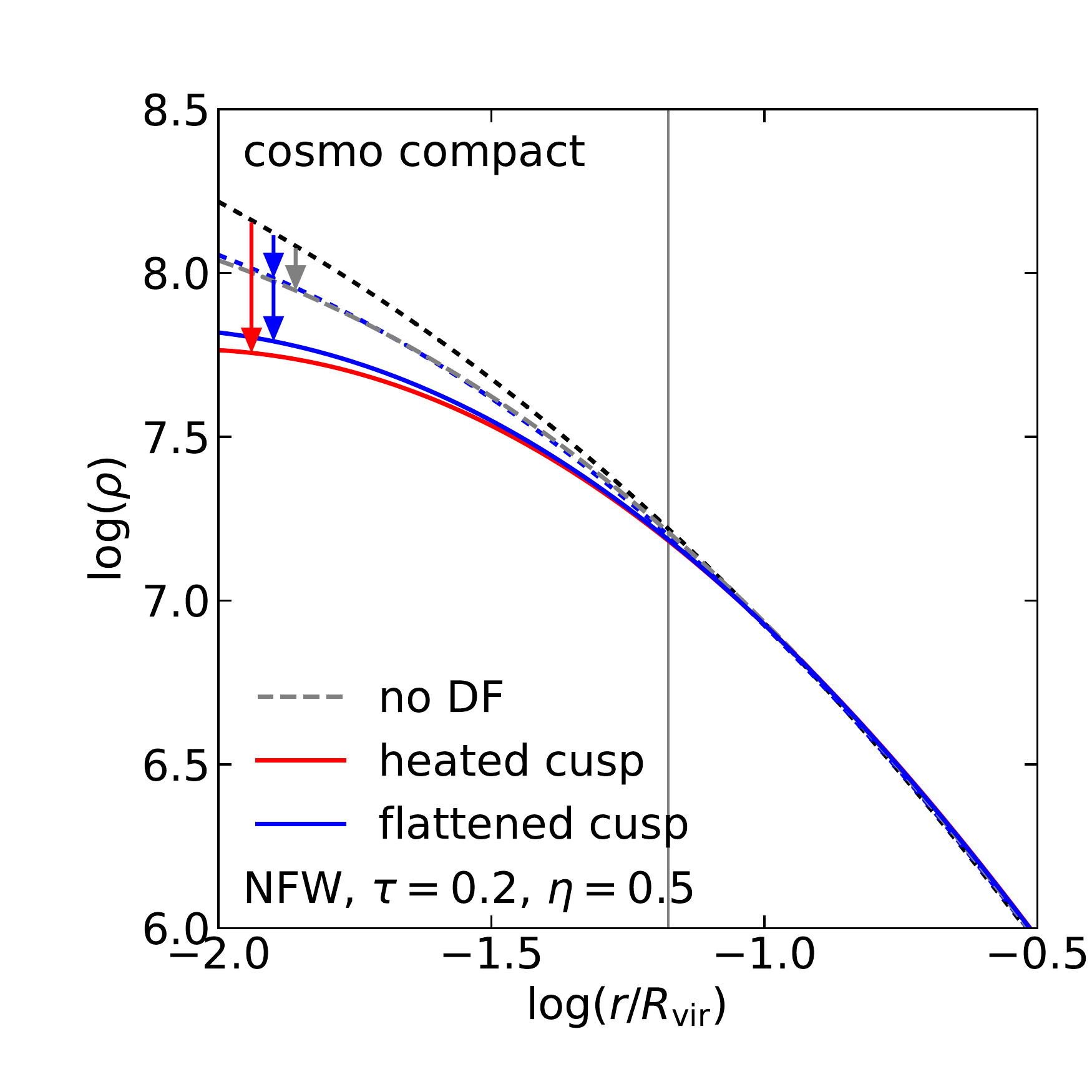}
\includegraphics[width=0.33\textwidth,trim={1.2cm 0.9cm 0.7cm 0.9cm},clip]
%{figs/DF_0525NFW_cosmo_heat_vs_flat_rho_compact_05_04W.pdf}
{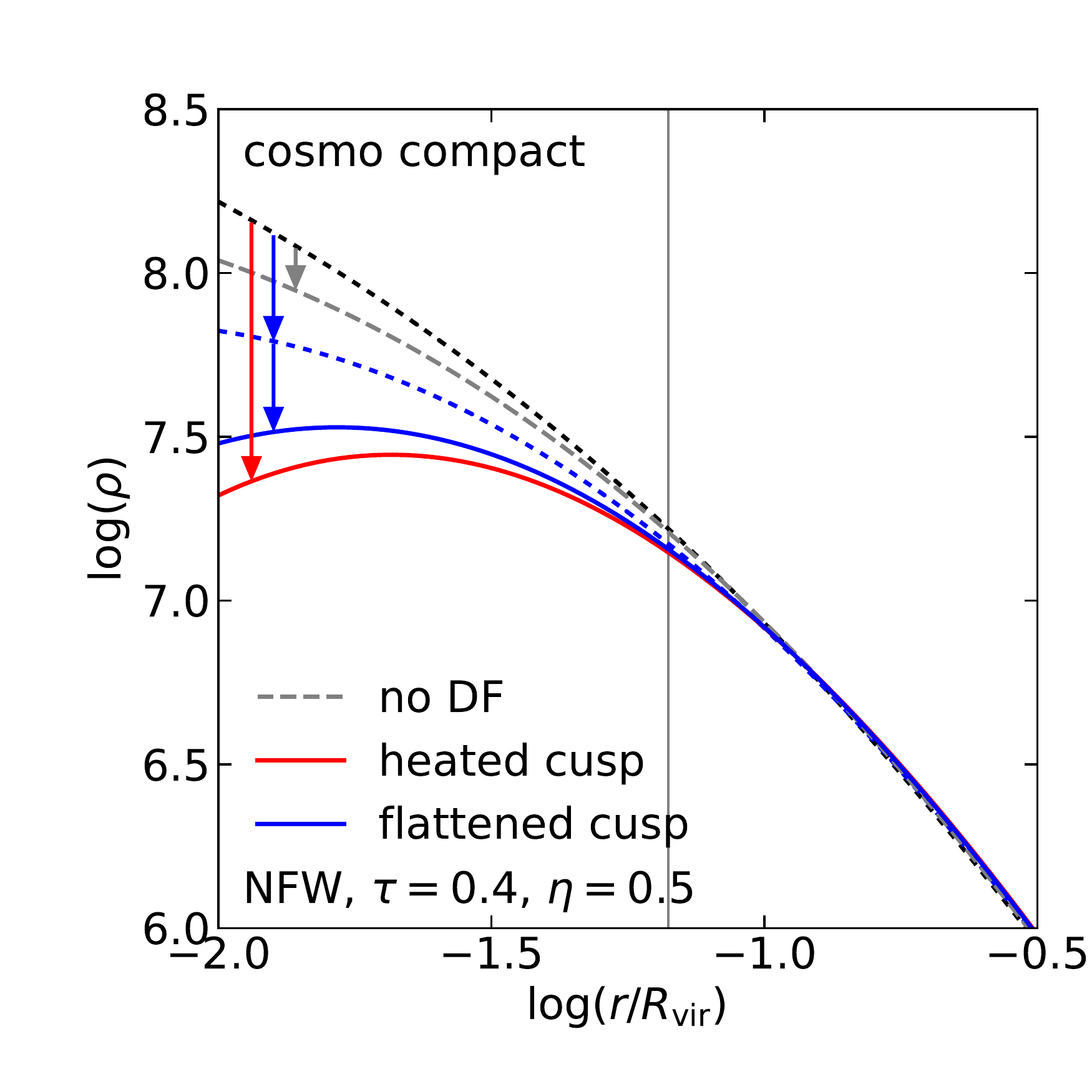}
\includegraphics[width=0.33\textwidth,trim={1.2cm 0.9cm 0.7cm 0.9cm},clip]
%{figs/DF_0525NFW_cosmo_heat_vs_flat_rho_diffuse_05_04W.pdf}
{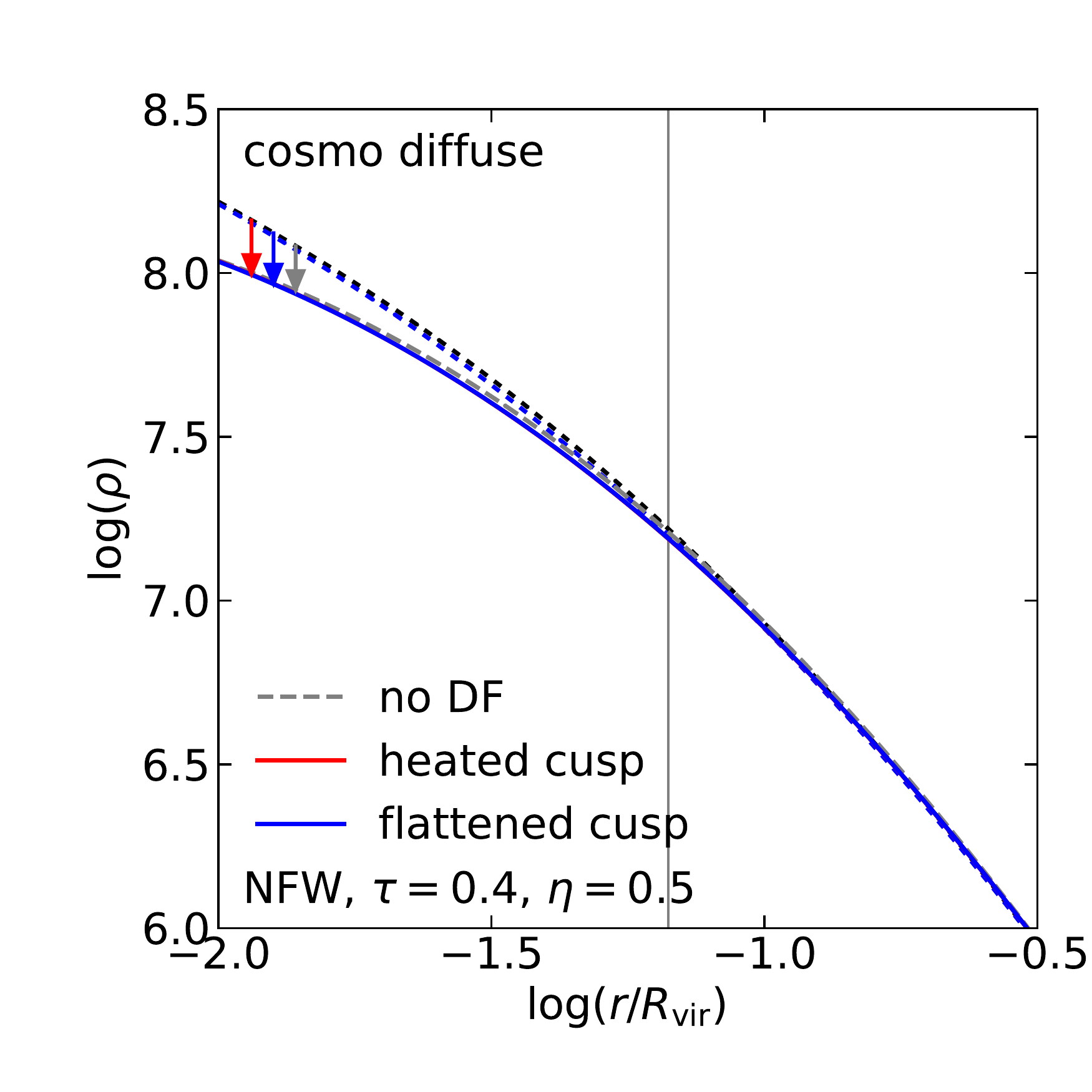}
\caption{
Similar to \fig{outflow_single}, with the same NFW initial host halo,
but for DF energy deposit by a cosmological sequence of satellites.
The energy deposited during $\tau\tvir$ is taken to be $\tau$ times the fit to
the median from the SatGen runs for $\tvir$ shown in \fig{satgen_cosmo} and
\tab{fits}.
{\bf Left:} compact satellites with $\tau\seq 0.2$.
{\bf Middle:} compact satellites with $\tau\seq 0.4$,
mimicking the typical duration of the hot phase
discussed in \se{relax}.
{\bf Right:} the same but for diffuse satellites.
The outflow is with a fraction $\eta\seq 0.5$ of the gas.
We learn that the effect of a typical sequence of satellites during
$\lsim\!0.5\tv$ is comparable to the effect of a single
$\mv\seq 0.14\Mv$ satellite.
This implies that the case of a pre-heated cusp (red) is realistic,
generating a flat core extending to $\sim\!10\kpc$
once the halo mass is above the golden mass, such that the
satellites are above the mass-threshold for compaction, and the halo allows
effective AGN feedback.
The diffuse satellites have a negligible effect, as predicted.
The corresponding results for a steep-cusp halo are shown in
\fig{app_outflow_steep_cosmo}
(available as supplementary material online).
}
\label{fig:outflow_cosmo}
\end{figure*}

\smallskip % 3 cases
Three cases are shown in each panel of \fig{outflow_single}, as follows.
The first (black) is a reference case of an outflow
ignoring any pre-heating by dynamical friction, 
starting from the initial cold NFW halo (short-dashed black).
One can see that 
the post-outflow cusp (long-dashed grey) is somewhat flatter than the original
cusp, but even with $\eta\seq 1$ it is not as extended and as flat as deduced 
from observations.

\smallskip % heated
In the second case in each panel (red), 
the pre-outflow configuration is set to mimic the hot phase
soon after the energy deposit by dynamical friction, namely the
density profile is the same initial NFW profile while all the 
input energy is deposited as kinetic energy.
We see that in this case the cusp heated by the compact satellite
generates an extended flat core, as desired (solid red).
On the other hand, the cusp heated by the diffuse satellite shows only a minor
flattening.
The unrealistic positive slope at small radii is an artifact of the 
CuspCore model, and should be interpreted as a flat inner core.

\smallskip % heat-flat
In the third case in each panel (blue), 
the pre-outflow configuration is set to mimic the cooled,
slightly flattened phase after relaxation, which may in principle
be reached a while ($\sim\!0.5\tvir$)
after the heating by dynamical friction (short dashed blue). 
Now, in the case of a compact satellite,
the outflow generates a flattened profile (solid blue), but not as flat and
extended as in the case of a hot pre-outflow cusp.
We learn that the response of a pre-heated inner DM halo to 
central mass ejection is indeed much stronger than that of a standard
cold NFW halo, and stronger than that of a cooled flattened cusp that may in
principle be reached late after the heating by dynamical friction.

%=================
\subsection{A cosmological sequence of satellites} 

%\smallskip % cosmo
\Fig{outflow_cosmo} shows the same as \fig{outflow_single} but now
starting with the DF energy deposit from the SatGen simulation of a 
cosmological sequence of satellites, the fiducial compact and diffuse, 
based on \fig{satgen_cosmo}, which refers to a duration of $\tau\tvir$ with
$\tau\seq 1$. The functional fits used here and shown in the figure
are listed in \tab{fits}.
%before, $\log_{10} W(\slt r) \seq A-B\,[\log_{10}(r/\Rm)]^2$.
%For the NFW host and a cosmological sequence of satellites the parameters are
%$(A,B,\Rm) \seq (16.8,0.24,692\kpc)$ and $(16.7,0.45,610\kpc)$ % May 7 by JF 
%for the compact and diffuse satellites respectively, where the energy is    
%in $\msun\kpc^2\Gyr^{-2}$.}
\Fig{outflow_cosmo} assumes $\tau \seq 0.5$, namely half the energy deposited
by DF at each radius in \fig{satgen_cosmo},
to match the duration of $\sim\!0.5\tvir$ for a heated cusp, as discussed in
\se{relax}, based on the estimates from
the N-body simulation described in \se{N-body}, \fig{NB_K(t)}.
The positive slope at small radii is an artifact of the CuspCore model, and
should be interpreted as a flat inner core.

\smallskip
We learn that the overall effect of an outflow after heating by a sequence of 
satellites over half a virial time is comparable to, and slightly stronger
than, the effect of an outflow after heating by a single satellite of 
$\mv/\Mv \ssim 0.14$ based on \fig{outflow_single}.
This implies that the case of a pre-heated cusp (red) is realistic
for the median galaxy, 
generating a flat core extending to $\sim\!10\kpc$
once the halo mass is above the golden mass, such that the
satellites are above the mass-threshold for compaction, and where
the halo allows effective AGN feedback.

%====================
\subsection{Other cases}

%\smallskip % steep
\Figs{app_outflow_steep_single} and \ref{fig:app_outflow_steep_cosmo}
(available as supplementary material online)
are analogous of \figs{outflow_single} and \ref{fig:outflow_cosmo}
but for an initial host halo with a cusp steeper than NFW, 
a DZ profile with  $(\c2h,\s1h)\seq(5,1.5)$, the same as studied
in \figs{app_frac_steep}, \ref{fig:app_satgen_steep_single},
and \ref{fig:app_satgen_steep_cosmo}.
We learn that core formation in response to AGN-driven outflow is more 
challenging in the steep-cusp halo than in the NFW halo,  
both in terms of less heating by dynamical friction and a weaker response
to an outflow. 
In order to demonstrate the formation of an extended core by compact satellites
in this case, one needs a larger energy deposit by dynamical friction and/or
a stronger outflow.
For a single satellite, \figs{app_outflow_steep_single}, 
we model a satellite of $\mv \seq 0.1\mu \Mv$ with $\mu \seq \sqrt{3}$
and an outflow that involves all the gas mass of $10^{10.6}\msun$, 
$\eta\seq 1$. 
In the case of DF heating by a cosmological sequence of compact satellites,
\fig{app_outflow_steep_cosmo}, we learn that an outflow of $\eta \seq 2$,
namely involving at least two succesive episodes of outflow,
is sufficient for producing a moderate core without DF heating.
A flatter core is produced with $\tau\seq 1$ and $\eta\seq 2$ if the slope of 
$K(\!<\!r)$ within the cusp is artificially flattened in order to make a better
fit to the DZ profile in CuspCore, and thus allow convergence with 
$\tau\seq 1$.
The high $\tau$ can be interpreted as
representing the top $1/3$ of the random realizations of the satellites
in the sequence (in terms of mass and circularity) during $\sim\!0.5\tvir$. 
Alternatively it can be interpreted as the median during
a longer period of $\sim\!\tvir$. 
Thus, we learn that an extended core can be reproduced even in an initially
steep cusp, at least in a fraction of the galaxies.

%\smallskip
%\adr{Jonathan: here or in future work,
%compute effect of an outflow on the profile, for the fiducial NFW, 
%a compact satellite, and heated phase, as a function of 
%(a) $\eta$, 
%(b) baryonic mass, 
%(c) baryonic-mass spatial distribution,
%(d) More than one AGN episode.}

\smallskip % tests
We explored the general dependence of the formation of a core on 
various parameters of the model.
This was done using the fiducial case of an NFW host of 
$\Mv\seq 10^{12.5}\msun$ and a single satellite 
of $\mv\seq 0.1\Mv$ with the fiducial compact or diffuse DM profile on a 
typical orbit, and with an outflow of $\eta\seq 0.5$ of the gas. 
The general conclusions are as follows.
(1) While a compact satellite with an outflow of $\eta\seq 0.5$ is needed for 
an extended core, even a diffuse satellite can lead to a core
with a stronger outflow of $\eta\ssim 1$, but it is at best a small core of 
a few kiloparsecs.  
(2) While a gas mass of $\sim\!10^{10.6}\msun$ leads to a core by
an $\eta\seq 0.5$ outflow even if it occurs after post-heating 
relaxation, \se{relax},
a gas mass smaller than $\sim\!10^{10}\msun$ may be enough for 
generating a core provided that the outflow occurs during the hot phase of the
DM.
It should be noted that the direction of the dependence on the baryonic mass
is not obvious a priori, because a lower baryonic mass is associated with 
both a shallower potential well and a weaker outflow, which work to strengthen
and weaken the formation of a core, respectively. 
(3) Spreading the baryons from a central point mass to an extended
configuration, out to $0.2\Rv$, makes only a small difference to the core
formation. This is due to the competing effects mentioned above, where a more
extended baryon distribution is associated with both, 
a shallower potential and a weaker effect of the outflow.

\smallskip % multipe outflows
%\adr{Jonathan: please check this:}
We also tested multiple episodes of outflow, each with $\eta\seq0.2$, following
one event of DF heating and no further supply of central baryons during the
sequence of outflows.
The total fractional outflown mass in $N\seq3$ such episodes is comparable 
to the outflown mass in a single episode of $\eta\seq 0.5$. 
We learn that under these conditions, the successive three weaker
episodes and the stronger single episode lead to a similar core.
%\adr{Jonathan: has this been tested with the latest fit to SatGen?
%In fact $N=4$ seems to be a better fit to the latest single result.}
This is line with the toy model of shell response to an outflow 
\citep[][Section 4]{dutton16b},
which predicts that the expansion of the shell radius is approximately
by a factor $1+Nf$, where $f\ll 1$ is the fraction of the total mass ejected 
in each episode, such that $Nf$ is the same for all $N$.
However, if each outflow is preceded by an inflow of the same gas mass, 
the net expansion factor (after certain contraction) becomes $1+Nf^2$. 
If the outflow events are preceded by a cosmological inflow of the same total
gas mass $f_{\rm inf}$, then $f\seq f_{\rm inf}/N$, such that the
single outflow is expected to be more effective than a sequence. 
If, on the other hand, each outflow is preceded by an inflow of recycled gas,
then $f$ is the same for all episodes, and a sequence of weak outflows is
expected to be more effective than a single outflow.

%%%%%%%%%%%%%%%%%%%%%%%%%%% 7
\section{Discussion}
\label{sec:disc}

%=================
\subsection{Cores at large masses and high redshifts}
\label{sec:mass_and_z}

% DF mass by compaction
We showed that the observed
preferred appearance of cores in haloes of $\Mv \sgt 10^{12}\msun$
is a natural outcome of the characteristic threshold for compaction events
at $\mv \sgt 10^{11.3}\msun$, as seen in simulations 
\citep{zolotov15,tomassetti16,tacchella16_prof}
and observations \citep{huertas18}.
Once the host halo is above $10^{12}\msun$, satellites above $10^{11.3}\msun$
constitute most of the accretion into the host halo, and these are 
the post-compaction compact satellites that allow $20\%$ or more of their mass 
to penetrate intact into the host cusp and significantly heat up the cusp by 
dynamical friction.
The transition in the satellite compactness as a function of 
mass translates to a threshold for haloes that can
generate extended DM cores.

\smallskip % DF redshift
The preference of cores at redshifts above $z\ssim 1$ arises from several 
different sources.
First, as estimated by our toy model, the relative energy deposited by
dynamical friction in the cusp 
in a Gyr is $\prop\! (1+z)^{5/2}$, \equ{wdot}, and in a halo crossing time it 
is $\prop\!(1+z)$, \equ{wvir}. With a typical bound fraction $f\ssim 0.2$ 
surviving stripping for a post-compaction satellite, the deposited energy 
is indeed comparable to the cusp energy near $z\ssim 2$, where the majority 
of the observed haloes show cores.
Second, the compaction events, which tend to occur near the characteristic mass,
tend to be deeper at higher redshifts due to the
higher gas fraction which allows more dissipation and angular-momentum loss
and therefore more compact satellites.
Third, the satellite orbits tend to be more radial at higher redshifts, where a
given halo mass represents a higher-sigma peak. 
According to \citet{wetzel11}, the mean orbit circularity is
$\epsilon \ssim 0.6,0.5,0.4$ at $z\ssim 0,2,4$ respectively.
% Danovich?, Dubois?) (higher sigma peaks).
The more radial orbits at higher redshift give rise to less tidal stripping,
deeper satellite penetration and larger energy deposit by DF in the host cusp.
Further discussion of the redshift and mass dependence of the orbit circularity
is in Appendix \se{app_circularity}
(available as supplementary material online).

\smallskip % AGN
The second-stage AGN-driven core formation gives rise to a similar mass 
threshold for cores. 
This is because the central black-hole growth is suppressed by supernova
feedback when the halo is less massive than the golden mass
$\sim\!10^{12}\msun$ \citep{dubois15,bower17,angles17},
and it turns into a rapid growth triggered by a compaction event
\citep{dekel19_gold,lapiner21} in the hot-CGM phase once above the golden mass,
consistent with observations \citep{kocevski17,forster19}.
In turn, 
the AGN-driven winds also contribute to the preference of cores at
high redshifts,
because the winds tend to be more effective at high redshifts  
due to the higher gas fraction, providing more gas for activating the AGN
and as a substance to be ejected.

%======================
\subsection{Gas availability for AGN-driven outflows}
\label{sec:gas}

\smallskip % eta
For a halo of mass $\Mv\seq 10^{12.5}\msun$, we assume 
a galaxy gas mass of $\Mg\!\simeq\! 4\times 10^{10}\msun$, 
following the average stellar-to-halo mass
ratio for such a halo mass at $z\seq2$ \citep{behroozi19} and assuming a gas
fraction of 0.5 \citep{tacconi18}. 
The outflowing mass in each ejection episode is assumed to be a fraction 
$\eta$ of the gas mass in the galaxy, between zero and unity, 
namely a fraction $0.5\eta$ of the central baryonic mass at $z\ssim 2$.
An upper limit on $\eta$ is provided by the gas mass available for being 
ejected, which we estimate from cosmological accretion plus recycling during 
a period of, e.g., $0.5\tvir$, the typical time for the hot phase as estimated
in \se{N-body}.
% recycling
The typical cosmological specific accretion rate in the EdS regime ($z\!>\!1$)
can be approximated by \equ{Mdot}, 
$\dot{M}/M \!\simeq\! 0.47\Gyr^{-1}(1+z)_3^{5/2}$,
where the mass can refer successfully
either to the total mass or to the baryonic mass \citep{dekel13}.
Assuming a cosmological baryon fraction $\fb \seq 0.17$, with an accreting 
gas fraction with respect to the total accreting baryons of $\fg \seq 0.5$ 
at $z\seq 2$, 
and assuming that the accretion rate of fresh and recycled gas is a factor 
$f_{\rm rec}$ times the fresh gas accretion rate, 
the total gas accretion rate into the galaxy inside a halo of
$\Mv\seq10^{12.5}\msun$ is 
\be
\dot{M}_{\rm g} \simeq f_{\rm rec}\, 1.3\times 10^{11} \msun \Gyr^{-1} \, .
\ee 
This provides $\Mg\ssim f_{\rm rec}\, 3.2\times 10^{10}\msun$ of gas that
is available for ejection during $0.5\tvir\!\simeq\!0.25\Gyr$.  
Given the comparable gas mass of $\sim\!4\times 10^{10}\msun$ in the galaxy, 
one can expect about $f_{\rm rec}$ outflow episodes of as much as 
$\eta \ssim 1$ each during that period. 
Based on FIRE simulations,
the recycling factor at $z\ssim 2$ is estimated to be as high as 
$f_{\rm rec}\ssim 3$ \citep[][Fig.~3]{angles17_recycling}.
We therefore assume that from the point of view of gas availability for
ejection, by modeling the response to one outflow episode 
of $\eta \seq 1$ we obtain a conservative lower limit to the effect of outflows 
during $0.5\tvir$.
This is provided that the duty cycle of AGN outbursts is appropriate for the
purpose, and that their energy/momentum and coupling to the available gas 
are sufficient for a removal of a significant fraction of the available gas.

%==========================
\subsection{Central stars, gas and dark matter}
\label{sec:dm_stars}

For the proposed scenario to produce DM cores with a low fraction of
dark matter and with a stellar system that remains intact, several conditions
have to be fulfilled. For example, 
(1) the accretion rate of dark matter to the cusp by satellites should not
overcome the expansion rate induced by DF heating and/or AGN-driven outflows,
(2) the stellar system should be more compact than the gas that is to be 
removed from the galaxy, or it has to be kinematically cooler than the dark
matter, and
(3) the dark matter that is to be pushed away from the cusp has to originally 
be at radii that are comparable to (or slightly larger than) those of the gas
to be removed.

%-------------
\subsubsection{Dark-matter accretion}

The galaxies with DM cores are observed to have a low 
fraction of dark matter with respect to the total core mass, 
$f_{\rm DM}\ssim 0\sdash 0.3$ 
within the effective radius $\Re$ \citep[][Fig.~10]{genzel21}.
The compact merging satellites that provide the DF heating bring into the 
cusp a mass that is comparable to the original cusp mass.
For example, according to \fig{satgen_single},
for a typical single compact satellite of $\mv\ssim0.1\Mv$,
the mass that enters the $10\kpc$ host cusp is 
$m_{\rm c}\ssim 0.4\mv \seq 0.04\Mv$, indeed comparable to the mass inside
the cusp.
For our scenario of core formation to work,
the dark matter that dominates this incoming mass has to be pushed away from 
the cusp, together with the dark matter that has already been in the cusp,
either by the post-heating expansion discussed in \se{relax}, 
and/or by the response to AGN-driven outflows discussed in \se{agn}.
In order to compute the timescale for doubling the cusp mass by incoming
satellites, we recall that the specific accretion rate into the halo is 
$\dot{M}_{\rm v}/\Mv \ssimeq 0.5 \Gyr^{-1}(1+z)_3^{5/2}$ \citep{dekel13}.
If about 60\% of this input mass is in compact satellites of $\mv \sgt 0.1\Mv$
\citep{neistein08_m}, and if the satellite mass
that penetrates to the cusp is $m_{\rm c}\ssim 0.4\mv$ while the cusp mass in 
an NFW host is $\Mc\sim\! 0.04\Mv$, 
we obtain a specific DM accretion rate into the cusp of 
$\dot{m}_{\rm c}/\Mc \ssim 3 \Gyr^{-1}(1+z)_3^{5/2}$, namely
a DM cusp doubling time of 
\be
t_{\rm c} \ssim 0.3\Gyr\, (1+z)_3^{-5/2}
\ssim 0.6 (1+z)_3^{-1}\tvir \, .
\ee
According to \fig{satgen_single}, 
this DM mass is mostly deposited in the outer cusp, at $3\sdash10\kpc$.
This timescale turns out to be comparable to the timescale for post-heating 
relaxation (\se{relax}) and to the timescale for fresh gas supply for an 
AGN-driven outflow that generates a core (\se{gas}).
This implies that the DM accretion rate into the cusp is slow enough for the 
dark matter to be maintained hot by dynamical friction,
and that the gas accretion rate (especially when adding recycling) 
is more than needed for producing AGN-driven outflows that push away the 
additional dark matter, once the AGN duty cycle, energy/momentum and coupling
to the gas are appropriate.

\smallskip
\Fig{app_outflow_msat_nfw_single_cosmo} 
(available as supplementary material online)
show the CuspCore results with the DM satellite mass that has been added to
the host according to SatGen, \fig{satgen_single} and \fig{satgen_cosmo}, 
prior to the outflow event.
Shown are cases with the same parameters as in \fig{outflow_single} and
\fig{outflow_cosmo}, confirming core formation in the presence of the added
mass. The core is slightly less flat than without the additional mass for a 
single satellite, but it is as flat for a sequence of satellites, where the
total added mass is only $0.2$ of the cusp mass.
This justifies ignoring the added mass in our main analysis.

%------------
\subsubsection{The stellar system}

 % stars - earlier compaction of host
What may lead to the initial compactness and the associated
coldness of the stellar systems?
The cored galaxies tend to have massive compact stellar bulges, 
with masses $10^{10.5}\sdash 10^{11}\msun$ \citep[][Fig.~11]{genzel21}.
These bulges have to be compact enough such that they are not significantly
heated by the merging satellites, and especially such that their compactness
compared to the gas and the (possibly heated) dark matter, 
and their self-gravity,
allow them to survive intact while the more extended dark matter expands 
in response to the AGN-driven outflows.
These bulges are likely to have partly formed by earlier dissipative compaction 
events that occurred when the host halo crossed the threshold mass, generating a
stellar nugget, blue that turned red, inside $\sim\! 1\kpc$
\citep{zolotov15,tacchella16_prof}. 

\smallskip % stars - earlier compaction in satellites
Furthermore, the stars that come in with the merging post-compaction
satellites are expected to be more compact than the gas and DM
at the centers of these satellites,
and thus to penetrate by DF deep into the host halo cusp with little stripping. 
The stellar-to-halo mass ratio at $z\ssim 2$, for our dominant post-compaction
satellite mass of $\mv \ssim 10^{11.5}\msun$, is expected to be
$\ms/\mv\ssim 0.004$ \citep{behroozi19}. 
According to \fig{satgen_single}, the bound mass of a typical satellite of
$\mv\seq0.1\Mv$ that enters the inner $1\kpc$ of the host
is $m_1\sim\!0.09\mv\seq 0.009\Mv$.
We learn that if the satellite stars are indeed confined to and dominate
the central regions of the satellites, they are all expected to penetrate 
intact into the inner $1\kpc$ of the host halo, while the mass that is 
stripped within the outer cusp is primarily dark matter. 

\smallskip % stars DF heating
The compact stellar nugget is not expected to heat up due to dynamical
friction.
The gravitational potential of the stars in the host central $\sim\!1\kpc$
nugget is deeper by a factor of a few than that of the dark matter at
several kiloparsecs.
According to \fig{satgen_single}, the energy deposited by dynamical friction
inside the inner $1\kpc$ of the host is two order of magnitude less than at a
a radius of a few kiloparsecs. 
These indicate that the heating effect of dynamical friction
on the inner nugget is negligible.

%-----------
\subsubsection{The gas distribution}

% gas
In order for the stellar nugget to remain intact while the dark matter is
expanding in response to AGN-driven outflows, the gas to be expelled
has to lie mostly outside the compact stellar system.
We learn from the VELA hydro cosmological simulations that in post-compaction
galaxies, even without AGN feedback,
the gas half-mass radius is typically $\sim\!3$ times the stellar
half-mass radius \citep{tacchella16_prof,kretschmer21_kvir}\footnote{For 
example, we read from Table A1 of \citet{kretschmer21_kvir} typical values of 
$\Rd/R_{\rm e,stars}\ssim 5$, and obtain the desired ratio of effective radii
using $R_{\rm e,gas}/\Rd \ssim 0.5$.}
With $R_{\rm e,stars} \ssim 1\kpc$, we estimate $R_{\rm e,gas}\ssim 3\kpc$,
which is outside most of the stellar system, but inside most of the dark-matter
cusp of $R_{\rm e,dm}\ssim 7\kpc$ that is to be pushed away, as required. 
The gas is more diffuse than the stars in these simulations
because of the depletion by central star formation and partial 
effects of supernova feedback when near the golden mass,
and because the gas might have been segregated from the stars already
in the merging satellites due to similar effects as well as ram-pressure 
stripping.
Considering the effect of earlier AGN-driven outflows may make the gas even 
more diffuse.

%\smallskip
%\adr{Jonathan: CuspCore of NFW single compact, compare to original with 
%$\eta\seq 0.5$ and with new $W$ and $\eta$ for same effect:  
%DM added $m(r)$.}
%(2) gas distributed $\Re \seq 3 \kpc$ (no DM added),
%(3) live stars $\Re \seq 1\kpc$ plus gas distributed. Stars could be identified
%with inner SDM in a new DZ fit.}

\smallskip %
Clearly, the above discussion is based on crude estimates.
A quantitative analysis of the stellar system, the gas distribution
and the dark-matter distribution in the presence of baryons
is deferred to a future study that will properly address the baryonic 
components of the satellites and the host halo.
In particular, the CuspCore model will be generalized to deal with a
three-component system.

%================
\subsection{Why do cosmological simulations fail?}
\label{sec:sims}

%\adr{Need to refer to each simulation - Fangzhou?}.

\smallskip % DF 
We can see several reasons for the failure of DF to heat up the host cusps
(and/or produce cores)
in current cosmological simulations 
\citep[e.g.][TNG and FIRE-2 respectively]{wang20,lazar20}.
First, the simulation resolution does not allow a full treatment of the 
compaction of satellites of $\sim\!10^{11.3}\msun$. While the $\sim\!10\kpc$
environment in the satellite is resolved, the inner $\sim\!1\kpc$ is not 
properly resolved in most simulations (e.g. EAGLE, Illustris-TNG). 
The compaction on these scales is 
important as it suppresses tidal heating in the inner satellites and 
reduces the overall tidal stripping of the satellites, which allows better
penetration and stronger DF heating of the host cusp. 

\smallskip
Second, the compaction process in the satellites may be suppressed
due to feedback that is too strong. This may be the case in some of the popular
cosmological simulations (e.g. FIRE, Illustris-TNG).

\smallskip
Third, if the satellites are not properly resolved by more
than $10^6$ particles and an optimal scale-dependent force softening length, 
there is an artificial tidal disruption of the satellites 
\citep{bosch18,errani21}. % \adr{Fangzhou: provide more details}.
This mostly affects satellites in the inner host halo, 
where the satellite abundance
could be artificially suppressed by a factor of two or more \citep{green21}, 
thus suppressing the DF heating of the host cusp. 

\smallskip % AGN
The fact that the AGN environment is unresolved may suppress its efficiency in
driving winds that could participate in generating DM cores.
First, the simulations do not resolve any clumpiness in the accretion onto the
black hole. Such clumpiness should boost the black-hole growth rate and AGN 
activity, and it can be incorporated as a subgrid model \citep{degraf17}. 
Second, the simulations may fail to resolve the coupling of the AGN with the
ISM around the black hole. For example, if the ISM is clumpy, the push of the
dense clumps may boost the generated wind.

\smallskip % future sims
The key test for the proposed scenario for core formation in massive galaxies
should eventually be cosmological simulations that reproduce such cores.
A maximum resolution of $\sim\! 20\pc$
will enable the formation of compact blue nuggets as merging satellites,
and $\sim\! 10^7\msun$ particles in their haloes will avoid artificial tidal
disruption.
Improvements of the black-hole and AGN subgrid recipes
could help generating the desired bursty duty cycle.
First, by introducing subgrid clumpiness in the accretion onto the black holes
\citep[e.g. following][]{degraf17}, and
Second, by limiting the AGN-feedback push to cold clumps in the ISM/CGM.
In addition, positive AGN feedback \citep{silk13} may trigger SN feedback
to assist the driving of the wind.

%=========================
\subsection{Caveats}
\label{sec:caveats}

In the current absence of proper simulations,
the analysis performed above is approximate, meant to provide a feasibility
test for the proposed hybrid scenario of massive core formation at high
redshift, so its quantitative predictions should be taken with a grain of salt.
It is yet to be verified with proper cosmological simulations that
resolve the satellite compactness and the processes of tidal stripping and
dynamical friction, and model realistic black-hole growth and AGN feedback.
We mention below certain caveats of our analysis.

\smallskip % compact satellites
We derived fiducial satellite profiles from cosmological simulations (VELA),
which indicate a compaction-driven transition from diffuse to compact when 
the galaxy is near a golden halo mass of $\Mv \ssim 10^{11.3}\msun$. 
While this transition is rather robust for most galaxies, 
in simulations and observations,
we adopted in our simplified analysis representative fiducial profiles below 
and above the golden mass, and ignored the large scatter in these profiles 
and in the threshold mass. This scatter may weaken the systematic
mass dependence of core formation.

\smallskip % toy, satgen
The analytic toy model for DF heating is clearly very
approximate, meant to motivate our qualitative expectations.
The SatGen semi-analytic simulations are a step forward in approximating the 
evolution of satellites, single and in a cosmological sequence, 
under tidal stripping and dynamical friction. It has been successfully
calibrated against cosmological simulations.
However, the way we translate the satellite evolution to local heating of the 
dark matter in the host halo is very crude, as dynamical friction is not a
local effect \citep{tremaine84}.

\smallskip % cuspcore
The CuspCore modeling of the dark-matter response to an outflow episode
is only an approximation.
In particular, the assumptions of a pre-outflow hot halo as the outcome of 
earlier DF heating, and of an instantaneous mass loss from a central
point mass, may be oversimplified. 
Furthermore, the implementation of energy conservation in the subsequent
relaxation process is not formally justified in the case of shell crossing. 
Here, the proof is in the pudding, in the sense that the model has been 
demonstrated to approximate the response of dark matter to outflow episodes 
in hydro cosmological simulations (NIHAO), except under merger situations. 
%\smallskip % AGN duty cycle
Finally, the CuspCore model as implemented here treats a single instantaneous 
mass loss, while realistic AGN feedback may involve a duty cycle of mass 
ejection episodes with mutual cross-talk.

\smallskip % baryons
In most of the above analysis, the focus was on the dark matter, ignoring
certain effects that may involve the baryons. In particular, for the scenario 
to work, the baryons in the satellites should penetrate into the halo center
such that the gas is available for ejection by AGN feedback while the stars,
incoming or forming in-situ, remain intact at the center as a visible galaxy. 
Clearly, a more sophisticated analysis that better addresses the baryons 
is required.

%%%%%%%%%%%%%%%%%%%%%%%%%%% 8
\section{Conclusion}
\label{sec:conc}

% observed cores
This work has been inspired by the puzzling observational indication for
a dearth of dark matter and 
flat-density cores extending to $\sim\!10\kpc$ in about one third of the 
DM haloes of mass $\Mv\!\geq\!10^{12}\msun$ at $z\ssim 2$
\citep{genzel21}.
This is puzzling because such cores, which are not detected at low redshifts,
are not reproduced by any current cosmological simulation, 
and because supernova feedback, 
the common process assumed responsible for core formation in low-mass galaxies,
is not expected to be energetic enough for producing cores in massive 
galaxies \citep{ds86}.
Despite potential uncertainties in the non-trivial interpretation of the 
observations, we take it as a non-trivial theoretical challenge.

\smallskip % hybrid scenario
We study a hybrid scenario where two processes combine to generate such
massive cores at high redshifts. These are ``heating" of the inner host-halo 
cusp by dynamical friction acting on {\it compact\,} merging satellites, 
followed by AGN-driven outflows that flatten the hot cusp into a core.
Each of these processes by itself seems to be incapable of forming 
extended-enough cores 
\citep{elzant01,elzant04,elzant08,martizzi12,martizzi13}. 
For the scenario to work, they should act in concert,
and for reproducing the desired combined effects in simulations, 
each should be treated properly.

\smallskip % key: compactness above critical mass
The key for efficient dynamical-friction heating is the compactness of the
incoming satellites. This turns out to be a natural outcome of the wet 
compaction process that galaxies typically undergo when their haloes are 
near or above a golden mass of $\mv \ssim 10^{11.3}\msun$ 
\citep{zolotov15,barro17,huertas18}. 
This is indeed the mass range for the dominant merging satellites in host 
haloes of $\Mv \!\geq\!10^{12}\msun$, in which the cores are observed.

\smallskip % toy, satgen
We use analytic toy modeling and dedicated semi-analytic simulations  
\citep[SatGen,][]{jiang21_satgen}
to explore the energy deposit by dynamical friction as a function of satellite
compactness, for a single satellite on a typical orbit and for a cosmological 
sequence of satellites during a period comparable to the halo virial crossing 
time at $z\ssim 2$.

\smallskip % Vela profiles
Zoom-in hydro cosmological simulations (VELA) allow us to determine fiducial 
density profiles for the post-compaction high-mass satellites in contrast to
the pre-compaction low-mass satellites.
The Dekel-Zhao functional form \citep{dekel17,freundlich20_prof}, with a 
flexible concentration and inner slope and analytic expressions for the
potential and kinetic energies, best-fit these with the DZ parameter values 
$(c,\alpha)\seq (7,1)$ and $(3,0.5)$ for the compact and diffuse satellites.
These correspond to the equivalent pairs of parameters 
$(c_2,s_1)\seq (15,1.5)$ and $(3,1)$, respectively,
where $c_2$ is the concentration where the density-profile slope is $-2$ and
$s_1$ is the density slope at $0.01\Rv$. 

\smallskip % finding DF compact vs diffuse
For a massive host halo of an initial NFW cusp,
we find using the toy model and SatGen simulations that the fiducial compact 
satellites, relevant for massive haloes, significantly heat up the host cusp. 
They typically penetrate into the cusp with one to a few tens of percents of 
their original mass, and by dynamical friction they deposit there an energy 
comparable to the initial kinetic energy in the cusp during one half of the
virial time at $z\ssim 2$, about $0.25\Gyr$. 
On the other hand, the diffuse satellites, relevant for less massive haloes,
penetrate with only a small fraction of their original mass, and deposit energy
that is a similarly small fraction of the cusp energy, causing a negligible 
effect.

\smallskip % heating vs flattening
We find, using toy models and N-body simulations, that
the ``hot" phase of the cusp, driven by dynamical friction acting
on a single compact
satellite, is temporary, lasting for about half a virial crossing time. 
Later, the cusp would have relaxed to a new Jeans equilibrium, by expanding 
and cooling into a cold and somewhat flatter cusp.
However,
at $z\ssim 2$ and for $\Mv \!\geq\! 10^{12}\msun$, 
given the energy deposited by dynamical friction in half a virial time,
the hot phase is expected to be maintained by the cumulative heating from 
a cosmological sequence of merging satellites.

\smallskip % AGN outflows
In the second stage,
using an analytic model for outflow-driven core formation
\citep[CuspCore,][]{freundlich20_cuspcore},
we demonstrate that the heated or flattened cusps respond strongly to removal
of about half the central gas mass.
This is a small fraction of the gas available, given the accretion and
recycling in the available time of half a virial time at $z\ssim 2$.
A pre-heated cusp develops into a flat core extending to $\sim\!10\kpc$,
while a cooled and somewhat flattened cusp generates a partly flattened
core.
AGN feedback is indeed expected to become effective in haloes of 
$\Mv\!\geq\!10^{12}\msun$,
where the black-hole growth is no longer suppressed by supernova feedback
and its rapid growth, triggered by a wet compaction event, is maintained
by a hot CGM \citep{bower17, dekel19_gold, lapiner21}.

\smallskip % steep cusp
Repeating the analysis for a host halo with a steeper initial cusp of slope 
$s_1\seq 1.5$, we find that core formation is more demanding, but plausible for
about one third of the galaxies.
For dynamical friction to sufficiently heat up the cusp in half a virial time,
the actual sequence of accreting satellites should include larger masses 
and/or more radial orbits than typical, in the top third of the distribution.
For the AGN feedback to then generate a core, the gas mass removed should be 
comparable to the instantaneous gas mass in the galaxy, which is still only a 
fraction of the gas supplied by accretion and recycling during half a virial 
time.

\smallskip % mass and redshift dependence
The preference for cores in haloes of mass above the golden mass of 
$\Mv\ssim 10^{12}\msun$ is primarily due to the threshold mass for compaction 
events that
(a) generate sufficiently compact satellites for significant DF 
cusp heating, and (b) trigger strong AGN-driven outflows in the host galaxies
\citep{dekel19_gold,lapiner21}.
The preference for cores at $z \!>\! 1$ arises from the more frequent mergers
that drive the DF heating, and the stronger compaction events
due to the higher gas fraction. The preferred mass and redshift are related
to each other through the fact that the typical haloes at the Press-Schechter 
nonlinear mass scale are $\sim\! 10^{12}\msun$ at $z\ssim 2$.

\smallskip % dm, stars, gas
We estimate that the accretion rate of dark matter into the cusp via the 
merging satellites is slow compared to the hot-cusp relaxation rate and the 
rate of gas supply for AGN-driven outflows. This allows the formation of 
a core with a low central DM fraction, as observed.
We anticipate that the central stellar nugget is likely to be more compact 
than the gas distribution such that the stellar system will not be
significantly heated by dynamical friction and will not be strongly affected 
by the AGN-driven gas removal that generates the DM core.
These should be verified via simulations including baryons.

\smallskip % future sims
For cosmological simulations to reproduce the required DF 
heating, they should resolve the wet compaction of the satellites of 
$\mv \!>\!  10^{11.3}\msun$ on scales below $\sim\! 100\pc$, and have
at least $\sim\! 10^7$ particles in each subhalo in order to avoid 
artificial tidal disruption.
AGN feedback could be boosted by resolving clumpy black-hole accretion
and clumpy response to outflows.
Simulations of this kind are not beyond reach in the near future,
and they should be performed for a more realistic test of the proposed scenario.

%%%%%%%%%%%%%
\section*{Acknowledgments}

We are grateful for stimulating interactions with Andrew Benson.  
This work was partly supported by the grants 
Germany-Israel GIF I-1341-303.7/2016 (AD, AB), 
Germany-Israel DIP STE1869/2-1 GE625/17-1 (AD, RG, AB), 
ISF 861/20 (AD), 
ERC GreatDigInTheSky 834148 (JF),
and a Troesh Scholarship (FJ).
The cosmological VELA simulations were performed at the National Energy
Research Scientific Computing Center (NERSC) at Lawrence Berkeley National
Laboratory, and at NASA Advanced Supercomputing (NAS) at NASA Ames Research
Center. Development and analysis have been performed in the astro cluster at
HU.

%%%%%%%%%%%%%%%%%%
\section*{DATA AVAILABILITY}

The codes used in this articles are available online, as referenced in the
article and in the online supplementary material. Data and results underlying 
this article will be shared on reasonable request to the corresponding author.

%%%%%%%%%%%%%%%%%%%%%%%%%%%%%%%%%%%%%%%%%%%%%
\bibliographystyle{mn2e}
\bibliography{core}

\begin{thebibliography}{111}
\expandafter\ifx\csname natexlab\endcsname\relax\def\natexlab#1{#1}\fi

\bibitem[{{Agertz} {et~al}\mbox{.}(2013){Agertz}, {Kravtsov}, {Leitner}, \&
  {Gnedin}}]{agertz13}
{Agertz} O., {Kravtsov} A.~V., {Leitner} S.~N., {Gnedin} N.~Y., 2013, \apj,
  770, 25

\bibitem[{{Angl{\'e}s-Alc{\'a}zar}
  {et~al}\mbox{.}(2017{\natexlab{a}}){Angl{\'e}s-Alc{\'a}zar},
  {Faucher-Gigu{\`e}re}, {Kere{\v{s}}}, {Hopkins}, {Quataert}, \&
  {Murray}}]{angles17_recycling}
{Angl{\'e}s-Alc{\'a}zar} D., {Faucher-Gigu{\`e}re} C.-A., {Kere{\v{s}}} D.,
  {Hopkins} P.~F., {Quataert} E., {Murray} N., 2017{\natexlab{a}}, \mnras, 470,
  4698

\bibitem[{{Angl{\'e}s-Alc{\'a}zar}
  {et~al}\mbox{.}(2017{\natexlab{b}}){Angl{\'e}s-Alc{\'a}zar},
  {Faucher-Gigu{\`e}re}, {Quataert}, {Hopkins}, {Feldmann}, {Torrey}, {Wetzel},
  \& {Kere{\v s}}}]{angles17}
{Angl{\'e}s-Alc{\'a}zar} D., {Faucher-Gigu{\`e}re} C.-A., {Quataert} E.,
  {Hopkins} P.~F., {Feldmann} R., {Torrey} P., {Wetzel} A., {Kere{\v s}} D.,
  2017{\natexlab{b}}, \mnras, 472, L109

\bibitem[{{Bardeen} {et~al}\mbox{.}(1986){Bardeen}, {Bond}, {Kaiser}, \&
  {Szalay}}]{bbks86}
{Bardeen} J.~M., {Bond} J.~R., {Kaiser} N., {Szalay} A.~S., 1986, \apj, 304, 15

\bibitem[{{Barro} {et~al}\mbox{.}(2017){Barro}, {Faber}, {Koo}, {Dekel},
  {Fang}, {Trump}, {P{\'e}rez-Gonz{\'a}lez}, {Pacifici}, {Primack},
  {Somerville}, {Yan}, {Guo}, {Liu}, {Ceverino}, {Kocevski}, \&
  {McGrath}}]{barro17}
{Barro} G. {et~al.}, 2017, \apj, 840, 47

\bibitem[{{Behroozi} {et~al}\mbox{.}(2019){Behroozi}, {Wechsler}, {Hearin}, \&
  {Conroy}}]{behroozi19}
{Behroozi} P., {Wechsler} R.~H., {Hearin} A.~P., {Conroy} C., 2019, \mnras,
  488, 3143

\bibitem[{{Behroozi}, {Wechsler} \& {Conroy}(2013){Behroozi}, {Wechsler}, \&
  {Conroy}}]{behroozi13}
{Behroozi} P.~S., {Wechsler} R.~H., {Conroy} C., 2013, \apjl, 762, L31

\bibitem[{Benson(2017)}]{benson17}
Benson A.~J., 2017, MNRAS, 467, 3454

\bibitem[{{Birnboim} \& {Dekel}(2003)}]{bd03}
{Birnboim} Y., {Dekel} A., 2003, \mnras, 345, 349

\bibitem[{{Bower} {et~al}\mbox{.}(2017){Bower}, {Schaye}, {Frenk}, {Theuns},
  {Schaller}, {Crain}, \& {McAlpine}}]{bower17}
{Bower} R.~G., {Schaye} J., {Frenk} C.~S., {Theuns} T., {Schaller} M., {Crain}
  R.~A., {McAlpine} S., 2017, \mnras, 465, 32

\bibitem[{{Bryan} \& {Norman}(1998)}]{bryan98}
{Bryan} G.~L., {Norman} M.~L., 1998, \apj, 495, 80

\bibitem[{{Bullock} {et~al}\mbox{.}(2001){Bullock}, {Kolatt}, {Sigad},
  {Somerville}, {Kravtsov}, {Klypin}, {Primack}, \& {Dekel}}]{bullock01_c}
{Bullock} J.~S., {Kolatt} T.~S., {Sigad} Y., {Somerville} R.~S., {Kravtsov}
  A.~V., {Klypin} A.~A., {Primack} J.~R., {Dekel} A., 2001, \mnras, 321, 559

\bibitem[{{Burkert}(1995)}]{burkert95}
{Burkert} A., 1995, \apjl, 447, L25

\bibitem[{{Burkert} {et~al}\mbox{.}(2010){Burkert}, {Genzel}, {Bouch{\'e}},
  {Cresci}, {Khochfar}, {Sommer-Larsen}, {Sternberg}, \& {et al.,}}]{burkert10}
{Burkert} A., {Genzel} R., {Bouch{\'e}} N., {Cresci} G., {Khochfar} S.,
  {Sommer-Larsen} J., {Sternberg} A., {et al.,}, 2010, \apj, 725, 2324

\bibitem[{{Ceverino}, {Dekel} \& {Bournaud}(2010){Ceverino}, {Dekel}, \&
  {Bournaud}}]{cdb10}
{Ceverino} D., {Dekel} A., {Bournaud} F., 2010, \mnras, 404, 2151

\bibitem[{{Ceverino} {et~al}\mbox{.}(2012){Ceverino}, {Dekel}, {Mandelker},
  {Bournaud}, {Burkert}, {Genzel}, \& {Primack}}]{ceverino12}
{Ceverino} D., {Dekel} A., {Mandelker} N., {Bournaud} F., {Burkert} A.,
  {Genzel} R., {Primack} J., 2012, \mnras,

\bibitem[{{Ceverino} {et~al}\mbox{.}(2015){Ceverino}, {Dekel}, {Tweed}, \&
  {Primack}}]{ceverino15_e}
{Ceverino} D., {Dekel} A., {Tweed} D., {Primack} J., 2015, \mnras, 447, 3291

\bibitem[{{Ceverino} \& {Klypin}(2009)}]{ceverino09}
{Ceverino} D., {Klypin} A., 2009, \apj, 695, 292

\bibitem[{{Ceverino} {et~al}\mbox{.}(2014){Ceverino}, {Klypin}, {Klimek},
  {Trujillo-Gomez}, {Churchill}, {Primack}, \& {Dekel}}]{ceverino14}
{Ceverino} D., {Klypin} A., {Klimek} E.~S., {Trujillo-Gomez} S., {Churchill}
  C.~W., {Primack} J., {Dekel} A., 2014, \mnras, 442, 1545

\bibitem[{{Ceverino}, {Primack} \& {Dekel}(2015){Ceverino}, {Primack}, \&
  {Dekel}}]{ceverino15_shape}
{Ceverino} D., {Primack} J., {Dekel} A., 2015, \mnras, 453, 408

\bibitem[{Chandrasekhar(1943)}]{chandrasekhar43}
Chandrasekhar S., 1943, ApJ, 97, 255

\bibitem[{{Danovich} {et~al}\mbox{.}(2015){Danovich}, {Dekel}, {Hahn},
  {Ceverino}, \& {Primack}}]{danovich15}
{Danovich} M., {Dekel} A., {Hahn} O., {Ceverino} D., {Primack} J., 2015,
  \mnras, 449, 2087

\bibitem[{{de Blok} {et~al}\mbox{.}(2001){de Blok}, {McGaugh}, {Bosma}, \&
  {Rubin}}]{deblok01}
{de Blok} W.~J.~G., {McGaugh} S.~S., {Bosma} A., {Rubin} V.~C., 2001, \apjl,
  552, L23

\bibitem[{{de Blok} {et~al}\mbox{.}(2008){de Blok}, {Walter}, {Brinks},
  {Trachternach}, {Oh}, \& {Kennicutt}}]{deblok08}
{de Blok} W.~J.~G., {Walter} F., {Brinks} E., {Trachternach} C., {Oh} S.~H.,
  {Kennicutt}, R.~C. J., 2008, \aj, 136, 2648

\bibitem[{{DeGraf} {et~al}\mbox{.}(2017){DeGraf}, {Dekel}, {Gabor}, \&
  {Bournaud}}]{degraf17}
{DeGraf} C., {Dekel} A., {Gabor} J., {Bournaud} F., 2017, \mnras, 466, 1462

\bibitem[{{Dekel} \& {Birnboim}(2006)}]{db06}
{Dekel} A., {Birnboim} Y., 2006, \mnras, 368, 2

\bibitem[{{Dekel}, {Devor} \& {Hetzroni}(2003){Dekel}, {Devor}, \&
  {Hetzroni}}]{dekel03}
{Dekel} A., {Devor} J., {Hetzroni} G., 2003, \mnras, 341, 326

\bibitem[{{Dekel} {et~al}\mbox{.}(2020{\natexlab{a}}){Dekel}, {Ginzburg},
  {Jiang}, {Freundlich}, {Lapiner}, {Ceverino}, \& {Primack}}]{dekel20_flip}
{Dekel} A., {Ginzburg} O., {Jiang} F., {Freundlich} J., {Lapiner} S.,
  {Ceverino} D., {Primack} J., 2020{\natexlab{a}}, arXiv e-prints

\bibitem[{{Dekel} {et~al}\mbox{.}(2017){Dekel}, {Ishai}, {Dutton}, \&
  {Maccio}}]{dekel17}
{Dekel} A., {Ishai} G., {Dutton} A.~A., {Maccio} A.~V., 2017, \mnras, 468, 1005

\bibitem[{{Dekel} \& {Krumholz}(2013)}]{dk13}
{Dekel} A., {Krumholz} M.~R., 2013, \mnras, 432, 455

\bibitem[{{Dekel}, {Lapiner} \& {Dubois}(2019){Dekel}, {Lapiner}, \&
  {Dubois}}]{dekel19_gold}
{Dekel} A., {Lapiner} S., {Dubois} Y., 2019, arXiv e-prints

\bibitem[{{Dekel} {et~al}\mbox{.}(2020{\natexlab{b}}){Dekel}, {Lapiner},
  {Ginzburg}, {Jiang}, {Ceverino}, \& {Primack}}]{dekel20_ring}
{Dekel} A., {Lapiner} S., {Ginzburg} O., {Jiang} F., {Ceverino} D., {Primack}
  J., 2020{\natexlab{b}}, arXiv e-prints

\bibitem[{{Dekel} {et~al}\mbox{.}(2019){Dekel}, {Sarkar}, {Jiang}, {Bournaud},
  {Krumholz}, {Ceverino}, \& {Primack}}]{dekel19_ks}
{Dekel} A., {Sarkar} K.~C., {Jiang} F., {Bournaud} F., {Krumholz} M.~R.,
  {Ceverino} D., {Primack} J.~R., 2019, arXiv e-prints

\bibitem[{{Dekel} \& {Silk}(1986)}]{ds86}
{Dekel} A., {Silk} J., 1986, \apj, 303, 39

\bibitem[{{Dekel} {et~al}\mbox{.}(2013){Dekel}, {Zolotov}, {Tweed}, {Cacciato},
  {Ceverino}, \& {Primack}}]{dekel13}
{Dekel} A., {Zolotov} A., {Tweed} D., {Cacciato} M., {Ceverino} D., {Primack}
  J.~R., 2013, \mnras, 435, 999

\bibitem[{{Dubois} {et~al}\mbox{.}(2015){Dubois}, {Volonteri}, {Silk},
  {Devriendt}, {Slyz}, \& {Teyssier}}]{dubois15}
{Dubois} Y., {Volonteri} M., {Silk} J., {Devriendt} J., {Slyz} A., {Teyssier}
  R., 2015, \mnras, 452, 1502

\bibitem[{{Dutton} {et~al}\mbox{.}(2016){Dutton}, {Macci{\`o}}, {Dekel},
  {Wang}, {Stinson}, {Obreja}, {Di Cintio}, {Brook}, {Buck}, \&
  {Kang}}]{dutton16b}
{Dutton} A.~A. {et~al.}, 2016, \mnras, 461, 2658

\bibitem[{Eddington(1916)}]{eddington16}
Eddington A.~S., 1916, Monthly Notices of the Royal Astronomical Society, 76,
  572

\bibitem[{{El-Zant}, {Shlosman} \& {Hoffman}(2001){El-Zant}, {Shlosman}, \&
  {Hoffman}}]{elzant01}
{El-Zant} A., {Shlosman} I., {Hoffman} Y., 2001, \apj, 560, 636

\bibitem[{{El-Zant}(2008)}]{elzant08}
{El-Zant} A.~A., 2008, \apj, 681, 1058

\bibitem[{{El-Zant} {et~al}\mbox{.}(2004){El-Zant}, {Hoffman}, {Primack},
  {Combes}, \& {Shlosman}}]{elzant04}
{El-Zant} A.~A., {Hoffman} Y., {Primack} J., {Combes} F., {Shlosman} I., 2004,
  \apjl, 607, L75

\bibitem[{{Errani} \& {Navarro}(2020)}]{errani21}
{Errani} R., {Navarro} J.~F., 2020, arXiv e-prints, arXiv:2011.07077

\bibitem[{{Errani}, {Pe{\~n}arrubia} \& {Walker}(2018){Errani},
  {Pe{\~n}arrubia}, \& {Walker}}]{errani18}
{Errani} R., {Pe{\~n}arrubia} J., {Walker} M.~G., 2018, \mnras, 481, 5073

\bibitem[{{Ferland} {et~al}\mbox{.}(1998){Ferland}, {Korista}, {Verner},
  {Ferguson}, {Kingdon}, \& {Verner}}]{ferland98}
{Ferland} G.~J., {Korista} K.~T., {Verner} D.~A., {Ferguson} J.~W., {Kingdon}
  J.~B., {Verner} E.~M., 1998, \pasp, 110, 761

\bibitem[{{Flores} \& {Primack}(1994)}]{flores94}
{Flores} R.~A., {Primack} J.~R., 1994, \apjl, 427, L1

\bibitem[{{F{\"o}rster Schreiber} {et~al}\mbox{.}(2019){F{\"o}rster Schreiber},
  {{\"U}bler}, {Davies}, {Genzel}, {Wisnioski}, {Belli}, {Shimizu}, {Lutz},
  {Fossati}, {Herrera-Camus}, {Mendel}, {Tacconi}, {Wilman}, {Beifiori},
  {Brammer}, {Burkert}, {Carollo}, {Davies}, {Eisenhauer}, {Fabricius},
  {Lilly}, {Momcheva}, {Naab}, {Nelson}, {Price}, {Renzini}, {Saglia},
  {Sternberg}, {van Dokkum}, \& {Wuyts}}]{forster19}
{F{\"o}rster Schreiber} N.~M. {et~al.}, 2019, \apj, 875, 21

\bibitem[{{Freundlich}, {Dekel} \& {Jiang}(2019){Freundlich}, {Dekel}, \&
  {Jiang}}]{freundlich19}
{Freundlich} J., {Dekel} A., {Jiang} F., 2019, in SF2A-2019: Proceedings of the
  Annual meeting of the French Society of Astronomy and Astrophysics, {Di
  Matteo} P., {Creevey} O., {Crida} A., {Kordopatis} G., {Malzac} J.,
  {Marquette} J.~B., {N'Diaye} M., {Venot} O., eds., p.~Di

\bibitem[{{Freundlich} {et~al}\mbox{.}(2020{\natexlab{a}}){Freundlich},
  {Dekel}, {Jiang}, {Ishai}, {Cornuault}, {Lapiner}, {Dutton}, \&
  {Macci{\`o}}}]{freundlich20_cuspcore}
{Freundlich} J., {Dekel} A., {Jiang} F., {Ishai} G., {Cornuault} N., {Lapiner}
  S., {Dutton} A.~A., {Macci{\`o}} A.~V., 2020{\natexlab{a}}, \mnras, 491, 4523

\bibitem[{{Freundlich} {et~al}\mbox{.}(2020{\natexlab{b}}){Freundlich},
  {Jiang}, {Dekel}, {Cornuault}, {Ginzburg}, {Koskas}, {Lapiner}, {Dutton}, \&
  {Macci{\`o}}}]{freundlich20_prof}
{Freundlich} J. {et~al.}, 2020{\natexlab{b}}, \mnras, 499, 2912

\bibitem[{{Genzel} {et~al}\mbox{.}(2020){Genzel}, {Price}, {{\"U}bler},
  {F{\"o}rster Schreiber}, {Shimizu}, {Tacconi}, {Bender}, {Burkert},
  {Contursi}, {Coogan}, {Davies}, {Davies}, {Dekel}, {Herrera-Camus}, {Lee},
  {Lutz}, {Naab}, {Neri}, {Nestor}, {Renzini}, {Saglia}, {Schuster},
  {Sternberg}, {Wisnioski}, \& {Wuyts}}]{genzel21}
{Genzel} R. {et~al.}, 2020, arXiv e-prints, arXiv:2006.03046

\bibitem[{Green, van~den Bosch \& Jiang(2021)Green, van~den Bosch, \&
  Jiang}]{green21}
Green S.~B., van~den Bosch F.~C., Jiang F., 2021, Monthly Notices of the Royal
  Astronomical Society, 503, 4075

\bibitem[{{Haardt} \& {Madau}(1996)}]{haardt96}
{Haardt} F., {Madau} P., 1996, \apj, 461, 20

\bibitem[{{Hopkins}, {Quataert} \& {Murray}(2012){Hopkins}, {Quataert}, \&
  {Murray}}]{hopkins12b}
{Hopkins} P.~F., {Quataert} E., {Murray} N., 2012, \mnras, 421, 3522

\bibitem[{{Huertas-Company} {et~al}\mbox{.}(2018){Huertas-Company}, {Primack},
  {Dekel}, {Koo}, {Lapiner}, {Ceverino}, {Simons}, {Snyder}, {Bernardi},
  {Chen}, {Dom{\'{\i}}nguez-S{\'a}nchez}, {Lee}, {Margalef-Bentabol}, \&
  {Tuccillo}}]{huertas18}
{Huertas-Company} M. {et~al.}, 2018, \apj, 858, 114

\bibitem[{{Inoue} {et~al}\mbox{.}(2016){Inoue}, {Dekel}, {Mandelker},
  {Ceverino}, {Bournaud}, \& {Primack}}]{inoue16}
{Inoue} S., {Dekel} A., {Mandelker} N., {Ceverino} D., {Bournaud} F., {Primack}
  J., 2016, \mnras, 456, 2052

\bibitem[{{Jiang} {et~al}\mbox{.}(2020){Jiang}, {Dekel}, {Freundlich}, {van den
  Bosch}, {Green}, {Hopkins}, {Benson}, \& {Du}}]{jiang21_satgen}
{Jiang} F., {Dekel} A., {Freundlich} J., {van den Bosch} F.~C., {Green} S.~B.,
  {Hopkins} P.~F., {Benson} A., {Du} X., 2020, arXiv e-prints, arXiv:2005.05974

\bibitem[{{Jiang} {et~al}\mbox{.}(2019){Jiang}, {Dekel}, {Kneller}, {Lapiner},
  {Ceverino}, {Primack}, {Faber}, {Macci{\`o}}, {Dutton}, {Genel}, \&
  {Somerville}}]{jiang19_spin}
{Jiang} F. {et~al.}, 2019, \mnras, 488, 4801

\bibitem[{{Jiang} \& {van den Bosch}(2014)}]{jiang14}
{Jiang} F., {van den Bosch} F.~C., 2014, \mnras, 440, 193

\bibitem[{Kazantzidis, Zentner \& Kravtsov(2006)Kazantzidis, Zentner, \&
  Kravtsov}]{kazantzidis05}
Kazantzidis S., Zentner A.~R., Kravtsov A.~V., 2006, Astrophys. J., 641, 647

\bibitem[{{King}(1962)}]{king62}
{King} I., 1962, \aj, 67, 471

\bibitem[{{Kocevski} {et~al}\mbox{.}(2017){Kocevski}, {Barro}, {Faber},
  {Dekel}, {Somerville}, {Young}, {Williams}, \& {et al.}}]{kocevski17}
{Kocevski} D.~D., {Barro} G., {Faber} S.~M., {Dekel} A., {Somerville} R.~S.,
  {Young} J.~A., {Williams} C.~C., {et al.}, 2017, \apj, 846, 112

\bibitem[{{Komatsu} {et~al}\mbox{.}(2009){Komatsu}, {Dunkley}, {Nolta},
  {Bennett}, {Gold}, {Hinshaw}, {Jarosik}, \& {et al.}}]{komatsu09}
{Komatsu} E., {Dunkley} J., {Nolta} M.~R., {Bennett} C.~L., {Gold} B.,
  {Hinshaw} G., {Jarosik} N., {et al.}, 2009, \apjs, 180, 330

\bibitem[{{Kravtsov}(2003)}]{krav03}
{Kravtsov} A.~V., 2003, \apjl, 590, L1

\bibitem[{{Kravtsov}, {Klypin} \& {Khokhlov}(1997){Kravtsov}, {Klypin}, \&
  {Khokhlov}}]{krav97}
{Kravtsov} A.~V., {Klypin} A.~A., {Khokhlov} A.~M., 1997, \apjs, 111, 73

\bibitem[{{Kretschmer} {et~al}\mbox{.}(2021){Kretschmer}, {Dekel},
  {Freundlich}, {Lapiner}, {Ceverino}, \& {Primack}}]{kretschmer21_kvir}
{Kretschmer} M., {Dekel} A., {Freundlich} J., {Lapiner} S., {Ceverino} D.,
  {Primack} J., 2021, \mnras, 503, 5238

\bibitem[{{Krumholz} \& {Thompson}(2013)}]{krum_thom13}
{Krumholz} M.~R., {Thompson} T.~A., 2013, \mnras, 434, 2329

\bibitem[{{Lacey} \& {Cole}(1993)}]{lacey93}
{Lacey} C., {Cole} S., 1993, \mnras, 262, 627

\bibitem[{{Lapiner}, {Dekel} \& {Dubois}(2020){Lapiner}, {Dekel}, \&
  {Dubois}}]{lapiner21}
{Lapiner} S., {Dekel} A., {Dubois} Y., 2020, arXiv e-prints, arXiv:2012.09186

\bibitem[{{Lazar} {et~al}\mbox{.}(2020){Lazar}, {Bullock}, {Boylan-Kolchin},
  {Chan}, {Hopkins}, {Graus}, {Wetzel}, {El-Badry}, {Wheeler}, {Straight},
  {Kere{\v{s}}}, {Faucher-Gigu{\`e}re}, {Fitts}, \&
  {Garrison-Kimmel}}]{lazar20}
{Lazar} A. {et~al.}, 2020, \mnras, 497, 2393

\bibitem[{{Macci{\`o}} {et~al}\mbox{.}(2020){Macci{\`o}}, {Crespi}, {Blank}, \&
  {Kang}}]{maccio20}
{Macci{\`o}} A.~V., {Crespi} S., {Blank} M., {Kang} X., 2020, \mnras, 495, L46

\bibitem[{{Mandelker} {et~al}\mbox{.}(2017){Mandelker}, {Dekel}, {Ceverino},
  {DeGraf}, {Guo}, \& {Primack}}]{mandelker17}
{Mandelker} N., {Dekel} A., {Ceverino} D., {DeGraf} C., {Guo} Y., {Primack} J.,
  2017, \mnras, 464, 635

\bibitem[{{Mandelker} {et~al}\mbox{.}(2014){Mandelker}, {Dekel}, {Ceverino},
  {Tweed}, {Moody}, \& {Primack}}]{mandelker14}
{Mandelker} N., {Dekel} A., {Ceverino} D., {Tweed} D., {Moody} C.~E., {Primack}
  J., 2014, \mnras, 443, 3675

\bibitem[{{Martizzi}, {Teyssier} \& {Moore}(2013){Martizzi}, {Teyssier}, \&
  {Moore}}]{martizzi13}
{Martizzi} D., {Teyssier} R., {Moore} B., 2013, \mnras, 432, 1947

\bibitem[{{Martizzi} {et~al}\mbox{.}(2012){Martizzi}, {Teyssier}, {Moore}, \&
  {Wentz}}]{martizzi12}
{Martizzi} D., {Teyssier} R., {Moore} B., {Wentz} T., 2012, \mnras, 422, 3081

\bibitem[{{Moody} {et~al}\mbox{.}(2014){Moody}, {Guo}, {Mandelker}, {Ceverino},
  {Mozena}, {Koo}, {Dekel}, \& {Primack}}]{moody14}
{Moody} C.~E., {Guo} Y., {Mandelker} N., {Ceverino} D., {Mozena} M., {Koo}
  D.~C., {Dekel} A., {Primack} J., 2014, \mnras, 444, 1389

\bibitem[{{Moster}, {Naab} \& {White}(2018){Moster}, {Naab}, \&
  {White}}]{moster18}
{Moster} B.~P., {Naab} T., {White} S. D.~M., 2018, \mnras, 477, 1822

\bibitem[{{Murray}, {Quataert} \& {Thompson}(2010){Murray}, {Quataert}, \&
  {Thompson}}]{murray10}
{Murray} N., {Quataert} E., {Thompson} T.~A., 2010, \apj, 709, 191

\bibitem[{{Navarro}, {Frenk} \& {White}(1997){Navarro}, {Frenk}, \&
  {White}}]{nfw97}
{Navarro} J.~F., {Frenk} C.~S., {White} S.~D.~M., 1997, \apj, 490, 493

\bibitem[{{Neistein} \& {Dekel}(2008)}]{neistein08_m}
{Neistein} E., {Dekel} A., 2008, \mnras, 388, 1792

\bibitem[{{Ogiya} {et~al}\mbox{.}(2019){Ogiya}, {van den Bosch}, {Hahn},
  {Green}, {Miller}, \& {Burkert}}]{ogiya19}
{Ogiya} G., {van den Bosch} F.~C., {Hahn} O., {Green} S.~B., {Miller} T.~B.,
  {Burkert} A., 2019, \mnras, 485, 189

\bibitem[{{Oh} {et~al}\mbox{.}(2011{\natexlab{a}}){Oh}, {Brook}, {Governato},
  {Brinks}, {Mayer}, {de Blok}, {Brooks}, \& {Walter}}]{oh11_sim}
{Oh} S.-H., {Brook} C., {Governato} F., {Brinks} E., {Mayer} L., {de Blok}
  W.~J.~G., {Brooks} A., {Walter} F., 2011{\natexlab{a}}, \aj, 142, 24

\bibitem[{{Oh} {et~al}\mbox{.}(2011{\natexlab{b}}){Oh}, {de Blok}, {Brinks},
  {Walter}, \& {Kennicutt}}]{oh11}
{Oh} S.-H., {de Blok} W.~J.~G., {Brinks} E., {Walter} F., {Kennicutt},
  Robert~C. J., 2011{\natexlab{b}}, \aj, 141, 193

\bibitem[{{Oh} {et~al}\mbox{.}(2015){Oh}, {Hunter}, {Brinks}, {Elmegreen},
  {Schruba}, {Walter}, {Rupen}, {Young}, {Simpson}, {Johnson}, {Herrmann},
  {Ficut-Vicas}, {Cigan}, {Heesen}, {Ashley}, \& {Zhang}}]{oh15}
{Oh} S.-H. {et~al.}, 2015, \aj, 149, 180

\bibitem[{Parkinson, Cole \& Helly(2008)Parkinson, Cole, \&
  Helly}]{parkinson08}
Parkinson H., Cole S., Helly J., 2008, MNRAS, 383, 557

\bibitem[{{Peirani} {et~al}\mbox{.}(2017){Peirani}, {Dubois}, {Volonteri},
  {Devriendt}, {Bundy}, {Silk}, {Pichon}, {Kaviraj}, {Gavazzi}, \&
  {Habouzit}}]{peirani17}
{Peirani} S. {et~al.}, 2017, \mnras, 472, 2153

\bibitem[{{Peirani} {et~al}\mbox{.}(2019){Peirani}, {Sonnenfeld}, {Gavazzi},
  {Oguri}, {Dubois}, {Silk}, {Pichon}, {Devriendt}, \& {Kaviraj}}]{peirani19}
{Peirani} S. {et~al.}, 2019, \mnras, 483, 4615

\bibitem[{Penarrubia {et~al}\mbox{.}(2010)Penarrubia, Benson, Walker, Gilmore,
  McConnachie, \& Mayer}]{penarrubia10}
Penarrubia J., Benson A.~J., Walker M.~G., Gilmore G., McConnachie A.~W., Mayer
  L., 2010, MNRAS, 406, 1290

\bibitem[{{Pontzen} \& {Governato}(2012)}]{pontzen12}
{Pontzen} A., {Governato} F., 2012, \mnras, 421, 3464

\bibitem[{{Roca-F{\`a}brega} {et~al}\mbox{.}(2018){Roca-F{\`a}brega}, {Dekel},
  {Faerman}, {Gnat}, {Strawn}, {Ceverino}, {Primack}, {Macci{\`o}}, {Dutton},
  {Prochaska}, \& {Stern}}]{roca19}
{Roca-F{\`a}brega} S. {et~al.}, 2018, arXiv e-prints

\bibitem[{{Rodr{\'{\i}}guez-Puebla}
  {et~al}\mbox{.}(2017){Rodr{\'{\i}}guez-Puebla}, {Primack}, {Avila-Reese}, \&
  {Faber}}]{rodriguez17}
{Rodr{\'{\i}}guez-Puebla} A., {Primack} J.~R., {Avila-Reese} V., {Faber} S.~M.,
  2017, \mnras, 470, 651

\bibitem[{{Sharma}, {Salucci} \& {van de Ven}(2021){Sharma}, {Salucci}, \& {van
  de Ven}}]{sharma21}
{Sharma} G., {Salucci} P., {van de Ven} G., 2021, arXiv e-prints,
  arXiv:2105.13684

\bibitem[{{Silk}(2013)}]{silk13}
{Silk} J., 2013, \apj, 772, 112

\bibitem[{{Snyder} {et~al}\mbox{.}(2015){Snyder}, {Lotz}, {Moody}, {Peth},
  {Freeman}, {Ceverino}, {Primack}, \& {Dekel}}]{snyder15}
{Snyder} G.~F., {Lotz} J., {Moody} C., {Peth} M., {Freeman} P., {Ceverino} D.,
  {Primack} J., {Dekel} A., 2015, \mnras, 451, 4290

\bibitem[{{Springel}(2005)}]{springel05_gadget2}
{Springel} V., 2005, \mnras, 364, 1105

\bibitem[{{Springel} \& {Hernquist}(2005)}]{springel05}
{Springel} V., {Hernquist} L., 2005, \apjl, 622, L9

\bibitem[{{Strawn} {et~al}\mbox{.}(2020){Strawn}, {Roca-F{\`a}brega},
  {Mandelker}, {Primack}, {Stern}, {Ceverino}, {Dekel}, {Wang}, \&
  {Dange}}]{strawn20}
{Strawn} C. {et~al.}, 2020, \mnras

\bibitem[{{Tacchella} {et~al}\mbox{.}(2016{\natexlab{a}}){Tacchella}, {Dekel},
  {Carollo}, {Ceverino}, {DeGraf}, {Lapiner}, {Mandelker}, \&
  {Primack}}]{tacchella16_prof}
{Tacchella} S., {Dekel} A., {Carollo} C.~M., {Ceverino} D., {DeGraf} C.,
  {Lapiner} S., {Mandelker} N., {Primack} J.~R., 2016{\natexlab{a}}, \mnras,
  458, 242

\bibitem[{{Tacchella} {et~al}\mbox{.}(2016{\natexlab{b}}){Tacchella}, {Dekel},
  {Carollo}, {Ceverino}, {DeGraf}, {Lapiner}, {Mandelker}, \& {Primack
  Joel}}]{tacchella16_ms}
{Tacchella} S., {Dekel} A., {Carollo} C.~M., {Ceverino} D., {DeGraf} C.,
  {Lapiner} S., {Mandelker} N., {Primack Joel} R., 2016{\natexlab{b}}, \mnras,
  457, 2790

\bibitem[{{Tacconi} {et~al}\mbox{.}(2018){Tacconi}, {Genzel}, {Saintonge},
  {Combes}, {Garc{\'{\i}}a-Burillo}, {Neri}, {Bolatto}, \& {et
  al.}}]{tacconi18}
{Tacconi} L.~J., {Genzel} R., {Saintonge} A., {Combes} F.,
  {Garc{\'{\i}}a-Burillo} S., {Neri} R., {Bolatto} A., {et al.}, 2018, \apj,
  853, 179

\bibitem[{{Tomassetti} {et~al}\mbox{.}(2016){Tomassetti}, {Dekel}, {Mandelker},
  {Ceverino}, {Lapiner}, {Faber}, {Kneller}, {Primack}, \&
  {Sai}}]{tomassetti16}
{Tomassetti} M. {et~al.}, 2016, \mnras, 458, 4477

\bibitem[{{Tremaine} \& {Weinberg}(1984)}]{tremaine84}
{Tremaine} S., {Weinberg} M.~D., 1984, \mnras, 209, 729

\bibitem[{{van den Bosch}(2017)}]{bosch17}
{van den Bosch} F.~C., 2017, \mnras, 468, 885

\bibitem[{{van den Bosch} \& {Ogiya}(2018)}]{bosch18}
{van den Bosch} F.~C., {Ogiya} G., 2018, \mnras, 475, 4066

\bibitem[{van Kampen(2000)}]{vankampen00}
van Kampen E., 2000, arXiv

\bibitem[{{Wang} {et~al}\mbox{.}(2020){Wang}, {Vogelsberger}, {Xu}, {Mao},
  {Springel}, {Li}, {Barnes}, {Hernquist}, {Pillepich}, {Marinacci}, {Pakmor},
  {Weinberger}, \& {Torrey}}]{wang20}
{Wang} Y. {et~al.}, 2020, \mnras, 491, 5188

\bibitem[{{Weinberger} {et~al}\mbox{.}(2018){Weinberger}, {Springel}, {Pakmor},
  {Nelson}, {Genel}, {Pillepich}, {Vogelsberger}, {Marinacci}, {Naiman},
  {Torrey}, \& {Hernquist}}]{weinberger18}
{Weinberger} R. {et~al.}, 2018, \mnras, 479, 4056

\bibitem[{{Wetzel}(2011)}]{wetzel11}
{Wetzel} A.~R., 2011, \mnras, 412, 49

\bibitem[{{Wuyts} {et~al}\mbox{.}(2016){Wuyts}, {F{\"o}rster Schreiber},
  {Wisnioski}, {Genzel}, {Burkert}, {Bandara}, {Beifiori}, {Belli}, {Bender},
  {Brammer}, {Chan}, {Davies}, {Fossati}, {Galametz}, {Kulkarni}, {Lang},
  {Lutz}, {Mendel}, {Momcheva}, {Naab}, {Nelson}, {Saglia}, {Seitz}, {Tacconi},
  {Tadaki}, {{\"U}bler}, {van Dokkum}, {Wilman}, \& {Wuyts}}]{wuyts16}
{Wuyts} S. {et~al.}, 2016, \apj, 831, 149

\bibitem[{{Zhao} {et~al}\mbox{.}(2009){Zhao}, {Jing}, {Mo}, \&
  {B{\"o}rner}}]{zhao09}
{Zhao} D.~H., {Jing} Y.~P., {Mo} H.~J., {B{\"o}rner} G., 2009, \apj, 707, 354

\bibitem[{{Zhao}(1996)}]{zhao96}
{Zhao} H., 1996, \mnras, 278, 488

\bibitem[{{Zolotov} {et~al}\mbox{.}(2015){Zolotov}, {Dekel}, {Mandelker},
  {Tweed}, {Inoue}, {DeGraf}, {Ceverino}, {Primack}, {Barro}, \&
  {Faber}}]{zolotov15}
{Zolotov} A. {et~al.}, 2015, \mnras, 450, 2327

\end{thebibliography}

%%%%%%%%%%%%%%%%%%%%%%%
\appendix

%A DZ
%B VELA
%C SatGen
%D NB
%E CuspCore

%F total sats
%G Steep host   Figs: Toy, SatGen single, SatGen cosmo, CuspCore
%H Toy s1,c2  Fig.
%%I CuspCore alpha Fig.

%%%%%%%%%%%%%%%%%%%%%%%%%%%
\section{The Dekel-Zhao Profile} % A
\label{sec:app_Dekel-Zhao}

The Dekel-Zhao halo profile 
\citep{dekel17,freundlich20_prof},\footnote{Available~for~implementation~in\\
https://github.com/JonathanFreundlich/Dekel\_profile .}
following a more general mathematical analysis by \citet{zhao96},
is a functional form for dark-matter haloes
with two free shape parameters, a concentration $c$ and 
an inner slope $\alpha$, allowing the central region to range continuously
from a steep cusp to a flat core.
It has been found to fit dark-matter haloes in cosmological hydro simulations 
better than other two-parameter profiles such as the Einasto and the 
generalized-NFW profiles with a flexible inner slope.
A unique feature is that it has analytic expressions not only for the density
and mass-velocity profiles but also for the potential and kinetic
energy profiles (as well as for gravitational lensing properties).
\citet{freundlich20_prof} 
also provide the typical profile parameters as a function of mass.

\smallskip
The profile of {\it mean} density within a sphere of radius $r$ is given by
\be
\brho(r) = \frac{{\brho}_{\rm c}}{x^{\alpha}\, (1+x^{1/2})^{2(3-\alpha)}} \, ,
\quad x= \frac{r}{\Rv}\, c \, ,
\label{eq:app_barrho}
\ee
\be
{\brho}_{\rm c} = c^3 \mu(c,\alpha)\, \brhov , \quad
\brhov = \frac{\Mv}{(4 \pi/3) \Rv^3} \, ,
\label{eq:app_rhoc}
\ee
\be
\mu(c,\alpha) = c^{\alpha-3}\, (1+c^{1/2})^{2(3-\alpha)} \, .
\ee
For completeness, the local density profile is
\be
\rho(r) = \frac{(1-\alpha/3)\,{\brho}_{\rm c}}
{x^\alpha\, (1+x^{1/2})^{2(3.5-\alpha)}}\, .
\ee

\smallskip
The associated mass profile is
\be
\frac{M(x)}{\Mv} = \frac{1}{c^3 \brhov} x^3 \brho(x) 
=\frac{\mu}{{\brho}_{\rm c}}\, x^3\, \brho(x)\, .
\label{eq:app_m}
\ee
%Note that by definition $\bsigv=\brhov$, the same for the host and for the
%satellite when it was still isolated.
The log slope of the mass profile is
\be
\nu(r) = \frac{3-\alpha}{1+x^{1/2}} \, .
\label{eq:app_nu}
\ee
The negative log slope of the density profile is
\be
s(r) = \frac{\alpha + 3.5 x^{1/2}}{1+ x^{1/2}} \, .
\label{eq:app_s}
\ee

\smallskip
In order to obtain the mass fraction $f=M(<r)/\Mv$ within a sphere of 
radius $r$, we express it using
\equ{app_m} as 
\be
f(r) = \mu\, \brho_{\rm c}^{-1}\, x^3\, \brho(x) \, .
\label{eq:app_f}
\ee
Combining \equ{app_f} and \equ{app_barrho}, we get
\be
f(r) = \mu\, x^{3-\alpha}\, (1+x^{1/2})^{-2(3-\alpha)} \, ,
\ee
from which we obtain
\be
x = [ (f/\mu)^{-1/[2(3-\alpha)]} -1]^{-2} \, .
\label{eq:app_xt}
\ee
% This is equ 17 of freundlich20b ?
Inserting \equ{app_xt} in \equ{app_f}, we obtain an equation for 
$f(r)$,
\be
(f/\mu)\, [ (f/\mu)^{-1/[2(3-\alpha)]} -1 ]^6
= \brho(r_{\rm t})/\brho_{\rm c} \, .
\label{eq:app_frac}
\ee

\smallskip
We recall from equations 11-14 of \citet{freundlich20_prof} that
a more physical pair of shape parameters, that refer to $\rho(r)$ rather to
$\bar{\rho}(r)$, may be $(c_2,s_1)$. The concentration $c_2$ refers to the
virial radius with respect to the radius where the log density slope is $-2$,
\be
c_2 = c\, \left( \frac{1.5}{2-\alpha} \right)^2 \, ,
\label{eq:c2}
\ee
valid for $\alpha\!<\!2$.
The inner slope $s_1$ is
minus the log slope of the local density profile $\rho(r)$
at a given radius $r_1$ (say $r_1\seq 0.01\Rv$),
\be
s_1 = \frac{\alpha+ 3.5 x_1^{1/2}}{1+x_1^{1/2}} \, , \quad
x_1=\frac{r_1}{\Rv}\, c \, .
\label{eq:s1}
\ee
For completeness, the inverse relations are
\be
c=\left(\frac{s_1-2}{(3.5-s_1)(r_1/\Rv)^{1/2} - 1.5 c_2^{-1/2}}\right)^2 \, ,
\label{eq:c}
\ee
\be
\alpha=\frac{1.5 s_1 - 2(3.5-s_1) x_{2,1}^{1/2}}{1.5-(3.5-s_1)x_{2,1}^{1/2}} 
\, , \quad x_{2,1} = \frac{r_1}{\Rv} c_2 \, .
\label{eq:alpha}
\ee
We note that a valid DZ solution is not guaranteed for any arbitrary
pair of values $(c_2,s_1)$, e.g., there is no solution where the denominator 
in either \equ{c} or \equ{alpha} vanishes.

\smallskip % NFW fit
A best fit for an NFW profile with a given concentration 
is obtained, e.g., by minimizing residuals in uniformly spaced log radii 
in the range $\log(r/\Rv) = (-2,0)$. 
For $c_{\rm NFW}\seq 5$ we obtain for the best-fit DZ parameters
$(c,\alpha)\seq(7.126,0.2156)$ or $(c_2,s_1)\seq(5.035,0.9076)$.
A slightly better fit near $0.01\Rv$ can be obtained with 
$(c_2,s_1)\!\simeq\!(4.8,1.0)$, but this is at the expense of slightly larger
deviations at large radii.

%\smallskip
%\adr{Possibly add $r_{\rm max}$ and $V_{\rm max}$}

\def\ta{\tilde\alpha}

\smallskip % \adr{Add energies}
The gravitational potential as given in eq. 19 of 
\citet[][]{freundlich20_prof} is  
\begin{equation}
\label{eq:U_DZ}
U(r) \seq -\Vv^2 \left[ 1 + 2c\mu 
\left( \frac{\chi_c^{\ta}-\chi^{\ta}}{\ta} - 
\frac{\chi_c^{\ta+1}-\chi^{\ta+1}}{\ta+1} \right)\right]\, , 
\end{equation}
%where 
%$\chi\seq x^{1/2}/(1+x^{1/2})$, $\chi_c \seq c^{1/2}/(1+c^{1/2})$,
%and $\ta\seq 2(2-\alpha)$.
\be
\chi = \frac{x^{1/2}}{1+x^{1/2}}, \quad \chi_c = \frac{c^{1/2}}{1+c^{1/2}},
\quad \ta = 2(2-\alpha) \, .
\ee

\smallskip % velocity dispersion
The velocity dispersion that stems from the Jeans equation 
\citep[][eq.~22]{freundlich20_prof}, providing the kinetic energy per unit
mass, is 
\begin{equation}
\label{eq:sigmar_DZ}
\sigma_r^2(r)
= 2 c\mu \Vv^2 \frac{\rho_{\rm c}}{\rho(r)}
\Big[ \mathcal{B}(4-4\alpha,\,9,\,\zeta)\Big]_\chi^{\chi_c} \, ,
\end{equation}
where $\mathcal{B}(a,b,x) \seq \int_0^x t^{a-1}(1-t)^{b-1} dt$ 
is the incomplete 
beta function, the brackets denote the difference of the enclosed function 
between 1 and $\chi$, i.e., 
$\left[ f(\zeta)\right]_\chi^{\chi_c} \equiv f(\chi_c)-f(\chi)$,
and $\rho_c \seq (1-\alpha/3)\bar{\rho}_{\rm c}$. 
The definition of the incomplete beta function has been extended to negative 
parameters since the bracketed term is well-defined. 
Equations B1 and B3 of \citet{freundlich20_prof} are the equivalent 
expressions in terms of finite series. 

\smallskip
Adding an additional point mass $M_{\rm b}$ at the halo center adds
the velocity dispersion term \citep[][eq.~C15]{freundlich20_prof}
\begin{equation}
\label{eq:sigmarm_DZ}
\sigma_{M_{\rm b}}^2 (r) = 2c\frac{G M_{\rm b}}{\Rv} \frac{{\rho_c}}{{\rho(r)}} 
\Big[ \mathcal{B}(-2-2\alpha,\,9,\,\zeta)\Big]_\chi^1 \, .
\end{equation}

%\smallskip
%\adr{Add Jeans equilibrium for relaxation after DF energy deposit}

\smallskip
For comparison, the NFW mass profile, with an inner cusp of negative log
density slope $\alpha\seq 1$ and a free NFW concentration parameter $\ch$, is 
\be
\frac{M(r)}{\Mv} = \frac{A(x)}{A(\ch)}\, , \quad
\frac{\brho(r)}{\brhov} = \frac{A(x)}{A(\ch)} \frac{\ch^3}{x^3}\, ,
\label{eq:app_nfw}
\ee
\be
A(x)=\ln(1+x)-\frac{x}{1+x}\, , \quad
x = \frac{r}{\Rv} \ch\, ,
\ee
The profile of the negative log slope of the mass profile is  
\be
\nu(r) = \frac{x^2}{(1+x)^2\,A(x)} \, .
\ee

%%%%%%%%%%%%%%%%%%%%%%%%%%%
\section{The VELA Simulations} % B
\label{sec:app_vela}

%\adr{shorten to relevant issues for this paper}

\smallskip % VELA suite
The VELA suite consists of hydro-cosmological simulations zooming-in on
34 moderately massive galaxies, presented in more detail in
\citet{ceverino14} and \citet{zolotov15}.
This suite has been used to study central issues in the evolution of galaxies
at high redshifts, including, e.g.,
compaction to blue nuggets and the trigger of quenching
\citep{zolotov15,tacchella16_ms,tacchella16_prof},
evolution of global shape
\citep{ceverino15_shape,tomassetti16},
violent disc instability \citep{mandelker14,mandelker17,inoue16},
the SFR-density relation by supernova feedback \citep{dekel19_ks},
post-compaction formation of discs and rings \citep{dekel20_flip,dekel20_ring},
OVI in the CGM \citep{roca19,strawn20},
and angular momentum and galaxy size \citep{jiang19_spin}.
Additional analysis of the same suite of simulations are discussed in
\citet{moody14,snyder15}.
This appendix provides an overview of the relevant features of these
simulations.
%and their limitations.

%--------------
%\subsection{The Cosmological Simulations}

\smallskip  % ART
The VELA simulations make use of the Adaptive Refinement Tree (ART) code
\citep{krav97,krav03,ceverino09}, which follows the
evolution of a gravitating N-body system and the Eulerian gas dynamics using
an adaptive mesh refinement. The maximum spatial 
resolution is $17-35\pc$ at all times.
%, which is achieved at densities of $\sim10^{-4}-10^3\cmc$.
The code incorporates subgrid recipes for physical process that are
relevant for galaxy formation, such as gas cooling by atomic hydrogen and
helium, metal and molecular hydrogen cooling, photoionization heating by the
UV background with partial self-shielding, star formation, stellar mass loss,
metal enrichment of the ISM and stellar feedback. Supernovae and stellar winds
are implemented by local injection of thermal energy as described in
\citet{ceverino09,cdb10} and \citet{ceverino12}. Radiation-pressure
stellar feedback is implemented at a moderate level, following
\citet{dekel13}, as described in \citet{ceverino14}.

\smallskip % cooling
Cooling and heating rates are 
%tabulated for a given gas density, temperature,
%metallicity and UV background 
based on the CLOUDY code \citep{ferland98}.
%assuming a slab of thickness $1\kpc$. 
A uniform UV background based on the
redshift-dependent \citet{haardt96} model is assumed, except at gas densities
higher than $0.1\cmc$, where 
%a substantially suppressed UV background is used
%($5.9\times10^6 \erg\, {\rm s}^{-1} \cms\, {\rm Hz}^{-1}$)
%in order to mimic the 
partial self-shielding allows dense gas
to cool down to $\sim300$K. The assumed equation of
state is that of an ideal mono-atomic gas. Artificial fragmentation on the cell
size is prevented by introducing a pressure floor, which ensures that the Jeans
scale is resolved by at least 7 cells \citep[see][]{cdb10}.
%
%\smallskip  % SFR
Star particles form in timesteps of $5 \Myr$ in cells where the gas density
exceeds $1~\cmc$ and the temperatures is below $10^4$K. 
%Most stars ($>90\%$) end up forming at temperatures well
%below $10^3$K, and more than half of the stars form near
%$300$K in cells where the gas density is higher than $10~\cmc$.
The code implements a stochastic star formation where a star particle with a
mass of $42\%$ of the gas mass forms with a probability
$P=(\rhog/10^3\cmc)^{1/2}$ but not higher than $0.2$.
%This corresponds to a local SFR that crudely mimics
%$\drhos \epsf \rhog/\tff$ with $\epsf \sim 0.02$.
%A stellar initial mass function of \citet{chabrier03} is assumed.

\smallskip  % feedback
Thermal feedback that mimics the energy release from stellar winds and
supernova explosions is incorporated as a constant heating rate over
the $40~\Myr$ following star formation.
A velocity kick of $\sim10\kms$ is applied
to $30~\%$ of the newly formed stellar particles -- this enables SN explosions
in lower density regions where the cooling may not overcome the heating without
implementing an artificial shutdown of cooling \citep{ceverino09}.
The code also incorporates the later effects of Type Ia supernova and
stellar mass loss, and it follows the metal enrichment of the ISM.
%
%\smallskip % rad pressure
Radiation pressure is incorporated through the addition of a non-thermal
pressure term to the total gas pressure in regions where ionizing photons
from massive stars are produced and may be trapped. This ionizing radiation
injects momentum in the cells neighbouring massive star particles younger than
$5\Myr$, and whose column density exceeds
$10^{21}\cms$, isotropically pressurizing the star-forming
regions \citep[see more details in][]{agertz13,ceverino14}.

\smallskip % halos
The initial conditions for the simulations are based on DM haloes that
were drawn from dissipationless N-body simulations at lower resolution in
cosmological boxes of $15-60\Mpc$. The $\Lambda$CDM cosmological model was
assumed with the WMAP5 values of the cosmological parameters,
$\omm=0.27$, $\oml=0.73$, $\omb=0.045$, $h=0.7$ and
$\sigma_8=0.82$ \citep{komatsu09}. Each halo was selected to have a
given virial mass at $z \seq 1$ and no ongoing major merger at $z \seq 1$.
This latter criterion eliminated less than $10~\%$ of the haloes, those
that tend to be in a dense, proto-cluster environment at $z\sim1$.
The virial masses at $z \seq 1$ were chosen to be in the range
$\Mv=2\times10^{11}-2\times10^{12}~M_{\odot}$, about a
median of $4.6\times10^{11}~M_{\odot}$. If left in isolation, the median mass
at $z=0$ was intended to be $\sim10^{12}~M_{\odot}$.

\smallskip % limitations
The VELA cosmological simulations are state-of-the-art in terms
of high-resolution adaptive mesh refinement hydrodynamics and the treatment of
key physical processes at the subgrid level.
In particular, they trace the cosmological streams that feed galaxies at high
redshift, including mergers and smooth flows, and they resolve the violent disc
instability that governs high-$z$ disc evolution and bulge formation
\citep{cdb10,ceverino12,ceverino15_e,mandelker14}.
To mention a few limitations,
like in other simulations, the treatments of star formation and feedback
processes are rather simplified. The code may assume a realistic SFR efficiency
per free fall time on the grid scale
but it does not follow in detail the formation of
molecules and the effect of metallicity on SFR.
The feedback is treated in a crude way, where the resolution does not allow
the capture of the Sedov-Taylor phase of supernova bubbles.
The radiative stellar feedback assumed
no infrared trapping, in the spirit of low trapping advocated by
\citet{dk13} based on \citet{krum_thom13},
which makes the radiative feedback weaker than in other simulations
that assume more significant trapping \citep{murray10,hopkins12b}.
AGN feedback, and feedback associated with cosmic rays and magnetic
fields, are not yet implemented. Nevertheless, as shown in
\citet{ceverino14}, the star-formation rates, gas
fractions, and stellar-to-halo mass ratio are all in the ballpark of the
estimates deduced from observations.

%-----------------
%\subsection{The Galaxy Sample and Measurements}
%\label{subsec:sample}

\smallskip % the sample of galaxies
The virial and stellar properties of the galaxies
are listed for example in Table 1 of \citet{dekel20_ring}.  
%Table~\ref{tab:sample}.
The virial mass $\Mv$ is the total mass within a sphere of radius
$\Rv$ that encompasses an overdensity of $\Delta(z)=
[18\pi^2-82\oml(z)-39\oml(z)^2]/\omm(z)$,
where $\oml(z)$ and $\omm(z)$ are the cosmological
parameters at $z$ \citep{bryan98,db06}. The stellar mass $\Ms$
is the instantaneous mass in stars within a radius of $0.2\Rv$,
accounting for past stellar mass loss.
%
%\smallskip  % sample
We start the analysis at the cosmological time corresponding to expansion
factor $a=0.125$ (redshift $z=7$), and most galaxies reach $a=0.50$ ($z=1$).
Each galaxy is analyzed at output times separated by a constant interval in
$a$, $\Delta a=0.01$, corresponding at $z=2$ to $\sim100~\Myr$
(roughly half an orbital time at the disc edge).
The sample consists of totally $\sim 1000$ snapshots in the redshift range
$z=7-0.8$ from 35 galaxies that at $z = 2$ span the stellar mass range
$(0.2-6.4)\times10^{11}\Msun$. The half-mass sizes
$\Re$ 
%are determined from the $\Ms$ that are measured within a
%radius of $0.2\Rv$ and they 
range $\Re\simeq0.4-3.2\kpc$ at $z=2$.
%
%\smallskip  % center
The determination of the centre of the galaxy is outlined in detail in
appendix B of \citet{mandelker14}.
Briefly, starting form the most bound star, the centre is refined iteratively
by calculating the centre of mass of stellar particles in spheres of
decreasing radii
%, updating the centre and decreasing the radius at each
%iteration.  We begin with an initial radius of 600 pc, and decrease the radius
%by a factor of 1.1 at each iteration. The iteration terminates when the radius
%reaches 
down to
$130\pc$ or when the number of stellar particles in the sphere drops below 20.

\smallskip % compaction and BN
We identify the major event of wet compaction to a blue nugget
for each galaxy. This is the one that leads to a significant central gas
depletion and SFR quenching, and marks the transition from dark-matter to
baryon dominance within $\Re$. Following \citet{zolotov15} and
\citet{tacchella16_prof}, the most physical way to identify the compaction
and blue nugget is by the steep rise of gas density (and SFR) within the inner
$1\kpc$ to the highest peak, as long as it is followed by a significant,
long-term decline in central gas mass density (and SFR).
The onset of compaction can be identified as the start of the steep rise
of central gas density prior to the blue-nugget peak.
An alternative identification is using the shoulder of the stellar mass density
within $1\kpc$ where its rise due to the starburst associated with the 
compaction turns into a plateau of maximum long-term compactness slightly after
the blue-nugget peak of gas density. This is a more practical way to identify
blue nuggets in observations \citep[e.g.][]{barro17}.

%%%%%%%%%%%%%%%%%%%%%%%%%%%% C
\section{SatGen - A Semi-Analytic Satellite Generator}
\label{sec:app_SatGen}

\def\SatGen{SatGen\ }

The semi-analytic model for satellite galaxies {SatGen} is presented in 
\citet{jiang21_satgen}.\footnote{Available~for~
implementation~in\\ https://github.com/shergreen/SatGen .}
It can generate statistical samples of satellite populations for a host halo 
of desired mass, redshift, and cosmological parameters. %assembly history. 
The model combines DM halo merger trees, empirical relations for the 
galaxy-halo connection, and simple analytical prescriptions for tidal effects, 
dynamical friction, and ram pressure stripping (if the satellites contains gas).
{SatGen} emulates cosmological zoom-in simulations in certain aspects. 
Satellites can reside in subhaloes of desired density profiles, with cores or
cusps, depending on the subhalo response to baryonic physics that are  
formulated from hydro-simulations or physical modeling. 
The host potential can be composed of a DM halo and baryonic components, such
as a disc and a bulge, each described by a density profile that allows analytic
integration of the satellite orbits.
The subhalo profile and the stellar mass and structure of a satellite evolves 
due to tidal heating and tidal mass loss, which depend on its initial 
structure. 
{SatGen} complements simulations by propagating the effect of halo response 
found in simulated central galaxies to satellites (which are typically
not properly resolved in simulations). It outperforms simulations by capturing 
the halo-to-halo variance of satellite statistics and overcoming artificial 
disruption due to insufficient resolution \citep{bosch18,green21}.
Certain features of SatGen that are relevant for our current study are
elaborated on below.

\smallskip
\SatGen generates halo merger trees using the algorithm of \citet{parkinson08}
as re-calibrated by  \citet{benson17}.
Merger trees are constructed using the time-stepping advocated in Appendix A of 
\citet{parkinson08}, which corresponds to $\Delta z \ssimeq 0.001$, 
but for book keeping the temporal resolution is down-sampled to
timesteps of $\Delta t \seq 0.1 \tdyn(z)$, 
where $\tdyn \seq \sqrt{3\pi/[16G\, \Delta\, \rho_{\rm crit}(z)]}$ 
is the instantaneous virial time of DM haloes.
In the EdS regime, approximately valid at $z \sgt 1$, 
$\Delta \ssimeq 200$ and the mean universal density approaches the critical 
cosmological density.

\smallskip
The structure of the host potential is determined in the following way.
First, the virial mass of the system $\Mv(t)$ is given by following the main
progenitor along the main branch of the merger tree.
The stellar mass $\Ms(t)$ is assigned according to the abundance matching 
relations of \citet{rodriguez17}.
Second, we determine the DZ profile of the halo, including the effect of
baryonic, as follows.
The concentration parameter in a DM-only scenario, $c_{2,\rm DMO}(\Mv, t)$, is 
obtained from the empirical relation of \citet{zhao09}.
We then consider the halo response to baryons following
\citet{freundlich20_cuspcore}, 
which provides empirically the ratio of the baryon-affected concentration and 
the DM-only concentration, $c_2/c_{2,\rm DMO}$, as a function of the 
stellar-to-halo-mass ratio $\Ms/\Mv$, and the inner logarithmic slope of the 
system $s_1=\dd\ln\rho/\dd\ln r$ at $r=0.01\Rv$.
Finally, we compute the DZ-profile parameters $(c,\alpha)$ using 
$c_2$ and $s_1$.
This procedure applies to both the host halo and the progenitors of satellites 
prior to infall.

\smallskip
The orbits of incoming satellites are initialized as follows.
We consider the infall locations to be isotropically distributed on the virial 
sphere of the host halo, for which we randomly draw an azimuthal angle 
$\phi$) from $[0,2\pi]$ and a cosine polar angle ($\cos\theta$) from $[0,1]$.
We assume that the orbital energy is the same as that of a circular orbit of 
of velocity $\Vv(t)$ at radius $\Rv(t)$, and randomly assign a circularity 
$\epsilon$ from a distribution, 
$\dd P/\dd \epsilon \seq \pi\sin(\pi\epsilon)/2$, which approximates the 
$\epsilon$ distribution of infalling satellites measured in cosmological 
simulations \citep{wetzel11,bosch17}.

\smallskip
We follow the orbits by treating satellites as point masses.
At each timestep, \SatGen solves the equations of motion
\be
\label{eq:EOM}
\ddot{\boldsymbol{r}} = -\nabla \Phi\ + a_{\rm DF}\, ,
\ee
where $\boldsymbol{r}$ is the position vector, $\Phi$ is the gravitational 
potential, and $a_{\rm DF}$ is the acceleration due to dynamical friction, 
modeled using the \citet{chandrasekhar43} formula as given in \equ{DF} and
\equ{H}.
%\be
%\label{eq:DF}
%\aDF = -4\pi G^2 m \ln\Lambda\, \rho(\boldsymbol{r}) F(<V_{{\rm rel}})
%\frac{\boldsymbol{V}_{{\rm rel}}}{V_{{\rm rel}}^3}\,.
%\ee
%Here $m$ is the instantaneous satellite mass, $\ln\Lambda$ is the Coulomb 
%logarithm, which we take to be $\ln\Lambda = \ln(\Mv/m)$, 
%$\boldsymbol{V}_{{\rm rel}}$ is the relative velocity of the satellite with 
%respect to the host, and 
%$F(<V_{{\rm rel}})$ is the fraction of local host particles contributing to DF.
%For simplicity, we assume that the velocity distributions of all of the host 
%components are Maxwellian and isotropic such that
%\be
%\label{eq:VelocityDistribution}
%F(<V_{{\rm rel}})= {\rm erf}(X) - \frac{ 2 X }{\sqrt{\pi}}e^{-X^2}\, ,
%\ee
%where $X\equiv V_{{\rm rel}} / (\sqrt{2}\sigma)$, 
%with $\sigma_i(\boldsymbol{r})$ the one-dimensional velocity dispersion of the 
%host.

\smallskip
We model the tidal mass loss using
\be
\label{eq:MassLossRate}
\dot{m} = -A \frac{m(>\ell_{\rm t})}{\tdyn(r)},
\ee
where we have introduced a fudge parameter $A$ as the stripping 
efficiency to encapsulate uncertainties in the definition of the tidal radius.
That is, the timescale on which stripping occurs is the local dynamical time 
$\tdyn(r) = \sqrt{3\pi/16G\bar{\rho}(r)}$ divided by $A$, 
with $\bar{\rho}(r)$ the average density of the host system within radius $r$.
We use $A\seq 0.55$ following the calibration by \citet{green21} 
from simulations.
The mass loss over a timestep $\Delta t$ is then given by 
$\Delta m =  \dot{m}\, \Delta t $.

\smallskip
To keep track of DF heating, we register the work done by DF on a satellite 
at each step, or equivalently the orbital energy change at each step,
\be
\label{eq:EnergyLoss}
\Delta W(t+\Delta t) = E(t) - E(t+\Delta t)\, ,
\ee
Note that the orbital energy $E$ at time $t+\Delta t$ includes the contribution
from the stripped mass $\Delta m$, which is assumed to be on
the same orbit of the satellite that it used to belong to.

\smallskip
The structural evolution of satellites in response to tidal mass loss, heating,
and re-virialization, is modeled using the empirical tidal tracks from
simulations \citep{penarrubia10}.
Note that the tidal track is conditioned on the initial structure of the 
satellites, which is important for capturing the difference in DF heating due 
to a compact satellite versus a diffuse one.
In the current study we do not explicitly include baryons within the 
satellites.

\smallskip
\citet{jiang21_satgen} used the model to study 
satellites of Milky-Way sized hosts, making it emulate
simulations of bursty or smooth star formation 
and experimenting with a disc potential in the host halo.
They found that the model reproduces the observed satellite statistics 
in the Milky Way and M31 reasonably well.
Different physical recipes make a difference in satellite abundance and spatial
distribution at the 25\% level, not large enough to be distinguished by 
current observations given the halo-to-halo variance. 
The MW/M31 disc depletes satellites by ${\sim} 20\%$ and has only a subtle 
effect of diversifying the internal structure of satellites, which may
be important for alleviating certain small-scale problems. 
We do not explicitly include in the current study
a central baryonic component.
%\adr{Fangzhou: true?}

%\be
%\begin{aligned}
%q(i,j)
%&=(-1)^j \frac{i!}{j!(1-j)!} , \quad i\geq j\geq 0 \\
%&=0 , \quad {\rm otherwise} .
%\end{aligned}
%\ee

%%%%%%%%%%%%%%%%%%%%%%%%%%% D
\section{N-body simulations of cusp heating and its relaxation}
\label{sec:app_N-body}

To test the impact of cusp heating by satellites and the following relaxation,
we run idealized N-body simulations, each with a host halo and a single merging
satellite.
At the beginning of the simulation, the two haloes are set in equilibrium, 
spherically symmetric and with an isotropic velocity dispersion.
In this case, the phase-space density of the particles is determined by
the particle specific energy $E$ and its radial distance from the halo center
$r$.
%In general, the phase-space distribution of N-body particles, $f$, is a 
%function of the particle specific energy $E$, angular momentum $L$, and radial 
%distance from the center $r$.
%For our purpose, we impose another assumption that the particle velocity 
%distribution
%is isotropic so that $f$ only depends on $E$ and $r$. 
%More general consideration s of
%anisotropic models have also been studied extensively, e.g. 
%with constant anisot
%ropy \citep{Henon:1973A,Kent:1982AJ},
%the Osipkov-Merritt model \citep{Osipkov:1979pv,Osipkov:1979sv,Merritt:1985} 
%with radial-dependent anisotropy. The anisotropy in velocity distribution 
%will have some minor effects on the tidal
%mass loss of the satellite and the dynamical friction effect, but these details
%are beyond the scope of this paper.
The halo density profile can be written as
\be
\rho(r) \seq m_{\textrm{p}} \int\! f(E) d^3 \mathbf{v}
\seq 4\pi\sqrt{2} m_{\textrm{p}} \int_0^{\Psi} \sqrt{\Psi-\mathcal{E}}
f(\mathcal{E}) d \mathcal{E},
\label{eq:rho_f}
\ee
where $\Psi \seq \Phi_0 \!-\! \Phi$, with $\Phi$ the gravitational potential 
and $\Phi_0$ its value at the boundary of the system, which we set at $4\Rv$.
The energy per unit mass is $\mathcal{E}\seq \Psi \!-\! (1/2)v^2$.
Given the density profile, the gravitational potential can be derived from the 
Poisson equation. For a realistic stationary halo,
$\Psi$ is a monotonically decreasing function of $r$, so $\rho$ can be  
written as a function of $\Psi$. Taking the derivative of both sides of 
\equ{rho_f} with respect to $\Psi$, one gets
\begin{equation}
\frac{d \rho}{d \Psi}=\sqrt{8}\pi m_{\textrm{p}}\int_0^{\Psi}
                 \frac{f(\mathcal{E})}{\sqrt{\Psi-\mathcal{E}}}\,d \mathcal{E},
\label{eq:drho}
\end{equation}
where $m_{\rm p}$ is the particle mass.
The above equation can be solved to give the Eddington's inversion formula 
\citep{eddington16}
\begin{equation}
f(\mathcal{E})=\frac{1}{\sqrt{8}\pi^2 m_{\textrm{p}}}
\frac{\mathrm{d}}{\mathrm{d}\mathcal{E}}
\int_0^{\mathcal E}\frac{\mathrm{d}\Psi}{\sqrt{\mathcal{E}-\Psi}}
\frac{\mathrm{d}\rho}\,{\mathrm{d}\Psi}.
\label{eq:eddington}
\end{equation}

\smallskip
The particle positions and velocities are randomly drawn from the density 
profile, \equ{rho_f}
and the velocity (energy) distribution, \equ{eddington}. For the host halo, 
an NFW profile is used with a sharp truncation at $4\,\Rv$. 
For the satellites, the density profile, either NFW profile
(compact satellite) or Burkert profile (diffuse satellite), is truncated 
exponentially at the virial radius following \cite{kazantzidis05} to roughly 
account for tidal truncation.

\smallskip
After generating the stationary halos, the satellite is put at the 
apocenter of an orbit with specified orbital parameters, e.g., 
the circularity and total energy. The system is then evolved using the
public N-body SPH code {\small GADGET}-2 \citep{springel05}.

\smallskip
The centers of the host and satellite are identified by searching for the 
most-bound particle, the particle that has the most negative total energy 
$E_i \seq \Phi(r_i) \!+\! (1/2)\,v_i^2$ within the corresponding halo. 
For the satellite, the bound mass is computed at each snapshot using an
iterative un-binding algorithm \citep{bosch18}.

%%%%%%%%%%%%%%%%%%%%%%%%%%% E
\section{CuspCore - An Analytic Model for DM Response to Outflows}
\label{sec:app_CuspCore}

%\adr{Jonathan: please elaborate.}

\smallskip
\citet{freundlich20_cuspcore}\footnote{Available for
application in\\ https://github.com/Jonathanfreundlich/CuspCore .}
 presents 
a simple analytic model for the response of a dissipationless spherical system 
to an instantaneous mass change at its center. 
It has been applied there to the formation of flat cores in low-mass 
dark-matter haloes and the origin of ultra-diffuse galaxies (UDGs) 
from outflow episodes driven by supernova feedback, but it is applicable for 
any rapid changes in the central mass. Here we use it for the dark-matter 
response to AGN-driven central gas ejection.
This model generalizes an earlier simplified analysis of an isolated shell 
\citep{dutton16b} into a system with continuous density, velocity and potential
profiles. 

\smallskip
The DM response is divided into two steps: 
an instantaneous change of potential at constant velocities due to a given 
rapid mass loss (or mass gain),
followed by energy-conserving relaxation to a new Jeans equilibrium. 
The halo profile is modeled by the two-parameter Dekel-Zhao profile described 
in \se{app_Dekel-Zhao}, using the analytic expressions for the 
associated potential and kinetic energies at equilibrium.
The way energy conservation is applied in the second stage of this model is not 
formally justified in the case of shell crossing, so its validity as an 
approximation should be based on testing against simulations.
In \citet{freundlich20_cuspcore}, the model has been tested against NIHAO 
cosmological zoom-in simulations, 
where it successfully predicts the evolution of the inner DM profile 
between successive snapshots in about 75\% of the cases, failing mainly in 
merger situations when the system strongly deviates from Jeans equilibrium.

\smallskip
The energy per unit mass of a shell at radius $r_{\rm i}$ in the initial halo
at Jeans equilibrium is the sum
\be
E_{\rm i}(r_i) = U(r_{\rm i};p_{\rm i}) + K(r_{\rm i};p_{\rm i}) \, , 
\ee
where $U(r_{\rm i};p_{\rm i})$ and $K(r_{\rm i};p_{\rm i})$ are functional 
forms for the potential and kinetic energies per unit mass, which depend on 
the parameters $p_{\rm i}$ that
characterize the initial halo density profile. 
We use the DZ profile with the parameters $c$ and $\alpha$, for which the 
potential $U(r_{\rm i};p_{\rm i})$ is given by \equ{U_DZ},
% plus a possible contribution of a baryonic component, 
and the kinetic energy
$K(r_{\rm i};p_{\rm i})$ derives from \equ{sigmarm_DZ}, stemming from Jeans 
equilibrium. For the two energies we may consider an additional baryonic 
component, characterized by additional parameters. 
In the temporary state immediately after the instantaneous mass change by $m$ 
(where $m \slt 0$ for an outflow and $m \sgt 0$ for an inflow), 
the energy becomes
\be
E_{\rm t}(r_{\rm i}) = U(r_{\rm i};p_{\rm i})-\frac{G m}{r_{\rm i}} + 
K(r_{\rm i};p_{\rm i})\, . 
\ee
After relaxation to the final Jeans equilibrium state of the halo, whose
profile is described by the parameters $p_{\rm f}$, the shell encompassing
a given mass has moved to a final radius $r_{\rm f}$ and its energy is
\be
E_{\rm f}(r_{\rm f}) = U(r_{\rm f};p_{\rm f})-\frac{G m}{r_{\rm f}} + 
K(r_{\rm f};p_{\rm f},m)\ , 
\ee
where the kinetic energy is again set by the Jeans equation but it also 
depends on the mass change $m$. 
The radius $r_{\rm f}$ is itself a function of the final 
parameters $p_{\rm f}$, given that the enclosed mass is constant,
$M(r_{\rm f};p_{\rm f})\seq M(r_{\rm i};p_{\rm i})$. 
%- [instead of the last paragraph] 
The assumed energy conservation during the relaxation phase corresponds to 
$E_{\rm f}(r_{\rm f}) \seq E_{\rm t}(r_{\rm i})$, which is solved numerically 
to obtain the final halo parameters $p_{\rm f}$. 
In practice, we minimize the difference 
$E_{\rm f}(r_{\rm f})-E_{\rm t}(r_{\rm i})$ 
for hundred shells equally spaced in 
$\log (r/R_{\rm v})$ from $-2$ to $0$ (thus giving more weight to central 
regions than linearly-spaced shells). 
The assumed energy conservation per shell that encompasses a given mass
is not formally justified in the case of shell crossing, and our use of it
is based on the success of this model in reproducing the results of 
simulations.
We refer to \citet{freundlich20_cuspcore}
for more details, and to \citet{freundlich19} 
for a brief presentation of the model.

%%%%%%%%%%%%%%%%%%% F
\section{A satellite with Central Baryons}  
\label{sec:app_tot}

\begin{figure*} % F1  4
\centering
\includegraphics[width=0.76\textwidth]
%{figs/test_evolve_CompactVersusDiffuse_NFWhost_tot.pdf}
{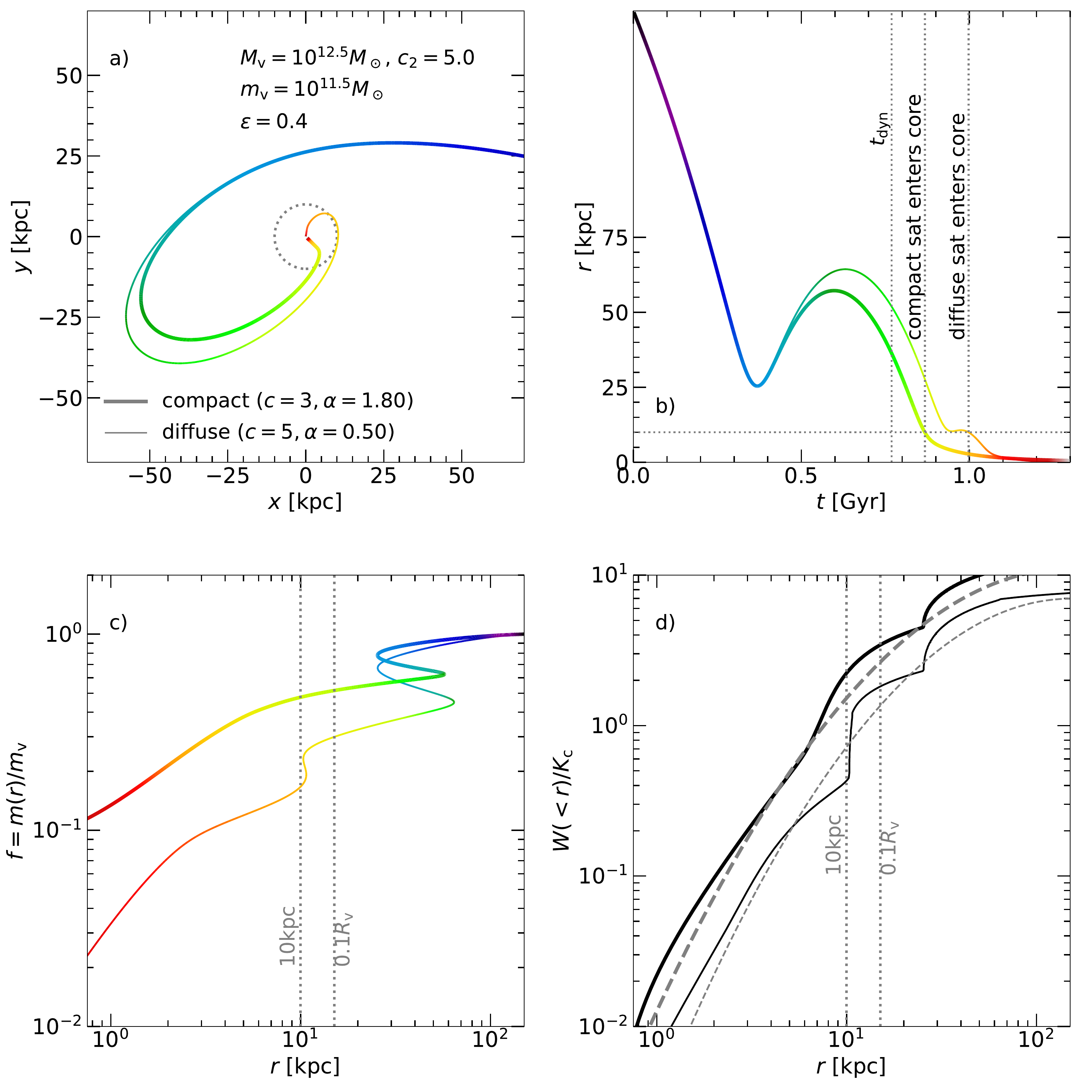}
\caption{
Semi-analytic SatGen simulations of single satellites, diffuse and compact,
similar to \fig{satgen_single}, but where the satellite profiles
are the best DZ fits to the {\it total} mass in the VELA pre and post
compaction galaxies from \fig{vela_prof}, including the baryons, with DZ
parameters
$(c,\alpha)\seq(5,0.5)$ and $(3,1.8)$,
respectively.
}
\label{fig:app_satgen_single_tot}
\end{figure*}

Complementing the main text, \fig{app_satgen_single_tot} is the analog of
\fig{satgen_single}, showing the result of a SatGen run with our fiducial NFW
host halo, but with the satellites following the more compact
DZ-profile fits to the VELA simulated galaxies using the {\it total}
mass including the baryons rather than the dark matter alone.
We learn that for the more compact satellites, as expected,
the penetrating mass to the host cusp is higher,
and the energy deposited in the cusp is higher accordingly. 
However, the difference is rather small, 
with $m/\mv \ssimeq 0.5$ compared to $0.4$,
and $\Wc/\Kc \ssimeq 2.5$ compared to $2$, for the compact satellites.

%%%%%%%%%%%%%%%%%%%% G
\section{A Steep-Cusp Host Halo}
\label{sec:app_steep}

\begin{figure*} % G1
\centering
\includegraphics[width=0.88\textwidth,trim={0.5cm 0.5cm 0.0cm 0.0cm},clip]
%{figs/a_vs_c_2Dhist_colored_f_ws_wc_steep_host_v11.pdf}
{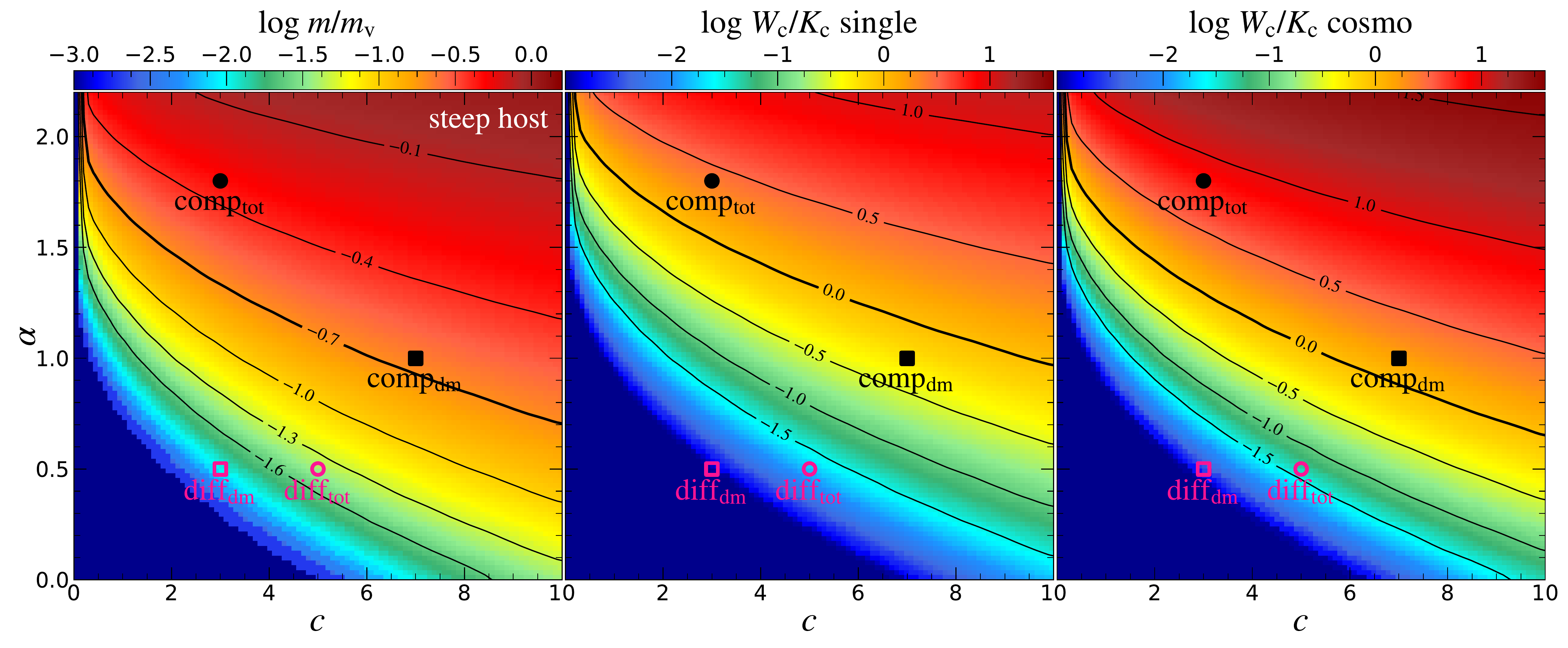}
\caption{
Same as \fig{frac} but for the fiducial steep-cusp host,
with $s_1\seq1.5$ and $c_2\seq 5$ in the DZ profile.
Shown are the
toy-model estimates for satellite penetration and energy deposited in the
host cusp by dynamical friction, as a function of the satellite compactness
via the Dekel-Zhao profile parameters of concentration and inner slope
$(c,\alpha)$.
For the steep-cusp host we read for the diffuse and compact satellites
respectively $m/\mv \ssim 0.01, 0.23$,
$\Wc/\Kc({\rm single}) \ssim 0.001, 0.56$
and $\Wc/\Kc({\rm cosmo}) \ssim 0.003, 1.53$.
The satellite stripping is stronger due to the steeper cusp, but the heating
by compact satellites is still significant during half a virial time.
}
\label{fig:app_frac_steep}
\end{figure*}

\begin{figure*} % G2
\centering
\includegraphics[width=0.74\textwidth]
%{figs/test_evolve_CompactVersusDiffuse_steephost.pdf}
{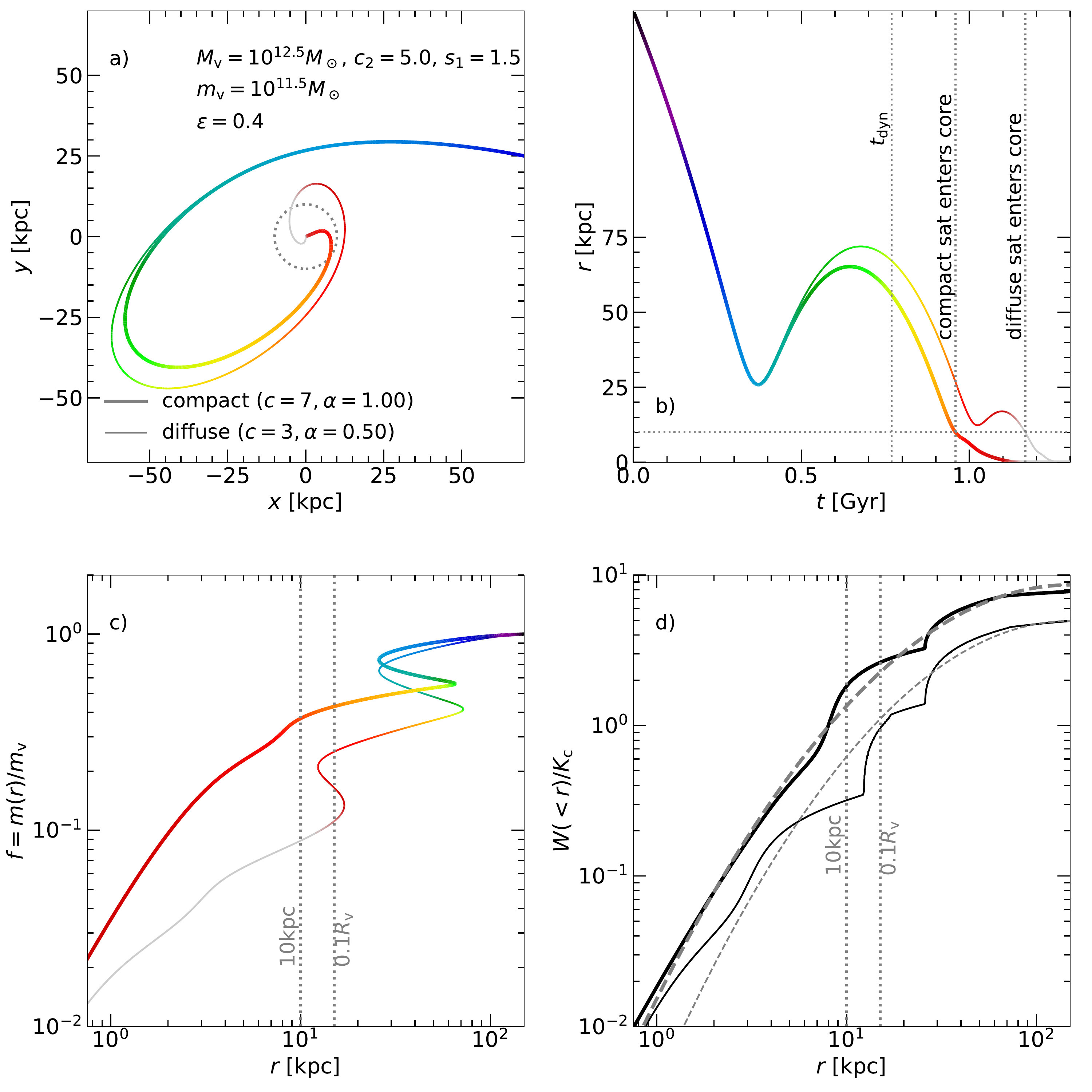}
\caption{
Same as \fig{satgen_single}, for an $\mv/\Mv\seq 0.1$ single satellite,
but for a steep-cusp host halo with DZ parameters $(c_2,s_1)\seq(5,1.5)$.
The penetrating satellite mass and the DF energy deposited in the steep cusp 
are similar to the case of an NFW halo, indicating significant heating by the
compact satellite and only partial heating by the diffuse 
satellite.
The functional fits (dashed) to be used by CuspCore are listed in \tab{fits}.
%\footnote{The functional fits used here are the same as
%before, $\log_{10} W(\slt r) \seq A-B\,[\log_{10}(r/\Rm)]^2$.
%For the steep-cusp host and a single satellite the parameters are
%\adr{$(A,B,\Rm) \seq (16.81,0.58,\Rv)$ and $(16.57,0.65,\Rv)$} % May 21 by FJ 
%for the compact and diffuse satellites respectively, where the energy is
%in $\msun\kpc^2\Gyr^{-2}$.}
}
\label{fig:app_satgen_steep_single}
\end{figure*}

\begin{figure*} % G3
\centering
\includegraphics[width=0.85\textwidth,trim={0.0cm 0.0cm 0.0cm 0.0cm},clip]
%{figs/compare_DFheating_CompactionVersusControl_WrKc_steep.pdf}
{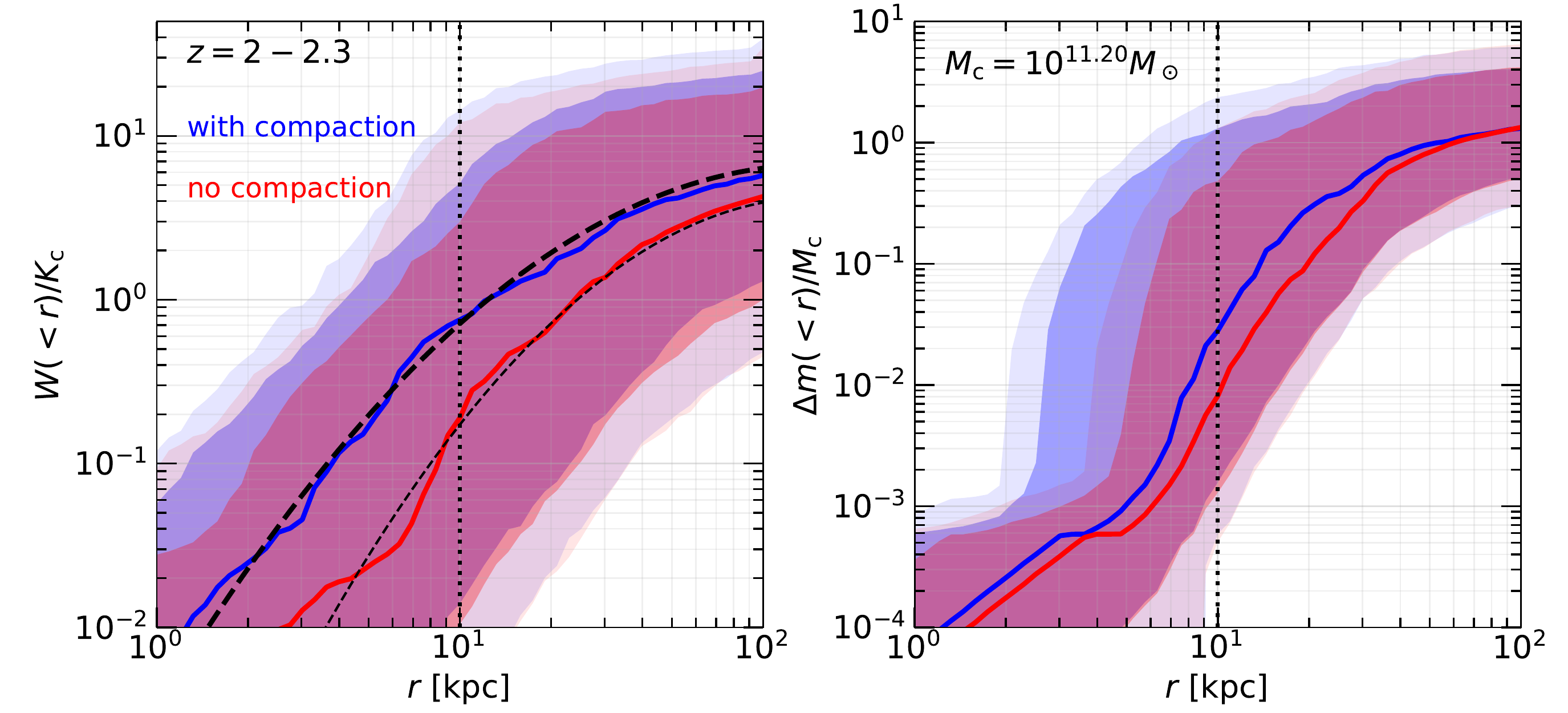}
\caption{
Same as \fig{satgen_cosmo} but for the steep-cusp host.
Shown are the results of a SatGen simulation of a cosmological sequence of
satellites during one halo virial time at $z\ssim 2$, 
where $\tvir \!\simeq\! 0.5\Gyr$.
The host halo starts with a DZ steep-cusp profile of
$(c_2,s_1)\seq(5,1.5)$.
The fits for the deposited energy by DF 
to be used in CuspCore is marked (dashed black), and listed in \tab{fits}.
%\footnote{The functional fits used here are the same as
%before, $\log_{10} W(\slt r) \seq A-B\,[\log_{10}(r/\Rm)]^2$.
%For the steep-cusp host and a cosmological sequence of satellites the 
%parameters are
%$(A,B,\Rm) \seq (16.7,0.7,\Rv)$ and $(16.5,1,\Rv)$ % May 20 by FJ 
%for the compact and diffuse satellites respectively, where the energy is
%in $\msun\kpc^2\Gyr^{-2}$.}
Most of the satellite mass is deposited near the outer edge of the cusp, with
only a small fraction of the mass penetrating to the inner cusp.
The energy deposited in the steep cusp by DF on a sequence of compact 
satellites is lower than that deposited in the NFW cusp (\fig{satgen_cosmo}) 
by a factor of $\sim\! 2.5$. However, it is 75\% of 
the cusp kinetic energy, implying heating also in the steep-cusp host.
The heating by diffuse satellites is weaker, only 20\% of the cusp energy.
}
\label{fig:app_satgen_steep_cosmo}
\end{figure*}

\begin{figure*} % G4 10
\centering
\includegraphics[width=0.44\textwidth]
%{figs/DF_0526steep_single_heat_vs_flat_rho_compact_1_3W.pdf}
{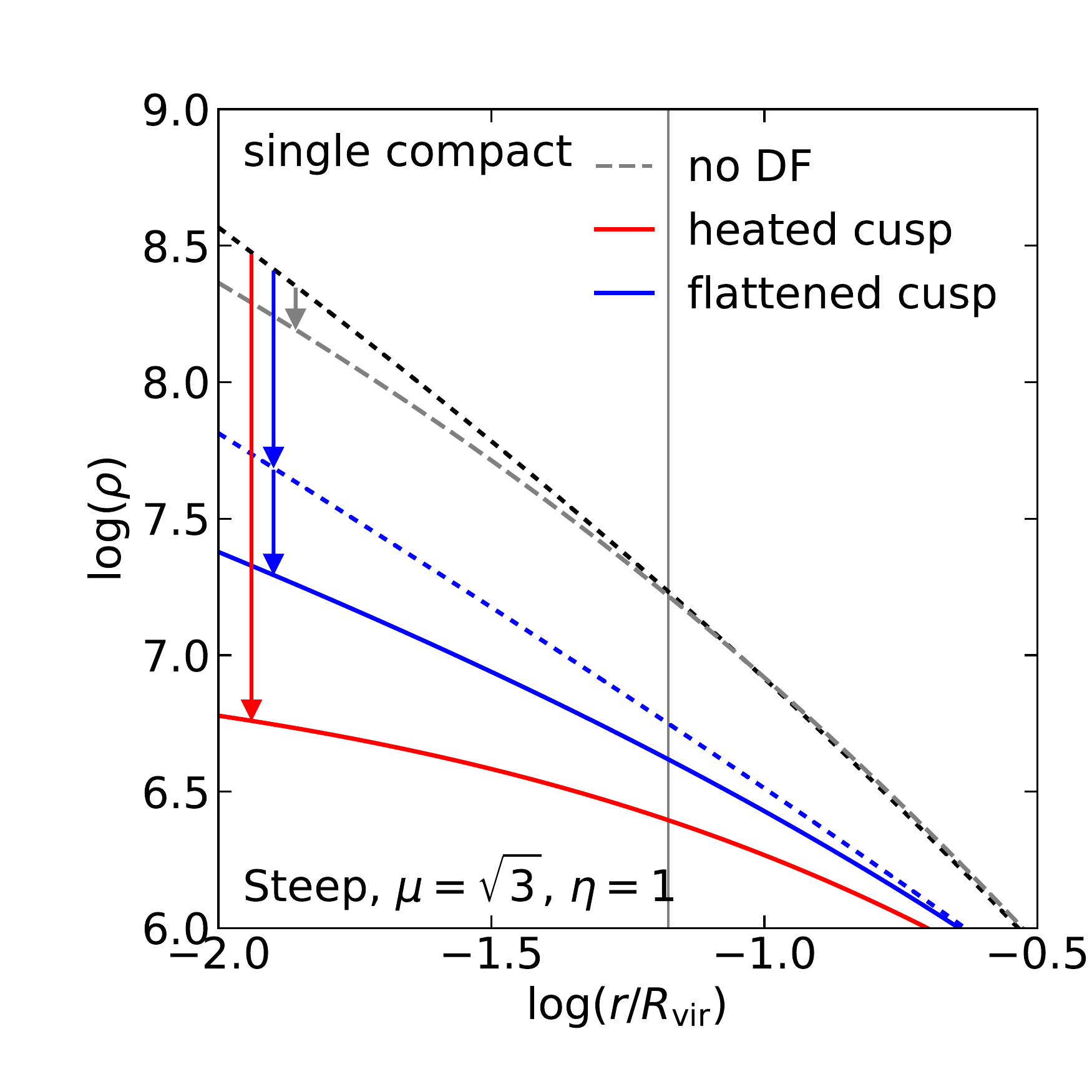}
\includegraphics[width=0.44\textwidth]
%{figs/DF_0526steep_single_heat_vs_flat_rho_diffuse_1_3W.pdf}
{figg4a.pdf}
\caption{
Density profiles as in \fig{outflow_single}, for DF heating by single
satellite, but for a steep-cusp initial host halo with
$(\c2h,\s1h)\seq(5,1.5)$ instead of NFW.
Here, in order to obtain a significant effect, the initial satellite mass
is $\mv\seq 0.1\mu\Mv$ with $\mu\seq \sqrt{3}$ and the outflow is with
$\eta\seq 1$, namely involving all the available gas of $10^{10.6}\msun$ 
(compared to $\mu\seq \sqrt{2}$ and $\eta\seq 0.5$ in \fig{outflow_single}).
The DF heating is 
based on the energy deposit profile by SatGen, bottom-left panel of
\fig{app_satgen_steep_single}, which turns out to be comparable to the energy
deposit of the NFW host.
We learn that the steep cusp is more resilient than the NFW cusp
both to DF heating and to outflows, requiring more massive satellite
and outflow for generating an extended core.
%\adr{Consider showing compact only.}
}
\label{fig:app_outflow_steep_single}
\end{figure*}

\begin{figure*} % G5 11
\centering
\includegraphics[width=0.33\textwidth,trim={1.2cm 0.9cm 0.7cm 0.9cm},clip]
%{figs/DF_0529steep_cosmo_heat_vs_flat_rho_compact_05_04W.pdf}
{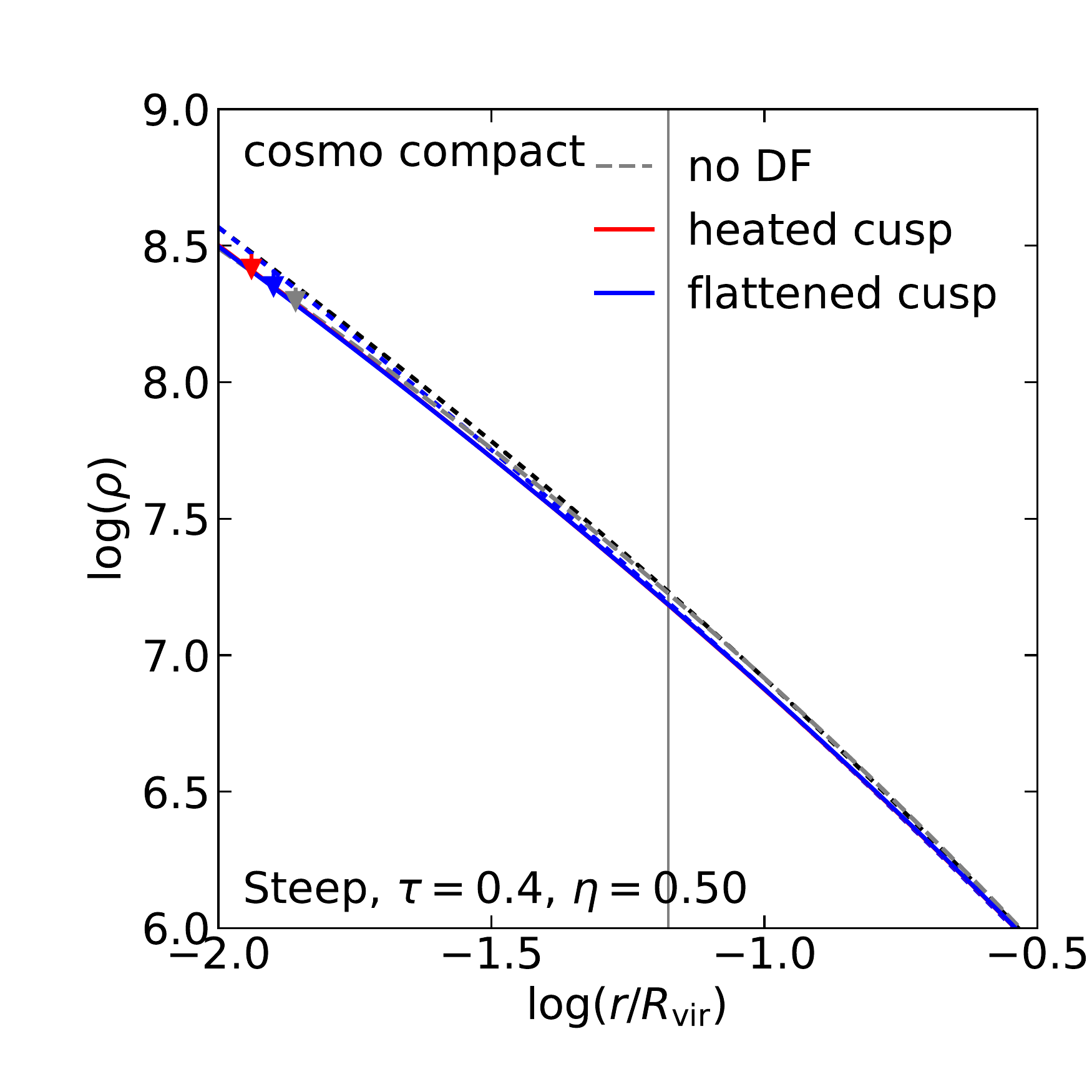}
\includegraphics[width=0.33\textwidth,trim={1.2cm 0.9cm 0.7cm 0.9cm},clip]
%{figs/DF_0529steep_cosmo_heat_vs_flat_rho_compact_2_04W.pdf}
{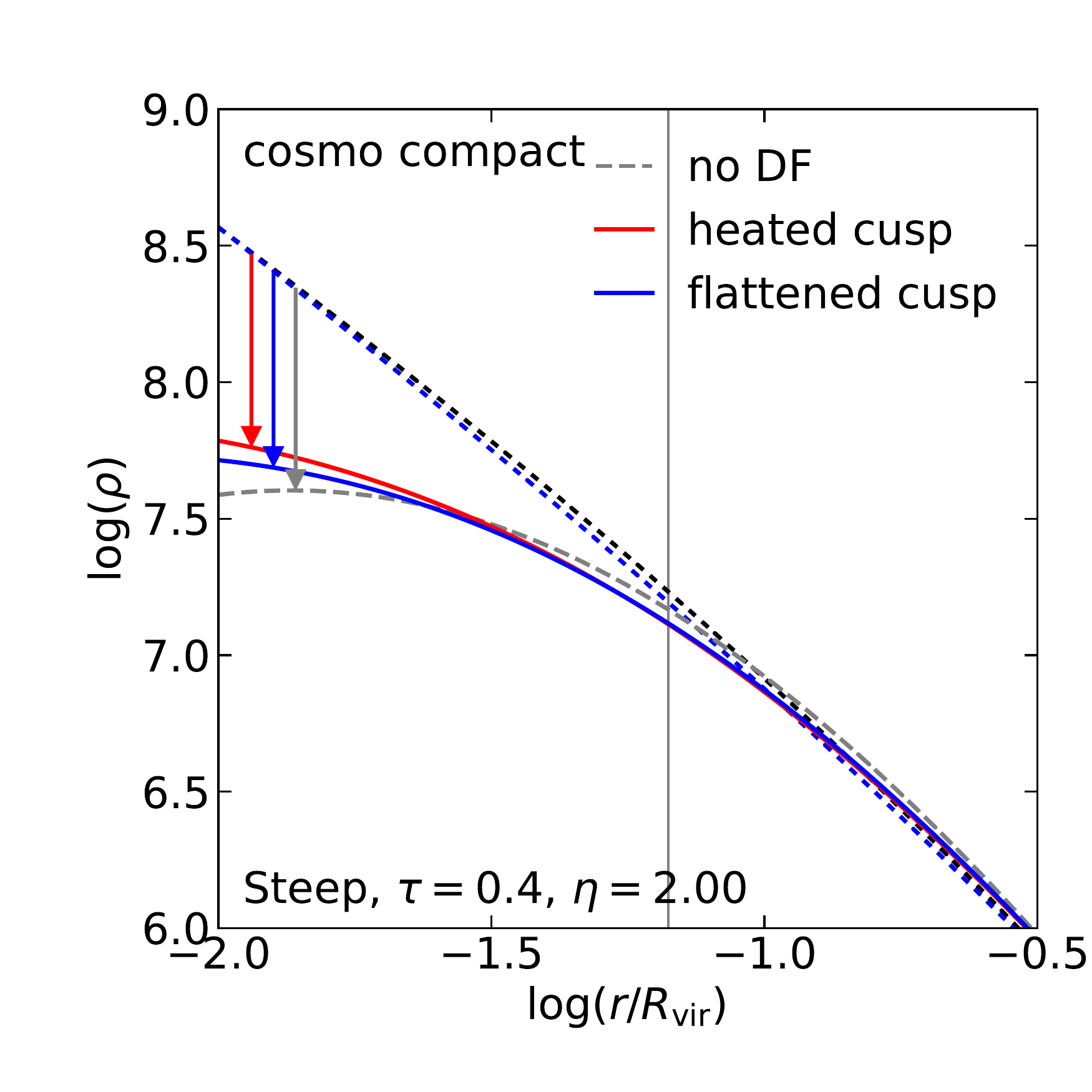}
\includegraphics[width=0.33\textwidth,trim={1.2cm 0.9cm 0.7cm 0.9cm},clip]
%{figs/DF_0528steep_cosmo_heat_vs_flat_rho_compact_2_1W_highfit.pdf}
{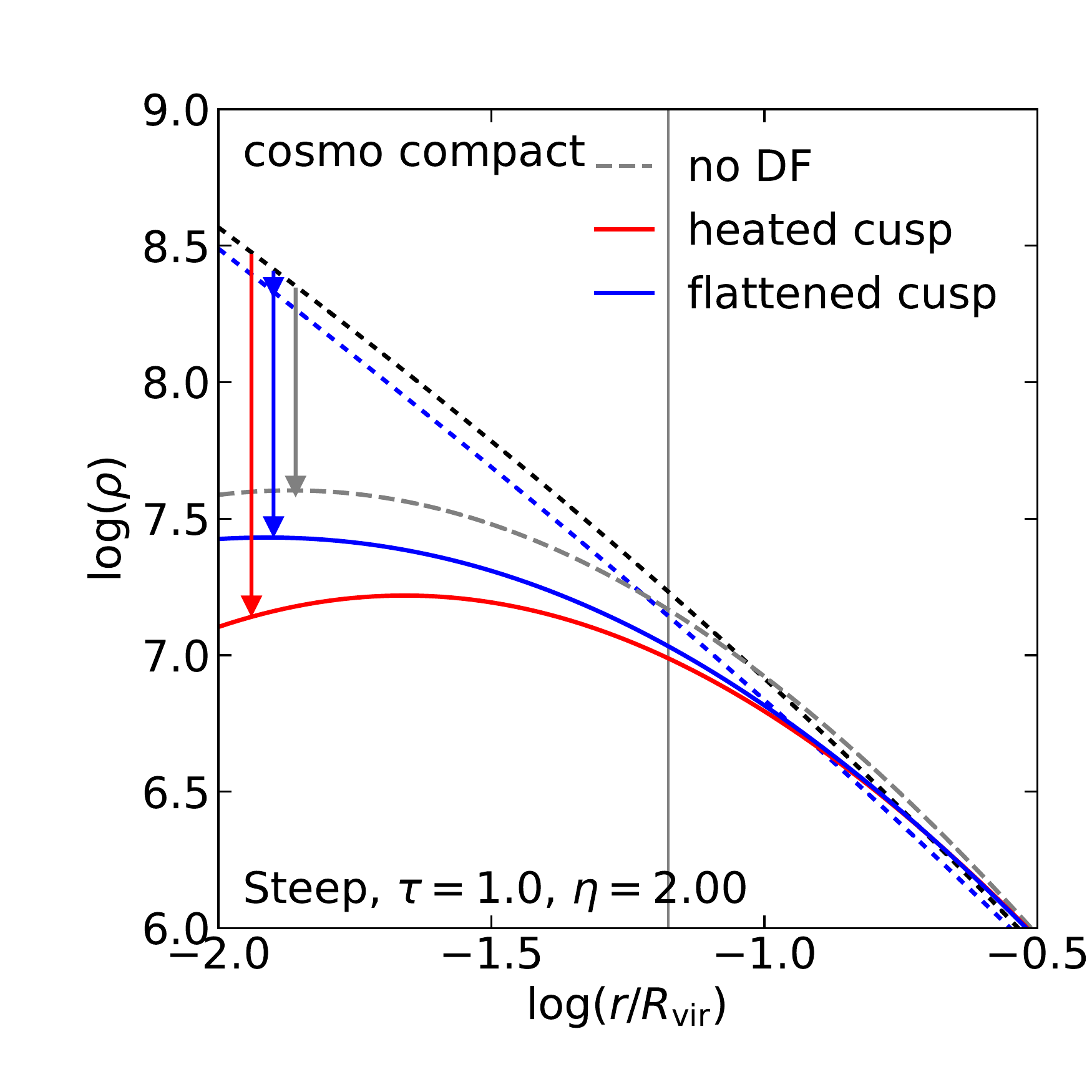}
\caption{
Same as \fig{outflow_cosmo}, for DF heating by a cosmological sequence of
satellites, but for a steep-cusp host halo with
$(\c2h,\s1h)\seq(5,1.5)$ instead of NFW.
The energy deposited in a virial time is based on the fit to the SatGen run 
shown in \fig{app_satgen_steep_cosmo}.
{\bf Left:} Compact satellites and outflows with the parameters that 
produced a core in the NFW cusp cause a negligible effect on the steep cusp.
{\bf Middle:} An outflow of $\eta \seq 2$ is sufficient for forming a moderate
core without DF heating. In this case, the DF actualy steepens the profile, 
making it a little harder to produce a core by inflow.
{\bf Right:} Similar but using a slightly flatter slope of
$K(<\! r)$ than produced by SatGen within the cusp, which enables a better
fit to the DZ profile and thus a convergence of CuspCore, leading to a flat
core after DF heating with $\tau\seq 1$. 
The higher value of $\tau$ can be interpreted as roughly representing 
the top $1/3$ of the random realizations drawn from the mass function and 
the circularity distribution of satellites during $0.4\,\tvir$. 
Alternatively it can be interpreted as the median energy during $\tvir$ or
or a longer duration.
}
\label{fig:app_outflow_steep_cosmo}
\end{figure*}

Here we show the same results that have been shown in the main text
for an NFW host halo with a moderately steep cusp,
but for a steep-cusp host of a DZ profile with $s_1\seq1.5$ and $c_2\seq 5$.

\smallskip
\Fig{app_frac_steep}, same as \fig{frac},
shows the toy model predictions as a function of the satellite compactness.
\Fig{app_satgen_steep_single}, same as \fig{satgen_single},
shows the results of a SatGen run with a single satellite.
\Fig{app_satgen_steep_cosmo}, same as \fig{satgen_cosmo},
refers to SatGen runs with a cosmological sequence of satellites.
\Fig{app_outflow_steep_single} and \fig{app_outflow_steep_cosmo},
same as \fig{outflow_single} and \fig{outflow_cosmo},
show the results of CuspCore for a single satellite and a cosmological sequence
of satellites, respectively.
The results for the steep-cusp host are discussed in comparison to the results
for the NFW host in the main text.

%%%%%%%%%%%%%%%%%%%%%
\section{Outflow with satellite mass added to host} % H

\begin{figure*} % H1
\centering
\includegraphics[width=0.33\textwidth,trim={1.2cm 0.9cm 0.7cm 0.9cm},clip]
%{figs/DF_0526NFW_single_msat_heat_vs_flat_rho_compact_1_1W.pdf}
{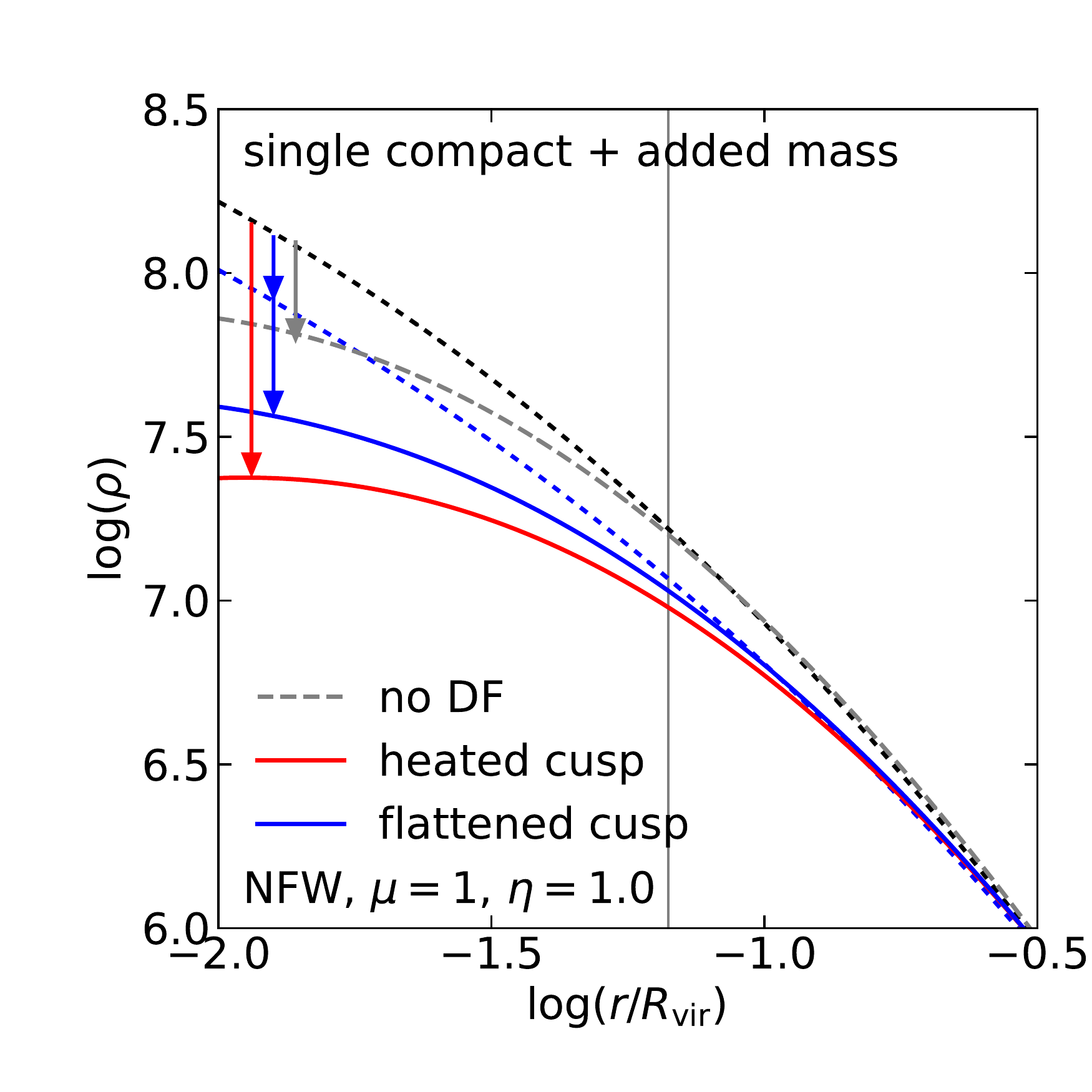}
\includegraphics[width=0.33\textwidth,trim={1.2cm 0.9cm 0.7cm 0.9cm},clip]
%{figs/DF_0526NFW_single_msat_heat_vs_flat_rho_compact_05_2W.pdf}
{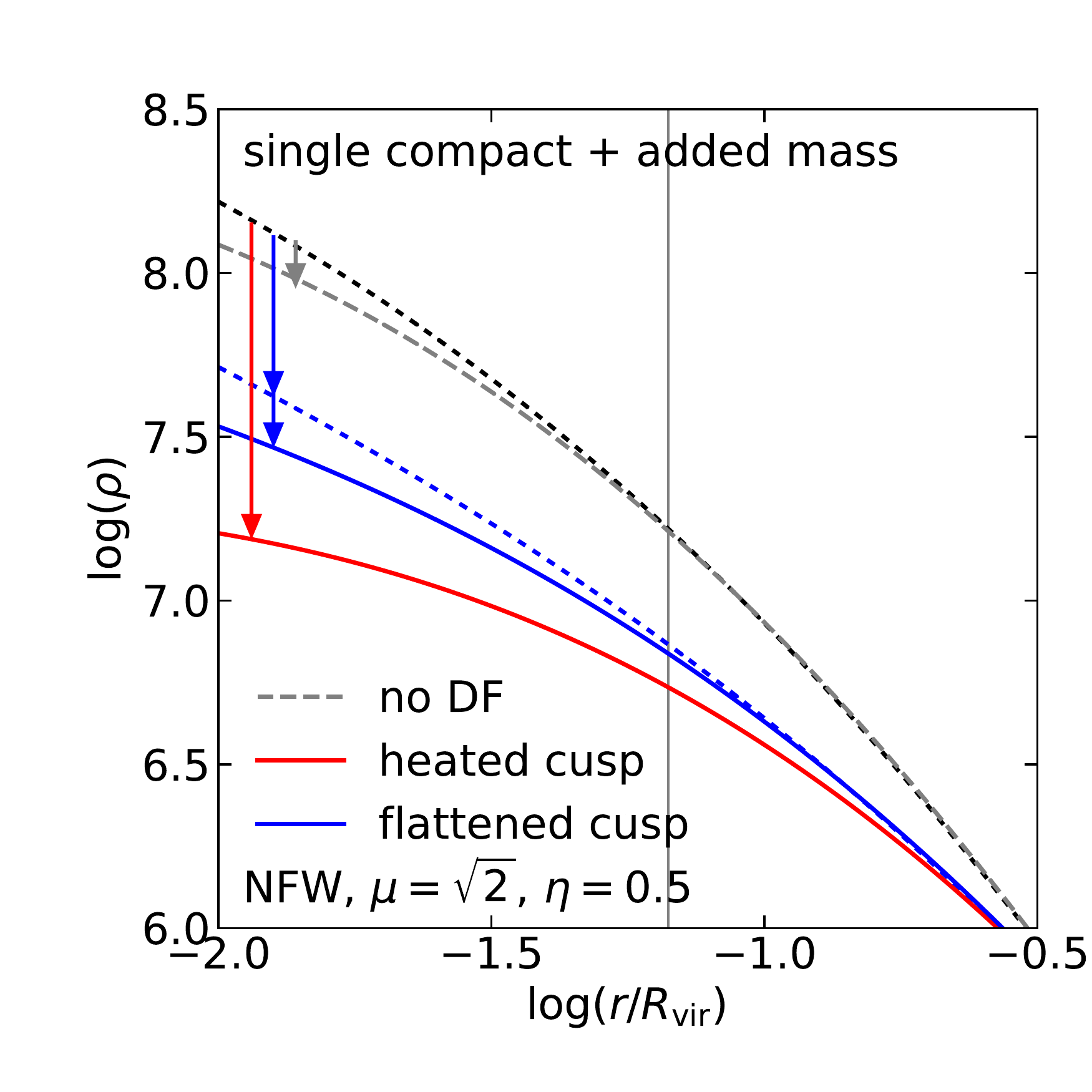}
\includegraphics[width=0.33\textwidth,trim={1.2cm 0.9cm 0.7cm 0.9cm},clip]
%{figs/DF_0526NFW_cosmo_msat_heat_vs_flat_rho_compact_05_04W.pdf}
{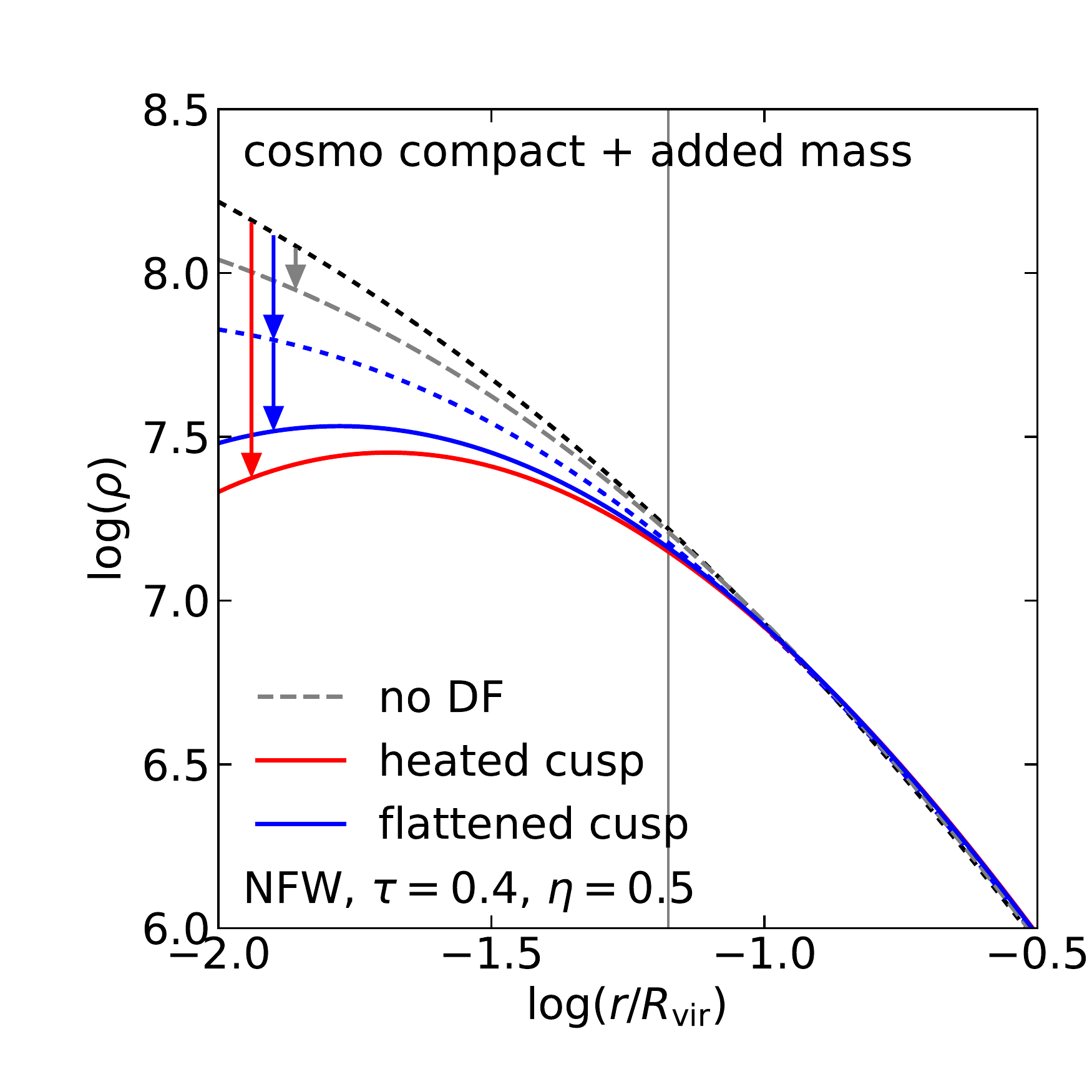}
\caption{
The effect of added satellite mass to the cusp.
Same as \fig{outflow_single} (left and middle) and \fig{outflow_cosmo}, 
for DF heating by a single compact satellite or a sequence of compact
satellites, in an NFW host, but with the mass of the satellite added to the 
host cusp where it is stripped or at the center,
based on the bottom-left panel of \fig{satgen_single} and the right panel of
\fig{satgen_cosmo}.
The difference from the results obtained without this additional mass in
\fig{outflow_single} and \fig{outflow_cosmo} is small. 
For the single satellite, the satellite mass deposited in the cusp
is comparable to the cusp mass, slightly steepening the fonal core.
For the sequence of satellites, the satellite mass is only $0.2\Mc$,
causing almost no change to the final core. 
This justifies ignoring the added mass in our main analysis.
}
\label{fig:app_outflow_msat_nfw_single_cosmo}
\end{figure*}

%%%%%%%%%%%%%%%%%%%% I
\section{Toy-Model $s_1$ and $c_2$}
\label{sec:app_s1c2}

\begin{figure*} % I1 
\centering
\includegraphics[width=0.90\textwidth,trim={0.5cm 0.0cm 0.0cm 0.0cm},clip]
%{figs/s1_vs_c2_2Dhist_colored_f_ws_wc_NFW_host_v11.pdf}
{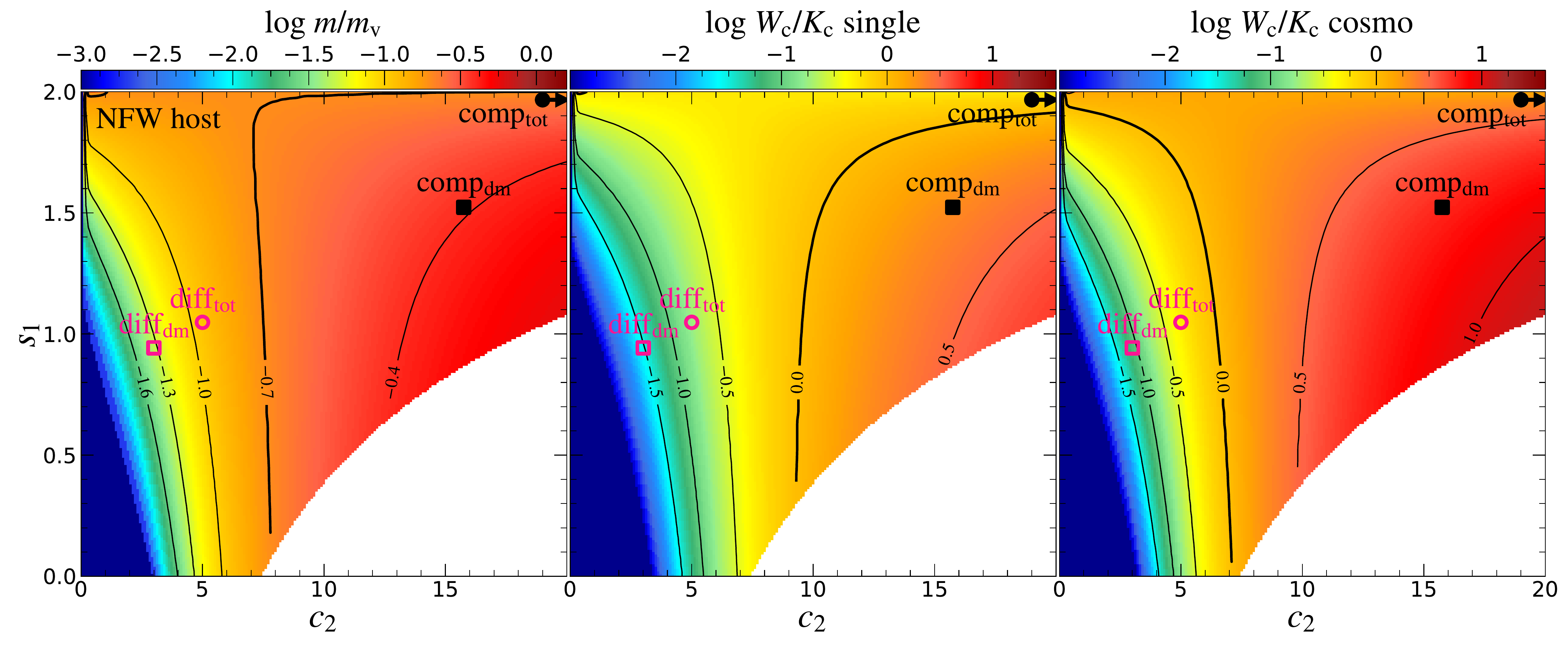}
\includegraphics[width=0.90\textwidth,,trim={0.5cm 0.0cm 0.0cm 0.0cm},clip]
%{figs/s1_vs_c2_2Dhist_colored_f_ws_wc_steep_host_v11.pdf}
{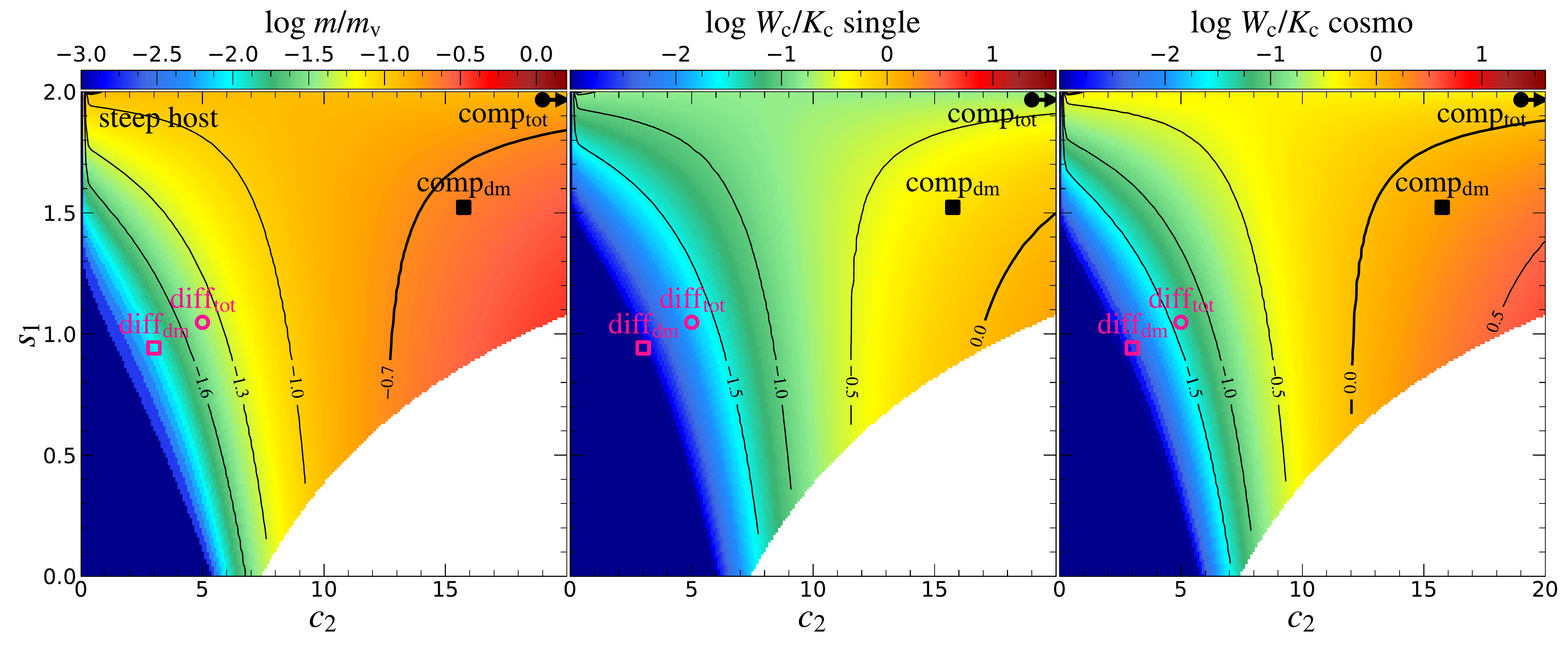}
\caption{ 
Toy-model estimates for satellite penetration and energy deposited in the
host cusp by dynamical friction, 
similar to \fig{frac}, but for the more accessible parameters
$(c_2,s_1)$ instead of the natural Dekel-Zhao parameters $(c,\alpha)$.
The concentration parameter $c_2$ refers to the radius where the
local log slope of
the density profile is $-2$ (as in the concentration of the NFW profile),
and $s_1$ is the minus the inner local log slope at $r=0.01\Rv$. 
The transformation between the two alternative pairs of parameters
is given in \equ{c2} to \equ{alpha}.
While there is a valid DZ profile for any values
of $c$ ($>\! 0$) and $\alpha$ ($<\!3$), a valid profile is not guaranteed for
arbitrary values of $c_2$ and $s_1$.
For example, $c_2 \rar \infty$ for $\alpha\seq 2$. The case of
compact total $(c,\alpha)\seq(3,1.8)$ has $(c_2,s_1)\seq (170,2.05)$, which is
outside the box of this figure.
We truncate the plot where $c\!>\!100$ (or where $\alpha \!\geq\! 2$).
}
\label{fig:app_frac}
\end{figure*}

\Fig{app_frac} is the analog of \fig{frac}, showing the toy-model estimates for
the satellite mass in the host cusp and the energy deposited there by dynamical
friction as a function of the satellite initial profile,
but here for the more accessible parameters
$(c_2,s_1)$ instead of the natural DZ parameters $(c,\alpha)$.
The conclusion is the same as in \fig{frac}.

%%%%%%%%%%%%%%%%%%%%%%%%%%%%%%%%%%%% J
\subsection{Orbit circularity as a function of redshift and halo mass}
\label{sec:app_circularity}

One may elaborate on the redshift and mass dependence of the satellite
orbit circularity as one of the factors in the tendency of
DF heating to be more effective at higher masses and higher redshifts.

\smallskip
The orbit is characterized at $\Rv$ by two parameters, e.g., energy and angular
momentum, or the velocity magnitude $V_{\rm in}$ and the circularity
$\epsilon \seq V_{\rm tan}/\Vv$.
For reference, the orbit eccentricity is $e^2 \seq 1- \epsilon^2$,
and the corresponding spin parameter is
\be
\lambda = \frac{V_{\rm tan}\Rv}{\sqrt{2}\Vv\Rv} = \frac{\epsilon}{\sqrt{2}}\, ,
\ee
independent of $V_{\rm in}$.
The orbit, and the effects of dynamical friction and tidal stripping,
depend in addition on $V_{\rm in}$, which for $\Mv\ssim 10^{12}\msun$ at
$z\ssim 2$ is roughly $V_{\rm in}/\Vv \simeq 1.15 \pm 0.15$
\citep[][Figs. 2,5,9]{wetzel11}.

\smallskip % epsilon(z,M)
According to the cosmological N-body simulations of \citet{wetzel11},
Figs.~5 and 8, for minor mergers of $\mv/\Mv \ssim 0.02$,
$\epsilon$ tends to decrease with increasing redshift,
where the average is
$\la \epsilon \ra \!\simeq\! 0.55$ and $0.45$
at $z\seq 0$ and $2.5$, respectively.
The corresponding median of the pericenter of the orbit is
roughly $\la r_{\rm peri} \ra /\Rv \!\simeq\! 0.24$ and $0.17$, respectively,
namely a deeper penetration at higher redshifts.
The distribution of $\epsilon$ is found to be approximately universal for a
given host halo mass when measured with respect to the non-linear
Press-Schechter mass $M_{\rm ps}(z)$.
For a given satellite mass, $\epsilon$ and $r_{\rm peri}$ tend to
decrease with increasing host halo mass.
These redshift and mass dependencies are in the desired sense,
but they are rather mild, possibly not sufficient by
themselves for explaining the redshift and mass dependencies of the
DM core phenomenon.

\smallskip % streams and spin
A qualitatively similar redshift dependence is obtained
from hydro cosmological simulations, via the analysis of the angular momentum
carried by the cosmic-web cold streams that build the galaxies at high
redshift, and contain the incoming satellites \citep{danovich15}.
According to their figure 15, the dominant stream carries on average $84\%$
of the angular-momentum inflow rate, and $64\%$ of the mass influx.
In order to relate the measured spin parameter to measured eccentricity,
we consider for an upper limit only one dominant stream, and
obtain $\lambda \ssim \epsilon/\sqrt{2} \sim 0.35$ for $\epsilon\seq0.5$.
For a lower limit, we consider three comparable streams with random
orientation and impact parameter,
and obtain $\lambda \ssim (\epsilon /\sqrt{2}) / \sqrt{3} \sim 0.2$ for
$\epsilon\seq0.5$.
We can therefore assume that the mean eccentricity measured by
\citet{wetzel11}, $\epsilon \sim 0.5$, would typically correspond to
$\lambda \ssim 0.3$.
% Danovich
However, from \citet[][Fig. 1]{danovich15}, at $\Rv$,
we read for the dark matter that $\lambda \!\simeq\! 0.13$ and it is not
varying with redshift,
while for the cold gas $\lambda \!\simeq\! 0.3, 0.2, 0.15$ at
$z \seq 1.5, 2.5. 3.5$, respectively.
Smaller $\lambda$ values are measured at higher redshifts also in their Fig.~7.
Similar results are obtained for cold gas at $z \seq 1.6 \sdash 3$ in their
Fig.~14, where $\lambda \ssim 0.2$.
This indicates that at $z\ssim 2.5$ one should assume $\epsilon \ssim 0.25$ for
the dark matter and $\epsilon \ssim 0.33$ for the cold gas.
These values are lower than the average value obtained for satellites
by \citet{wetzel11}, indicating more radial orbits and thus stronger dynamical
friction.
The redshift dependence in \citet{danovich15} for the cold gas
is stronger than in \citet{wetzel11}, but this may be balanced by the
weaker redshift dependence for the dark matter.

\smallskip
Qualitatively similar redshift and mass dependencies can be deduced from the
analysis of random Gaussian fluctuation fields by \citet{bbks86},
who predict that higher-sigma density peaks have lower $\lambda$ values and
more radial orbits.
%\adr{Search reference}.
This is consistent with the trends found in \citet{wetzel11} and
\citet{danovich15},
and with the core phenomenon being more pronounced at higher
redshifts and masses, being higher-sigma peaks.

%%%%%%%%%%%%%%%%%%%%%%%%%%%%%%%%%%%%%%%%%%%%%

\label{lastpage}
\end{document}